\documentclass[a4paper,11pt]{article}
\usepackage{jheppub}

\usepackage{array}
\usepackage{arydshln}

\usepackage[utf8]{inputenc}
\usepackage{setspace}
\usepackage{graphicx}
\graphicspath{ {./figures/} }
\usepackage{subcaption}
\usepackage{lineno}
\usepackage{braket}
\usepackage{physics}
\usepackage{amsmath,amsfonts,amssymb,amsthm,mathtools}
\usepackage{slashed}
\usepackage{comment}

\usepackage{booktabs}
\usepackage{multirow}

\usepackage{xcolor}
\usepackage{dsfont}

\usepackage[multiple]{footmisc} 

\usepackage{tikz} 
\tikzset{
  every node/.append style={
    execute at begin node={\everymath{\displaystyle}}
  }
}

\usepackage{pgfplots} 
\usepackage{pgfplotstable} 
\usetikzlibrary{fadings}
\usetikzlibrary{patterns}
\usetikzlibrary{shadows.blur}
\usetikzlibrary{shapes}

\makeatletter
\newcommand{\ogeneric}[2][0.7]{%
  \vphantom{\oplus}\mathpalette\o@generic{{#1}{#2}}%
}
\newcommand{\o@generic}[2]{\o@@generic#1#2}
\newcommand{\o@@generic}[3]{%
  \begingroup
  \sbox\z@{$\m@th#1\oplus$}%
  \dimen@=\dimexpr\ht\z@+\dp\z@\relax
  \savebox\tw@[\totalheight]{$\m@th#1\bigcirc$}%
  \makebox[\wd\z@]{%
    \ooalign{%
      $#1\vcenter{\hbox{\resizebox{\dimen@}{!}{\usebox\tw@}}}$\cr
      \hidewidth
      $#1\vcenter{\hbox{\resizebox{#2\dimen@}{!}{$#1\vphantom{\oplus}{#3}$}}}$%
      \hidewidth
      \cr
    }%
  }%
  \endgroup
}
\makeatother

\usepackage{graphicx}

\renewcommand\labelenumi{(\roman{enumi})}
\renewcommand\theenumi\labelenumi

\usepackage{pdfpages}
\usepackage{svg}

\title{Thermal Order in the Biconical Model}

\author{Michael Smolkin}
\author{and Lev Yung}

\affiliation{The Racah Institute of Physics, The Hebrew University of Jerusalem, \\ Jerusalem 91904, Israel \\}

\emailAdd{michael.smolkin@mail.huji.ac.il}
\emailAdd{lev.yung@mail.huji.ac.il}

\abstract{Thermal fluctuations are generally expected to destroy order and restore symmetries at sufficiently high temperatures. Recently, however, a family of scalar theories in $2+1$ dimensions was shown to exhibit $\mathbb Z_2$ symmetry breaking that persists to arbitrarily high temperatures using a variety of approaches, including the $\epsilon$-expansion, the FRG, and large-$N$ techniques. Although the similarities among these theories suggest that they belong to the same universality class, this connection has not been established explicitly. In this work, we fill this gap. Using large-$N$ methods, we analytically study the biconical vector model in a range of spacetime dimensions, including $2+1$, at leading and next-to-leading order in the $1/N$ expansion. We determine the RG flow, fixed-point structure, and the CFT data of the infrared theory, reproducing and extending previous results and thereby unifying the apparently distinct constructions within a common analytic framework. We derive the effective potential, establish its stable minima at zero and finite temperature, and demonstrate spontaneous $\mathbb{Z}_2$ symmetry breaking at arbitrarily high temperatures for large finite $N$.}

\begin{document}
\maketitle

\section{Introduction}

The interplay between symmetry and temperature lies at the heart of our understanding of phase transitions. In the conventional picture, thermal fluctuations drive systems toward increasingly symmetric phases by restoring spontaneously broken symmetries. This behavior reflects the familiar thermodynamic tendency for disordered phases to possess higher entropy than ordered ones. As the temperature increases, the entropic contribution to the free energy therefore typically favors the symmetric phase, leading to symmetry restoration. Although this paradigm underlies most theoretical descriptions and experimental observations, it is not universal.

A classic example is provided by Rochelle salt, whose symmetry is lowered upon heating. Below $-18^\circ\mathrm{C}$ the crystal has an orthorhombic structure, whereas in the temperature range $-18^\circ\mathrm{C}<T<+24^\circ\mathrm{C}$ it transforms into a monoclinic phase of lower symmetry~\cite{landau1980statistical}. Another well-known example is the Pomeranchuk effect in $^3\mathrm{He}$. For temperatures $T<0.32\,\mathrm{K}$ and within a certain pressure range, the solid phase possesses a larger entropy than the liquid phase, so that heating induces solidification rather than melting~\cite{Pomeranchuk:1950}.

Quantum Field Theory (QFT) also admits striking departures from this conventional paradigm. The first such example was discovered by Weinberg~\cite{weinberg1974gauge}, who showed perturbatively that an $O(N)\times O(N)$ scalar theory in four spacetime dimensions can undergo symmetry non-restoration: as the temperature increases, the theory evolves from a symmetric phase to one with spontaneously broken symmetry.

These examples illustrate that order and entropy need not always compete. The entropy of an ordered phase can exceed that of a more symmetric phase, or, more generally, the free-energy landscape can favor an ordered state through entropic effects. Consequently, increasing the temperature does not necessarily restore the symmetry.

Although these examples challenge the conventional expectation that heating restores symmetry, none demonstrates an ordered phase that survives to arbitrarily high temperatures. Indeed, both Rochelle salt and $^3\mathrm{He}$ display this unconventional behavior only within a finite region of their phase diagrams before eventually transitioning to more symmetric phases as the temperature is further increased. Weinberg's four-dimensional scalar model faces a different limitation. Because it is not asymptotically safe, its quartic couplings grow with the characteristic energy scale $E\sim T$, causing the theory to leave the perturbative regime and ultimately reach the Landau pole. Consequently, it cannot establish symmetry non-restoration in the asymptotically high-temperature limit. This naturally raises the question of whether there exist systems in which an ordered phase persists up to arbitrarily high temperatures.

In recent years, a class of relativistic, UV-complete QFTs exhibiting spontaneous symmetry breaking that persists to arbitrarily high temperatures has been introduced~\cite{Chai:2020zgq,Chai:2020onq,Liendo:2022bmv,Chaudhuri:2020xxb,Bajc:2020gpa,Chai:2021djc,Chaudhuri:2021dsq,Chai:2021tpt,Hawashin:2024dpp,Komargodski:2024zmt,SmolkinYung:2025parity,Bajc:2026ppk,Chaudhuri:2026}. By contrast, existing holographic constructions with thermal order have so far yielded only metastable ordered phases at nonzero temperature, while the true thermal equilibrium state in the holographic dual is always the disordered phase~\cite{Buchel:2020thm,Buchel:2020jfs,Buchel:2021ead,Buchel:2022zxl,Buchel:2023zpe,Buchel:2025cve,Buchel:2025jup}. Following the existing literature on persistent symmetry breaking (PSB), we will use the more explicit term Persistent Spontaneous Symmetry Breaking (PSSB) to emphasize that the symmetry breaking is spontaneous rather than explicit.

Most of the QFT constructions achieve UV completeness through conformal invariance \cite{Chai:2020zgq,Chai:2020onq,Liendo:2022bmv,Chaudhuri:2020xxb,Chai:2021djc,Chai:2021tpt,Hawashin:2024dpp,Komargodski:2024zmt}, asymptotic freedom, or asymptotic safety~\cite{Bajc:2020gpa,Chaudhuri:2021dsq,SmolkinYung:2025parity,Bajc:2026ppk}. Early examples include models with global symmetries formulated in fractional dimensions using the $\epsilon$-expansion~\cite{Chai:2020zgq,Chai:2020onq}, models in $3+1$ dimensions in the limit of infinitely many fields~\cite{Chaudhuri:2020xxb,Bajc:2020gpa,Chaudhuri:2021dsq}, and non-local models in $2+1$ dimensions~\cite{Chai:2021djc,Chai:2021tpt}.

The limitations associated with fractional dimensions, non-locality, and infinitely many fields were subsequently overcome. In $2+1$ dimensions, numerical studies based on the truncated Functional Renormalization Group (FRG)~\cite{Hawashin:2024dpp} together with an analytic large-$N$ analysis~\cite{Komargodski:2024zmt} established persistent breaking of an internal $\mathbb{Z}_2$ symmetry in a family of local, unitary, UV-complete Conformal Field Theories (CFTs) with a finite number of fields. Furthermore, \cite{SmolkinYung:2025parity} constructed an asymptotically safe $2+1$-dimensional QFT exhibiting PSSB of spacetime parity rather than an internal symmetry.

In $2+1$ dimensions, the Coleman--Mermin--Wagner theorem restricts PSSB to discrete symmetries by forbidding spontaneous breaking of continuous global symmetries at finite temperature. In contrast, in $3+1$ dimensions, PSSB of continuous global symmetries is possible. This phenomenon was first demonstrated in the infinite-field limit~\cite{Chaudhuri:2020xxb,Chaudhuri:2021dsq}, whereas the finite-field case was established by \cite{Bajc:2026ppk}, who showed that the asymptotically free model of \cite{Bajc:2020gpa} exhibits a Higgs phase that persists to arbitrarily high temperatures even with a finite number of fields. More recently, developments in lattice systems~\cite{Huang:2025gvi,Han:2025eiw,Andriolo:2026udg} have clarified the mechanism underlying entropic order: ordering one sector can increase the entropy available to another, thereby making the ordered phase thermodynamically favored even in the high-temperature limit.

In this paper, we analytically study the $O(N)\times \mathbb{Z}_2$ scalar biconical model in $2+1\le d<3+1$ dimensions at large but finite $N$. The model was originally introduced in the context of multicritical phenomena~\cite{Fisher_mcl:1974} and subsequently explored in~\cite{Vicari:2003}. Only recently was it recognized \cite{Chai:2020zgq,Chai:2020onq} that, in $2+1$ dimensions, it provides a promising candidate for realizing PSSB in a local, unitary, UV-complete relativistic QFT with a finite number of fields. Because the theory is strongly coupled, previous studies relied on the $\epsilon$-expansion \cite{Chai:2020zgq,Chai:2020onq} and on numerical FRG calculations in $2+1$ dimensions~\cite{Hawashin:2024dpp}. Both approaches provided strong evidence for the existence of an RG fixed point corresponding to an interacting CFT, which we refer to as the critical biconical model, in which the $\mathbb{Z}_2$ symmetry remains spontaneously broken at arbitrarily high temperatures.

A closely related scalar field construction was introduced in \cite{Komargodski:2024zmt}. There, the authors coupled a free scalar field to the strongly coupled critical $O(N)$ vector model, formulated in terms of the original $O(N)$ vector field and the Hubbard--Stratonovich $O(N)$ singlet, in such a way that the resulting theory possesses an $O(N)\times\mathbb{Z}_2$ symmetry. In the large, but finite, $N$ limit, the interaction becomes weakly relevant, allowing for a controlled perturbative expansion in $1/N$. In $2+1$ dimensions, the critical model also contains a nearly marginal sextic interaction, which ultimately plays a crucial role in establishing PSSB of the $\mathbb{Z}_2$ symmetry. While the similarities between this construction and the models studied in \cite{Chai:2020zgq,Chai:2020onq,Hawashin:2024dpp} suggest that they belong to the same universality class, this connection has never been established explicitly. In this work, we use large-$N$ methods to show that the critical biconical model and the construction of \cite{Komargodski:2024zmt} are equivalent, thereby unifying various approaches in the literature within a common analytic framework.

We formulate the biconical model in the UV and determine its IR dynamics from the effective potential at leading and next-to-leading order in the $1/N$ expansion, working entirely with the original field content. Imposing scale invariance identifies the critical biconical model as an IR fixed point. We show that its critical data agree with those of the conformal theory constructed in~\cite{Komargodski:2024zmt}, establishing the latter as an alternative formulation of the same fixed point. We further classify all relevant and weakly irrelevant deformations of the critical theory, derive their beta functions, and determine the critical surface and relevant scaling exponents, finding good agreement with numerical FRG results. Finally, the effective-potential analysis establishes the stability of the critical theory and shows that its $\mathbb{Z}_2$ field acquires a nonzero thermal expectation value that persists to arbitrarily high temperatures.

The paper is organized as follows. Section~\ref{sec: model} introduces the biconical model and computes the field anomalous dimensions at criticality. Section~\ref{sec: LO} derives the effective potential at leading order (LO) in the $1/N$ expansion and identifies the critical biconical model, while Section~\ref{sec: d3_chi6_addition} explains the necessity of the $\chi^6$ interaction in $d=3$. Section~\ref{sec: NLO} extends the effective-potential analysis to next-to-leading order (NLO) and identifies the critical theory at this order. Section~\ref{sec: PSSB} establishes spontaneous $\mathbb{Z}_2$ breaking at finite temperature. Section~\ref{sec: beta_functions} enlarges the coupling space to include all relevant and weakly irrelevant deformations at the IR fixed points and derives the corresponding beta functions and relevant scaling exponents. Section~\ref{sec: discussion} summarizes our results and outlines future directions. 

\section{The Biconical Model} \label{sec: model}
Consider a $d$-dimensional Euclidean field theory containing an $N$-component real scalar field $\phi_i$ and a single real scalar field $\chi$. We begin with the following $O(N)\times\mathbb{Z}_2$ symmetric
action defined at a UV cutoff scale $\Lambda$
\begin{equation} \label{eq: bicon_uv_action}
S[\phi,\chi]= \int d^d x \left\{
\frac12(\partial\phi)^2
+\frac12(\partial\chi)^2
+\frac{\lambda_\phi}{8N}\,(\phi^2)^2
+\frac{\lambda_{\phi\chi}}{4N}\,\phi^2\,\chi^2
+\frac{\lambda_\chi}{8N}\,\chi^4 \right\},
\end{equation}
where $3\leq d <4$, and $\lambda_\phi, \lambda_{\phi\chi}$, and $ \lambda_{\chi}$ are classically relevant quartic couplings, each of mass dimension $(4-d)$. Since we are ultimately interested in the deep IR, well below both the cutoff and the scales set by these couplings, we avoid introducing an additional dimensionful parameter and set $\lambda_a=\bar{\lambda}_a\Lambda^{4-d}$, with $\bar{\lambda}_a$ held fixed as both $N$ and $\Lambda$ are taken large. We restrict our analysis to $\lambda_\phi > 0$, while boundedness of the UV potential requires $\lambda_\chi \geq 0$ and $\lambda_{\phi\chi} \geq -\sqrt{\lambda_\phi \lambda_\chi}$. The quadratic mass terms are tuned to zero.

The model was studied perturbatively in $d=4-\epsilon$ at leading order
in the $\epsilon$-expansion~\cite{Chai:2020onq, Chai:2020zgq}. The beta functions were found to admit an IR fixed point, identifying a CFT with a stable potential, whose thermal state exhibits PSSB of the $\mathbb{Z}_2$ symmetry for $N\geq 18$. Away from $d\approx 4$, the theory is strongly coupled in the IR, and the perturbative expansion is no longer parametrically controlled. Instead, we determine the infrared behavior of the theory using the large-$N$ expansion. We will show that the IR physics of \eqref{eq: bicon_uv_action} is governed by a CFT when the UV couplings are tuned to satisfy $\lambda_{\phi\chi}^2 = \lambda_\phi \lambda_\chi$. At large finite $N$, for $\lambda_{\phi\chi} > 0$, this CFT is the critical $O(N+1)$ model, whereas for $\lambda_{\phi\chi} < 0$, it is the critical biconical model, whose thermal state we demonstrate exhibits PSSB. At $d=3$, we will see that the RG flow toward the biconical IR fixed point generates a $\chi^6$ interaction. 

Our approach to the IR physics of \eqref{eq: bicon_uv_action} is the following. Given the UV action, we compute the effective potential $V(\phi,\chi)$ for IR field values $\phi, \chi \ll \Lambda^{\frac{d-2}{2}}$, thereby determining the vacuum structure of the theory.  At zero temperature, we require the IR effective potential to be scale-invariant. This imposes necessary conditions on the UV couplings for \eqref{eq: bicon_uv_action} to flow to a scale-invariant IR theory.
A more comprehensive analysis is presented in Section \ref{sec: beta_functions}, where we enlarge the coupling space to include all relevant and weakly irrelevant couplings at the IR fixed points and derive their beta functions by requiring the effective potential to satisfy the Callan--Symanzik equation. The simultaneous zeros of these beta functions determine the critical couplings of the biconical IR fixed point, while linearizing the RG flow around it yields the scaling exponents. Under the standard assumption that scale invariance implies conformal invariance, this fixed-point theory is a CFT.

\subsection{Anomalous dimensions} \label{sec: anomalous_dimensions}

We conclude this section by computing the anomalous dimensions of the $\phi$ and $\chi$ fields in the IR. The self-energy of the $\phi_i$ field up to NLO in $1/N$ is given by the following diagrams
\begin{center}
\tikzset{every picture/.style={line width=0.75pt}} 

\begin{tikzpicture}[x=0.75pt,y=0.75pt,yscale=-1,xscale=1]

\draw    (297.43,53.62) -- (242.59,53.4) ;
\draw    (261.89,37.85) .. controls (262.91,25.51) and (278.37,27.53) .. (278.34,37.85) ;
\draw    (278.34,37.85) .. controls (277.32,50.18) and (261.86,48.17) .. (261.89,37.85) ;
\draw  [dash pattern={on 0.84pt off 2.51pt}]  (261.89,37.85) -- (249.59,53.71) ;
\draw  [dash pattern={on 0.84pt off 2.51pt}]  (278.34,37.85) -- (289.01,52.99) ;
\draw    (391.95,53.9) -- (313.21,53.9) ;
\draw    (327.4,35.9) .. controls (328.3,24.88) and (342.11,26.68) .. (342.09,35.9) ;
\draw    (342.09,35.9) .. controls (341.18,46.91) and (327.37,45.12) .. (327.4,35.9) ;
\draw  [dash pattern={on 0.84pt off 2.51pt}]  (327.4,35.9) -- (322.82,54.54) ;
\draw  [dash pattern={on 0.84pt off 2.51pt}]  (374.74,35.9) -- (379.63,53.82) ;
\draw    (360.05,35.9) .. controls (360.95,24.88) and (374.76,26.68) .. (374.74,35.9) ;
\draw    (374.74,35.9) .. controls (373.83,46.91) and (360.02,45.12) .. (360.05,35.9) ;
\draw  [dash pattern={on 0.84pt off 2.51pt}]  (360.05,35.9) -- (342.09,35.9) ;
\draw    (483.57,53.9) -- (404.84,53.9) ;
\draw    (414.54,35.9) .. controls (415.45,24.88) and (429.26,26.68) .. (429.23,35.9) ;
\draw    (429.23,35.9) .. controls (428.32,46.91) and (414.51,45.12) .. (414.54,35.9) ;
\draw  [dash pattern={on 0.84pt off 2.51pt}]  (414.54,35.9) -- (409.96,54.54) ;
\draw  [dash pattern={on 0.84pt off 2.51pt}]  (472.76,35.9) -- (477.65,53.82) ;
\draw    (458.07,35.9) .. controls (458.98,24.88) and (472.79,26.68) .. (472.76,35.9) ;
\draw    (472.76,35.9) .. controls (471.85,46.91) and (458.04,45.12) .. (458.07,35.9) ;
\draw  [dash pattern={on 0.84pt off 2.51pt}]  (436.3,35.26) -- (429.23,35.9) ;
\draw    (436.3,35.26) .. controls (437.21,24.24) and (451.02,26.04) .. (451,35.26) ;
\draw    (451,35.26) .. controls (450.09,46.27) and (436.28,44.48) .. (436.3,35.26) ;
\draw  [dash pattern={on 0.84pt off 2.51pt}]  (458.07,35.9) -- (451,35.26) ;
\draw    (104.84,53.57) -- (74.63,53.57) ;
\draw    (79.8,28.39) .. controls (80.99,13.95) and (99.09,16.31) .. (99.06,28.39) ;
\draw    (99.06,28.39) .. controls (97.87,42.83) and (79.77,40.47) .. (79.8,28.39) ;
\draw  [dash pattern={on 0.84pt off 2.51pt}]  (89.33,38.69) -- (89.74,53.57) ;
\draw    (189.89,53.31) -- (175.53,53.31) ;
\draw  [draw opacity=0] (189.89,53.31) .. controls (189.02,51.7) and (188.52,49.85) .. (188.52,47.88) .. controls (188.52,41.58) and (193.63,36.47) .. (199.93,36.47) .. controls (206.24,36.47) and (211.35,41.58) .. (211.35,47.88) .. controls (211.35,49.8) and (210.87,51.61) .. (210.04,53.19) -- (199.93,47.88) -- cycle ; \draw   (189.89,53.31) .. controls (189.02,51.7) and (188.52,49.85) .. (188.52,47.88) .. controls (188.52,41.58) and (193.63,36.47) .. (199.93,36.47) .. controls (206.24,36.47) and (211.35,41.58) .. (211.35,47.88) .. controls (211.35,49.8) and (210.87,51.61) .. (210.04,53.19) ;  
\draw    (224.4,53.19) -- (210.04,53.19) ;
\draw  [dash pattern={on 0.84pt off 2.51pt}]  (210.04,53.19) -- (189.89,53.31) ;
\draw    (159.41,53.57) -- (129.2,53.57) ;
\draw [color={rgb, 255:red, 255; green, 0; blue, 0 }  ,draw opacity=1 ]   (134.37,42.93) .. controls (135.56,28.49) and (153.66,30.84) .. (153.63,42.93) ;
\draw [color={rgb, 255:red, 255; green, 0; blue, 0 }  ,draw opacity=1 ]   (153.63,42.93) .. controls (152.44,57.36) and (134.34,55.01) .. (134.37,42.93) ;
\draw    (240.07,102.57) -- (175.85,102.32) ;
\draw [line width=3]  [dash pattern={on 7.88pt off 4.5pt}]  (189.89,102.53) .. controls (191.99,77.02) and (225.88,75.02) .. (226.87,102.51) ;
\draw    (105.71,101.73) -- (75.5,101.73) ;
\draw    (80.67,76.55) .. controls (81.86,62.11) and (99.96,64.47) .. (99.93,76.55) ;
\draw    (99.93,76.55) .. controls (98.74,90.99) and (80.64,88.63) .. (80.67,76.55) ;
\draw  [dash pattern={on 0.84pt off 2.51pt}]  (90.61,86.73) -- (91.01,101.6) ;
\draw    (159.39,102.57) -- (129.17,102.57) ;
\draw [color={rgb, 255:red, 255; green, 0; blue, 0 }  ,draw opacity=1 ]   (134.34,91.93) .. controls (135.53,77.49) and (153.63,79.84) .. (153.6,91.93) ;
\draw [color={rgb, 255:red, 255; green, 0; blue, 0 }  ,draw opacity=1 ]   (153.6,91.93) .. controls (152.41,106.36) and (134.31,104.01) .. (134.34,91.93) ;

\draw (483.05,35.5) node [anchor=north west][inner sep=0.75pt]    {$+\ ...$};
\draw (297.94,35.66) node [anchor=north west][inner sep=0.75pt]    {$+$};
\draw (390.79,35.71) node [anchor=north west][inner sep=0.75pt]    {$+$};
\draw (13,32.79) node [anchor=north west][inner sep=0.75pt]    {$\Sigma _{\phi }( p) =$};
\draw (110,35.66) node [anchor=north west][inner sep=0.75pt]    {$+$};
\draw (226.13,35.66) node [anchor=north west][inner sep=0.75pt]    {$+$};
\draw (73.23,39.77) node [anchor=north west][inner sep=0.75pt]    {$i$};
\draw (101.08,16) node [anchor=north west][inner sep=0.75pt]    {$j$};
\draw (175.8,39.52) node [anchor=north west][inner sep=0.75pt]    {$i$};
\draw (207.69,27.69) node [anchor=north west][inner sep=0.75pt]    {$i$};
\draw (160.48,35.66) node [anchor=north west][inner sep=0.75pt]    {$+$};
\draw (54.02,85.5) node [anchor=north west][inner sep=0.75pt]    {$\equiv $};
\draw (160.07,84.66) node [anchor=north west][inner sep=0.75pt]    {$+$};
\draw (109.97,84.66) node [anchor=north west][inner sep=0.75pt]    {$+$};
\draw (239.05,80.5) node [anchor=north west][inner sep=0.75pt]    {$+\ O\left( N^{-2}\right) .$};

\end{tikzpicture}
\end{center}
In the first line, the first diagram is $O(N^0)$, while all subsequent diagrams are $O(N^{-1})$. Black and red solid lines denote the $\phi$ and $\chi$ propagators, respectively. For clarity, we formally split each local $(\phi^2)^2$ vertex into two $\phi^2$ factors joined by a dotted line. Its endpoints are understood to coincide in spacetime, so the dotted line carries no propagation and serves only to display the $O(N)$ index-contraction structure. For example, the first and third diagrams both contain a loop, but only the first has a free index sum; in the third, the loop index is fixed by the external field. The thick dashed line in the rainbow diagram denotes the geometric resummation of the bubble chain,
\begin{center}
    \tikzset{every picture/.style={line width=0.75pt}} 

\begin{tikzpicture}[x=0.75pt,y=0.75pt,yscale=-1,xscale=1]

\draw  [dash pattern={on 0.84pt off 2.51pt}]  (108.77,29.27) -- (85.3,29.27) ;
\draw [line width=3]  [dash pattern={on 7.88pt off 4.5pt}]  (60.54,29.28) -- (17.01,29.28) ;
\draw    (149.01,29.52) .. controls (150.04,16.96) and (165.79,19.01) .. (165.76,29.52) ;
\draw [shift={(165.76,29.52)}, rotate = 90.16] [color={rgb, 255:red, 0; green, 0; blue, 0 }  ][fill={rgb, 255:red, 0; green, 0; blue, 0 }  ][line width=0.75]      (0, 0) circle [x radius= 1.34, y radius= 1.34]   ;
\draw [shift={(149.01,29.52)}, rotate = 274.71] [color={rgb, 255:red, 0; green, 0; blue, 0 }  ][fill={rgb, 255:red, 0; green, 0; blue, 0 }  ][line width=0.75]      (0, 0) circle [x radius= 1.34, y radius= 1.34]   ;
\draw    (165.76,29.52) .. controls (164.73,42.08) and (148.98,40.03) .. (149.01,29.52) ;
\draw    (219.01,29.52) .. controls (220.04,16.96) and (235.79,19.01) .. (235.76,29.52) ;
\draw [shift={(235.76,29.52)}, rotate = 90.16] [color={rgb, 255:red, 0; green, 0; blue, 0 }  ][fill={rgb, 255:red, 0; green, 0; blue, 0 }  ][line width=0.75]      (0, 0) circle [x radius= 1.34, y radius= 1.34]   ;
\draw [shift={(219.01,29.52)}, rotate = 274.71] [color={rgb, 255:red, 0; green, 0; blue, 0 }  ][fill={rgb, 255:red, 0; green, 0; blue, 0 }  ][line width=0.75]      (0, 0) circle [x radius= 1.34, y radius= 1.34]   ;
\draw    (235.76,29.52) .. controls (234.73,42.08) and (218.98,40.03) .. (219.01,29.52) ;
\draw    (250,29.52) .. controls (251.03,16.96) and (266.78,19.01) .. (266.75,29.52) ;
\draw [shift={(266.75,29.52)}, rotate = 90.16] [color={rgb, 255:red, 0; green, 0; blue, 0 }  ][fill={rgb, 255:red, 0; green, 0; blue, 0 }  ][line width=0.75]      (0, 0) circle [x radius= 1.34, y radius= 1.34]   ;
\draw [shift={(250,29.52)}, rotate = 274.71] [color={rgb, 255:red, 0; green, 0; blue, 0 }  ][fill={rgb, 255:red, 0; green, 0; blue, 0 }  ][line width=0.75]      (0, 0) circle [x radius= 1.34, y radius= 1.34]   ;
\draw    (266.75,29.52) .. controls (265.71,42.08) and (249.97,40.03) .. (250,29.52) ;
\draw  [dash pattern={on 0.84pt off 2.51pt}]  (149.01,29.52) -- (134.77,29.52) ;
\draw  [dash pattern={on 0.84pt off 2.51pt}]  (180,29.52) -- (165.76,29.52) ;
\draw  [dash pattern={on 0.84pt off 2.51pt}]  (219.01,29.52) -- (204.77,29.52) ;
\draw  [dash pattern={on 0.84pt off 2.51pt}]  (250,29.52) -- (235.76,29.52) ;
\draw  [dash pattern={on 0.84pt off 2.51pt}]  (280.99,29.52) -- (266.75,29.52) ;
\draw    (321.01,29.52) .. controls (322.04,16.96) and (337.79,19.01) .. (337.76,29.52) ;
\draw [shift={(337.76,29.52)}, rotate = 90.16] [color={rgb, 255:red, 0; green, 0; blue, 0 }  ][fill={rgb, 255:red, 0; green, 0; blue, 0 }  ][line width=0.75]      (0, 0) circle [x radius= 1.34, y radius= 1.34]   ;
\draw [shift={(321.01,29.52)}, rotate = 274.71] [color={rgb, 255:red, 0; green, 0; blue, 0 }  ][fill={rgb, 255:red, 0; green, 0; blue, 0 }  ][line width=0.75]      (0, 0) circle [x radius= 1.34, y radius= 1.34]   ;
\draw    (337.76,29.52) .. controls (336.73,42.08) and (320.98,40.03) .. (321.01,29.52) ;
\draw    (352,29.52) .. controls (353.03,16.96) and (368.78,19.01) .. (368.75,29.52) ;
\draw [shift={(368.75,29.52)}, rotate = 90.16] [color={rgb, 255:red, 0; green, 0; blue, 0 }  ][fill={rgb, 255:red, 0; green, 0; blue, 0 }  ][line width=0.75]      (0, 0) circle [x radius= 1.34, y radius= 1.34]   ;
\draw [shift={(352,29.52)}, rotate = 274.71] [color={rgb, 255:red, 0; green, 0; blue, 0 }  ][fill={rgb, 255:red, 0; green, 0; blue, 0 }  ][line width=0.75]      (0, 0) circle [x radius= 1.34, y radius= 1.34]   ;
\draw    (368.75,29.52) .. controls (367.71,42.08) and (351.97,40.03) .. (352,29.52) ;
\draw  [dash pattern={on 0.84pt off 2.51pt}]  (321.01,29.52) -- (306.77,29.52) ;
\draw  [dash pattern={on 0.84pt off 2.51pt}]  (352,29.52) -- (337.76,29.52) ;
\draw  [dash pattern={on 0.84pt off 2.51pt}]  (382.99,29.52) -- (368.75,29.52) ;
\draw    (382,29.52) .. controls (383.03,16.96) and (398.78,19.01) .. (398.75,29.52) ;
\draw [shift={(398.75,29.52)}, rotate = 90.16] [color={rgb, 255:red, 0; green, 0; blue, 0 }  ][fill={rgb, 255:red, 0; green, 0; blue, 0 }  ][line width=0.75]      (0, 0) circle [x radius= 1.34, y radius= 1.34]   ;
\draw [shift={(382,29.52)}, rotate = 274.71] [color={rgb, 255:red, 0; green, 0; blue, 0 }  ][fill={rgb, 255:red, 0; green, 0; blue, 0 }  ][line width=0.75]      (0, 0) circle [x radius= 1.34, y radius= 1.34]   ;
\draw    (398.75,29.52) .. controls (397.71,42.08) and (381.97,40.03) .. (382,29.52) ;
\draw  [dash pattern={on 0.84pt off 2.51pt}]  (412.99,29.52) -- (398.75,29.52) ;

\draw (65,26) node [anchor=north west][inner sep=0.75pt]    {$=$};
\draw (115,23) node [anchor=north west][inner sep=0.75pt]    {$+$};
\draw (185,23) node [anchor=north west][inner sep=0.75pt]    {$+$};
\draw (34,10) node [anchor=north west][inner sep=0.75pt]    {$\vec{p}$};
\draw (65,47) node [anchor=north west][inner sep=0.75pt]    {$=\frac{1}{N} \lambda _{\phi } -\frac{1}{N} \lambda _{\phi }^{2} \Pi _{m^{2}}( p) +\frac{1}{N} \lambda _{\phi }^{3} \Pi _{m^{2}}^{2}( p) +\dotsc =\frac{1}{N}\frac{\lambda _{\phi }}{1+\lambda _{\phi } \Pi _{m^{2}}( p)} ,$};
\draw (287,23) node [anchor=north west][inner sep=0.75pt]    {$+$};
\draw (421,23) node [anchor=north west][inner sep=0.75pt]    {$+\ \dotsc $};

\end{tikzpicture}
\end{center}
where $\Pi_{m^2}(p)$ denotes a single bubble, also known as the fish diagram,
\begin{equation} \label{eq: bubble}
    \Pi_{m^2}(p) = \frac{1}{2}\int_k \frac{1}{k^2 + m^2} \frac{1}{(k+p)^2 + m^2}, \qquad \int_k \equiv \int^\Lambda \frac{d^d k}{(2\pi)^d}.
\end{equation}
In all subsequent diagrams, we disregard the dotted lines to avoid clutter. The quadratic mass terms are fine-tuned to zero in \eqref{eq: bicon_uv_action},
so the bubble is massless. Including local counterterm contributions ($\text{c.t.}$), the $\phi$ self-energy reads
\begin{equation} \label{eq: sigma_phi_nlo}
    \begin{split}
    \Sigma_\phi(p)=\,  
\frac{\lambda_\phi}{2}\,\int_q \frac{1}{q^2 + \Sigma_\phi(q)} + &\frac{\lambda_{\phi\chi}}{2N}\,\int_q \frac{1}{q^2 +\Sigma_\chi(q)}\\  +&\frac{1}{N}\int_q \frac{1}{(p+q)^2}
\frac{\lambda_\phi}{1+\lambda_\phi \,\Pi_0(q)} + \text{c.t.} + O(N^{-2}).
    \end{split}
\end{equation}
We choose the local mass and wavefunction counterterms so that the self-energy contains neither momentum-independent divergent terms $\propto \Lambda^2$ nor finite terms $\propto p^2$. In particular, the first two integrals are independent of the external momentum $p$ and are removed by the local mass counterterm. Consider the third integral. For momenta $q\ll \Lambda$ and couplings scaling as $\lambda_a \sim \Lambda^{4-d}$, the bubble and the resummed bubble chain approach their IR limits 
\begin{equation}
    \Pi_{0}(q) = \frac{1}{2}\int_k \frac{1}{k^2} \frac{1}{(k+q)^2} \rightarrow A_0^{-1} q^{d-4},\qquad \frac{\lambda_\phi}
    {1+\lambda_\phi\Pi_0(q)} \rightarrow
    \frac{1}{\Pi_0(q)},
\end{equation}
where 
\begin{equation}
    \quad A_0
=\frac{2(4\pi)^{d/2}\Gamma(d-2)}
{\Gamma\left(2-\frac d2\right)\Gamma\left(\frac d2-1\right)^2}.
\end{equation}
Replacing the finite-cutoff bubble and the resummed bubble chain by their IR expressions changes the self-energy by a quadratically divergent $\propto\Lambda^2$ term, a finite $\propto p^2$ term, and cutoff-suppressed corrections. The first two are absorbed into the mass and finite wavefunction counterterms, respectively, while the cutoff-suppressed corrections vanish in the continuum limit. None of these changes modifies the coefficient of the universal logarithmic divergence. Hence, with our choice of counterterms, the self-energy reads
\begin{equation} \label{eq: bicon_gammaphi}
    \Sigma_\phi (p) = \frac{1}{N}\int_q \frac{1}{(p+q)^2}
\frac{1}{\Pi_0(q)}  + \widetilde{ \text{c.t.}} +O(N^{-2}) = 2\gamma_\phi p^2 \ln \frac{\Lambda}{p} + O(N^{-2}),
\end{equation}
with 
\begin{equation}
    \gamma_\phi = \frac{1}{N}\frac{2^d \sin(\tfrac{\pi d}{2}) \Gamma(\frac{d-1}{2})}{\pi^{3/2} (d-2) d \Gamma(\frac{d}{2}-2)} +O(N^{-2}),
\end{equation}
where $\gamma_\phi$ is the anomalous dimension of the $\phi$ field, and it agrees with the known result from the critical $O(N)$ model \cite{Goykhman:2019kcj}. Indeed, the coupling to the $\chi$ field does not affect $\gamma_\phi$ at this order.

As for the anomalous dimension $\gamma_\chi$ of the $\chi$ field, its self-energy reads
\begin{center}
\tikzset{every picture/.style={line width=0.75pt}} 

\begin{tikzpicture}[x=0.75pt,y=0.75pt,yscale=-1,xscale=1]

\draw [color={rgb, 255:red, 255; green, 0; blue, 0 }  ,draw opacity=1 ]   (456.57,49.33) -- (380.04,49.03) ;
\draw [line width=3]  [dash pattern={on 7.88pt off 4.5pt}]  (399.16,49.45) .. controls (401.3,23.47) and (435.82,22.45) .. (436.82,50.45) ;
\draw [color={rgb, 255:red, 255; green, 0; blue, 0 }  ,draw opacity=1 ]   (302.16,49.38) -- (266.2,49.38) ;
\draw [color={rgb, 255:red, 0; green, 0; blue, 0 }  ,draw opacity=1 ]   (272.35,36.71) .. controls (273.77,19.53) and (295.31,22.34) .. (295.27,36.71) ;
\draw [color={rgb, 255:red, 0; green, 0; blue, 0 }  ,draw opacity=1 ]   (295.27,36.71) .. controls (293.86,53.9) and (272.31,51.09) .. (272.35,36.71) ;

\draw [color={rgb, 255:red, 255; green, 0; blue, 0 }  ,draw opacity=1 ]   (359.16,49.38) -- (323.2,49.38) ;
\draw [color={rgb, 255:red, 255; green, 0; blue, 0 }  ,draw opacity=1 ]   (329.35,36.71) .. controls (330.77,19.53) and (352.31,22.34) .. (352.27,36.71) ;
\draw [color={rgb, 255:red, 255; green, 0; blue, 0 }  ,draw opacity=1 ]   (352.27,36.71) .. controls (350.86,53.9) and (329.31,51.09) .. (329.35,36.71) ;

\draw (210,27.27) node [anchor=north west][inner sep=0.75pt]  [font=\normalsize]  {$\Sigma _{\chi }( p) =$};
\draw (305.4,29.27) node [anchor=north west][inner sep=0.75pt]    {$+$};
\draw (363.4,29.38) node [anchor=north west][inner sep=0.75pt]    {$+$};
\draw (459.4,25.38) node [anchor=north west][inner sep=0.75pt]    {$+\ O\left( N^{-2}\right) .$};

\end{tikzpicture}
\end{center}
The first two diagrams are independent of the external momentum and are therefore absorbed by the local mass counterterm. The third is the same rainbow diagram as in $\Sigma_\phi$, except that the thick dashed line is attached to the external $\chi$ line via two $\lambda_{\phi\chi}$ couplings rather than $\lambda_\phi$ couplings. Hence, 
\begin{equation} \label{eq: bicon_gammachi}
    \Sigma_\chi (p) = \frac{1}{N} \frac{\lambda_{\phi\chi}^2}{\lambda_\phi^2}\int_q \frac{1}{(p+q)^2}
\frac{1}{\Pi_0(q)}  + \text{c.t.}  +O(N^{-2})  \,\,\Rightarrow\,\, \gamma_\chi = \alpha^2 \gamma_\phi  +O(N^{-2}),
\end{equation}
where $\alpha = \lambda_{\phi\chi}/\lambda_\phi$. As we later show in our NLO analysis in Section \ref{sec: NLO}, only particular values of $\alpha$ correspond to a CFT in the IR.

The discussion above fixes a convention used throughout the remainder of the paper. In subsequent IR calculations, we work directly with the continuum IR expressions for the bubble $\Pi_{m^2}(p)$ and the resummed bubble
chain, taking $\Lambda\to\infty$ and couplings $\lambda_a\sim\Lambda^{4-d}\to\infty$. When these expressions occur inside loop integrals, the difference from the corresponding finite-cutoff expressions is absorbed by finite or power-law divergent local counterterms. These counterterms are non-universal and depend on the details of the regularization scheme. For example, power-law divergences are absent in dimensional regularization. The particular form of such counterterms is unimportant, as they contain no logarithmic divergences and therefore do not introduce a renormalization-scale dependence in the minimal subtraction scheme.

\section{The Effective Potential at LO in $1/N$} \label{sec: LO}
Consider a general Euclidean QFT with fields $\phi_i$ and UV action $S[\phi]$. The effective action $\Gamma [\phi]$ of this theory is defined as the Legendre transform of the generating functional of connected correlation functions. Equivalently, $\Gamma[\phi]$ may be defined implicitly by  
\begin{equation} \label{eq: effective_action_path_integral_definition}
    \exp(-\Gamma [\phi]) =\int D\hat{\phi } \exp\left(-S[\phi +\hat{\phi } ]+\int_x \, \frac{\delta \Gamma [\phi ]}{\delta \phi }\hat{\phi}\right).
\end{equation}
The effective action admits an expansion in terms of one-particle-irreducible (1PI) vertex functions,
\begin{equation} \label{eq: effective_action_1PI_expansion}
    \Gamma[\phi] = \sum_{n\geq 0} \frac{1}{n!} \sum_{i_1,\dots, i_n}\int_{p_1, \dots, p_{n}} \Gamma^{(n)}_{i_1, \dots, i_n}(p_1, \dots, p_n)\, \phi_{p_1}^{i_1} \dots \phi_{p_{n}}^{i_n}\, (2\pi)^d\delta^{(d)} ({\sum\nolimits_j^n} p_j),
\end{equation}
where $\Gamma^{(n)}_{i_1,\dots, i_n}(p_1, \dots, p_n)$ is a 1PI vertex function of $n$ fields $\phi_{i_1}, \dots, \phi_{i_n}$ with momenta $p_1, \dots, p_n$. Note the absence of any constant normalization factors in the definition of $\Gamma[\phi]$ in \eqref{eq: effective_action_path_integral_definition}. This implies that all terms in the UV action, including constant and quadratic in fields, are treated as interactions. By this definition, a zero-dimensional Gaussian theory $S[\phi] = \frac{1}{2}\phi^T A \phi$ has an effective action $\Gamma[\phi] = \frac{1}{2}\phi^T A \phi + \frac{1}{2} \ln \det \left(A/2\pi\right)$, where the constant $\frac{1}{2}\ln \det A$ comes from quadratic interactions in 1PI vacuum diagrams. Evaluating \eqref{eq: effective_action_1PI_expansion} for constant field configurations $\phi(x) = \phi$ gives us the effective potential, 
\begin{equation}
    V(\phi) \equiv \frac{1}{\text{Vol}} \Gamma[\phi_i(x)=\phi_i].    
\end{equation}
In other words, the effective potential is obtained by setting all the integration momenta in \eqref{eq: effective_action_1PI_expansion} to zero and summing the series. In a CFT, however, the absence of scales implies that 1PI vertices with nonzero scaling dimension vanish or become singular at zero momentum. This creates an obstacle for a straightforward series resummation.

A different route to calculating the effective potential is to define a new shifted theory with a UV action $\tilde S[\eta] = S[\phi + \eta]$, where $\phi$ is some constant background field configuration, and $\eta$ plays the role of a dynamical field \cite{Coleman:1985rnk}. By the definition in \eqref{eq: effective_action_path_integral_definition}, the effective actions of the original and shifted theories are related by a shift $\tilde \Gamma[\eta] = \Gamma[\phi + \eta]$. The 1PI expansion also applies to the shifted theory
\begin{equation}
    \tilde \Gamma[\eta]= \sum_{n\geq 0} \frac{1}{n!} \sum_{i_1,\dots, i_n}\int_{p_1, \dots, p_{n}} \tilde\Gamma^{(n)}_{i_1, \dots, i_n}(p_1, \dots, p_n)\, \eta_{p_1}^{i_1} \dots \eta_{p_{n}}^{i_n}\, (2\pi)^d\delta^{(d)} ({\sum\nolimits_j^n} p_j),
\end{equation}
where $\tilde{\Gamma}^{(n)}_{i_1,\dots, i_n}(p_1, \dots, p_n)$ is a 1PI vertex function of $n$ fields $\eta_{i_1}, \dots, \eta_{i_n}$ with momenta $p_1, \dots, p_n$ in the shifted theory. Since $\Gamma[\phi] = \tilde \Gamma[0]$,
the effective potential of the original theory is
\begin{equation}
    V(\phi) = \tilde\Gamma^{(0)} \equiv \text{1PI vacuum diagrams of the shifted theory } \tilde S[\eta].
\end{equation} 
Taking two derivatives of 
$\tilde\Gamma[\eta] = \Gamma[\phi+\eta]$ with respect to $\eta_i$ and then setting $\eta=0$, we obtain
\begin{equation}
    \tilde\Gamma^{(2)}_{ii}(0,0) = \partial_{\phi_i}^2 V(\phi),\quad\phi\in \mathbb R^N.
\end{equation}
If the theory is $O(2)$ symmetric in $\phi_1, \phi_2$ fields, the effective potential depends on these fields only through $(\phi_1^2+\phi_2^2)$. Consequently, 
\begin{equation} \label{eq: two_point_function_effective_potential}
    \tilde\Gamma^{(2)}_{22}(0,0) = 2\partial_{\phi_{1}^2} V(\phi), \quad \phi=(\phi_1, 0, \phi_3, \dots,\phi_N).
\end{equation}

Following this general discussion, we turn to calculating the effective potential of the theory defined by the UV action \eqref{eq: bicon_uv_action} at LO in $1/N$ at zero temperature. We shift the UV action by the substitutions $\phi_i(x)\rightarrow N^{1/2}\phi \,\delta_{i 1}+\eta_i(x)$ and $\chi(x) \rightarrow N^{1/2} \chi+\xi(x)$ with powers of $N$ chosen to yield a consistent $1/N$ expansion. We treat $\eta$ and $\xi$ as new dynamical field variables, while $\phi$ and $\chi$ are constant background field configurations, with $\phi$ aligned along the first direction by $O(N)$ symmetry. Hence, the action $\tilde S [\eta, \xi]$ is given by 
\begin{align} \label{eq: bicon_shited_action}
\tilde S[\eta,\xi]
&=\int d^dx\left[\frac12(\partial_\mu\eta_a)^2+\frac12(\partial_\mu\xi)^2+\tilde U(\eta,\xi)\right],
\\[4pt]
\tilde U(\eta,\xi)
&= N\left\{\frac{\lambda_\phi}{8}\phi^4+\frac{\lambda_{\phi\chi}}{4}\phi^2\chi^2+\frac{\lambda_\chi}{8}\chi^4\right\} + \qquad\text{constant}
\nonumber\\
\text{linear} &\quad+N^{\frac{1}{2}}\left\{\left(\frac{\lambda_\phi}{2}\phi^3+\frac{\lambda_{\phi\chi}}{2}\phi\chi^2\right)\eta_1
+\left(\frac{\lambda_{\phi\chi}}{2}\phi^2\chi+\frac{\lambda_\chi}{2}\chi^3\right)\xi\right\}
\nonumber\\
\text{quadratic} &\quad+N^{0}\left\{\left(\frac{\lambda_\phi}{4}\phi^2+\frac{\lambda_{\phi\chi}}{4}\chi^2\right)\eta^2
+\frac{\lambda_\phi}{2}\phi^2\,\eta_1^2
+\left(\frac{\lambda_{\phi\chi}}{4}\phi^2+\frac{3\lambda_\chi}{4}\chi^2\right)\xi^2
+\lambda_{\phi\chi}\phi\chi\,\eta_1\xi\right\}
\nonumber\\
\text{cubic} &\quad+N^{-\frac{1}{2}}\left\{\frac{\lambda_\phi}{2}\phi\,\eta_1\eta^2
+\frac{\lambda_{\phi\chi}}{2}\chi\,\xi\,\eta^2
+\frac{\lambda_{\phi\chi}}{2}\phi\,\eta_1\xi^2
+\frac{\lambda_\chi}{2}\chi\,\xi^3\right\}
\nonumber\\
\text{quartic}&\quad+N^{-1}\left\{\frac{\lambda_\phi}{8}(\eta^2)^2+\frac{\lambda_{\phi\chi}}{4}\eta^2\xi^2+\frac{\lambda_\chi}{8}\xi^4\right\},
\end{align}
where each line in $\tilde U(\eta, \xi)$ contains interactions of a given order in the dynamical fields $\eta, \xi$. We denote the part independent of $\eta,\xi$ by $U(\phi,\chi) \equiv \tilde U(0, 0)$, which coincides with the UV potential of the unshifted theory up to the large-$N$ field rescaling. 

To compute the LO vacuum 1PI diagrams, we first need to obtain the LO 1PI two-point function of the $\eta_{2\dots N}$ fields. Note that the shifted action retains an $O(N-1)$ symmetry with respect to these fields. Define self-energies as $\Sigma_{i}(p) = \tilde \Gamma^{(2)}_{ii}(p) - p^2$. At LO in $1/N$, the diagrammatic equation for $\Sigma_{\eta_2}$ is given by
\begin{center}
    \input{pics/bicon_lo_sigma_eta2}
\end{center}
In the first line, the blue lines denote massless $\eta_{i=2\dots N}$ propagators $1/k^2$, while the double-spines denote the quadratic mass-like $\eta^2$ interactions. Each loop has a sum over $\sim N$ fields, whereas the attachment of a loop costs a coupling multiplication $\sim1/N$, so adding new loops is $O(N^0)$. Adding spines is also $O(N^0)$. Therefore, there is an infinite series of contributions to the $\eta_2$ self-energy of order $O(N^0)$ composed of various combinations of cacti with spines. The second line shows that this infinite series can be iteratively reproduced using a single double-spine and a single loop with the dressed $\eta_{2}$ propagator denoted by the black line. This propagator does not include tadpole insertions involving $\eta_1$ and $\xi$ linear interactions, since they would generate one-particle-reducible contributions. This gives a self-consistency equation for $\Sigma_{\eta_2}$, assuming a field-space-localized state of the shifted theory centered at $\eta,\xi =0$
\footnote{\phantomsection\label{footnote: field-space-localized_states}Following Weinberg and Wu \cite{Weinberg:1987vp}, we restrict to states whose wave functional is concentrated around a single macroscopic field configuration. We refer to such states as field-space localized; Weinberg and Wu call them “homogeneous” states. The form of the propagator used in the equation for $\Sigma_{\eta_2}$ assumes a state $\ket{\Omega}$ which satisfies cluster decomposition, $$\langle \Omega | \eta_i(x)  \eta_i(y)|\Omega \rangle_{c} \rightarrow 0,\quad \text{as}\,\,\,|x-y|\rightarrow \infty .$$
In an infinite-volume scalar QFT, a vacuum in a definite superselection sector is field-space localized. A state with support near several macroscopically separated configurations instead represents a mixture of sectors and does not satisfy cluster decomposition. To illustrate this, consider two field-space-localized states $\ket{\Omega_\pm}$ satisfying $\langle \Omega_\pm | \eta_i(x)|\Omega_\pm \rangle = \pm\delta_{i1}\sqrt{N} a$, and construct their superposition $\ket{\Omega'} = \frac{1}{\sqrt{2}}(\ket{\Omega_+}+\ket{\Omega_-})$. One finds, in violation of cluster decomposition, $$\langle \Omega' |\eta_i(x)|\Omega' \rangle = 0, \quad\langle \Omega' | \eta_1(x) \eta_1(y)|\Omega' \rangle \rightarrow a^2 N, \quad \text{as}\,\,\, |x-y|\rightarrow \infty.$$ 
If we were to include states like $\ket{\Omega'}$ in our calculation, the $\sim N$ contribution to the $\eta_1$ propagator would modify \eqref{eq: bicon_sigma_eta2_eq}. 

Although the exact effective potential is convex, restricting its calculation to field-space-localized states may lead to non-convex, complex, or undetermined values in some regions of field space. In the effective potential, such regions can be interpreted as having potential walls for field-space-localized states: there they are physically unstable and are subject to decay.
}

\begin{equation} \label{eq: bicon_sigma_eta2_eq}
    \Sigma_{\eta_2} = \frac{\lambda_{\phi}}{2} \int_k \frac{1}{k^2 + \Sigma_{\eta_2}} + \left(\frac{\lambda_\phi}{2} \phi^2 + \frac{\lambda_{\phi\chi}}{2}\chi^2\right) +O(N^{-1}).
\end{equation}
In the hard-cutoff regularization, the integral contains a power-law divergence of order $\sim \lambda_{\phi} \Lambda^{d-2}$, which is removed by a mass counterterm in the original theory. Power-law divergences depend on the regularization scheme, and, for example, are absent in dimensional regularization. Hence, we simply ignore such divergences in what follows to reduce clutter in the equations.

We are interested in the IR physics of the theory, and therefore the background fields are taken at their IR values $\phi, \chi \ll \Lambda^{\frac{d-2}{2}}$. The UV quartic couplings scale with the cutoff $\lambda_a \sim \Lambda^{4-d}$, and we assume $\Sigma_{\eta_2}\ll \Lambda^2$ with an \textit{a posteriori} consistency check after solving the equation \eqref{eq: bicon_sigma_eta2_eq}. We refer to this joint limit throughout the paper as the IR limit; at fixed IR quantities, it is equivalently realized by taking $|\lambda_a|, \Lambda \rightarrow \infty$. 
Dividing \eqref{eq: bicon_sigma_eta2_eq} by $\lambda_\phi > 0$, the left-hand side can be neglected, and the integral is a standard one. This leads to 
\begin{align} \label{eq: sigma_eta2_lo}
        &0=\frac{\Gamma(1-\frac{d}{2})}{(4\pi)^{d/2}} \Sigma_{\eta_2}^{\frac{d-2}{2}} + (\phi^2 + \alpha \chi^2)  + O(N^{-1}), \quad \alpha = \frac{\lambda_{\phi\chi}}{\lambda_\phi} \nonumber\\
        & \Rightarrow\Sigma_{\eta_2} = \left(\frac{(4\pi)^{d/2}}{-\Gamma(1-\frac{d}{2})}\right)^{\frac{2}{d-2}}\left(\phi^2 + \alpha \chi^2\right)^{\frac{2}{d-2}} + O(N^{-1}), \quad \phi^2 + \alpha \chi^2 \geq 0,
\end{align}
where, indeed, $\Sigma_{\eta_2}\ll \Lambda^2$. The Euclidean integral in \eqref{eq: bicon_sigma_eta2_eq} forces the domain of $\Sigma_{\eta_2}$ to be the principal branch, with a cut along the negative real axis. For $\alpha < 0$, the bracket $(\phi^2 + \alpha \chi^2)$ can be negative, and then the equation enforces $\Sigma_{\eta_2}^{(d-2)/2}$ to be negative as well. In such cases, since $0<(d-2)/2 < 1$, there is no solution for $\Sigma_{\eta_2}$.

Now we are ready to sum all the LO vacuum diagrams of the shifted theory to obtain the effective potential restricted to field-space-localized states,
\begin{center}
    \input{pics/bicon_lo_vacuum}
\end{center}
where the big black dot represents the constant UV potential contribution $\tilde U(0,0) \equiv U(\phi,\chi)$. To fully account for the combinatorics of the diagrams above, it is easiest to note that the number of loops $L(D)$ in each diagram $D$ is one more than the number of quartic vertices $V(D)$, i.e., $L(D)=V(D)+1$. Hence, the LO effective potential $V_{\text{LO}}$ reads
\begin{equation} \label{eq: effective_potential_sum_diagrams}
    V_{\text{LO}}(\phi, \chi) = U (\phi,\chi) + \sum_D A(D) = U (\phi,\chi) + \sum_D L(D) A(D) - \sum_D V(D) A(D),
\end{equation}
where $A(D)$ is the amplitude of a diagram $D$. The first sum on the right-hand side allows us to mark a single loop and attach to it various interactions, which are encapsulated in~$\Sigma_{\eta_2}$. Accounting for the cyclic and inversion symmetry factors, we find 
\begin{center}
    \tikzset{every picture/.style={line width=0.75pt}} 

\begin{tikzpicture}[x=0.75pt,y=0.75pt,yscale=-1,xscale=1]

\draw [color={rgb, 255:red, 0; green, 0; blue, 255 }  ,draw opacity=1 ]   (160.41,46.33) .. controls (161.83,29.12) and (183.4,31.93) .. (183.36,46.33) ;
\draw [color={rgb, 255:red, 0; green, 0; blue, 255 }  ,draw opacity=1 ]   (183.36,46.33) .. controls (181.94,63.53) and (160.37,60.72) .. (160.41,46.33) ;
\draw  [fill={rgb, 255:red, 0; green, 0; blue, 0 }  ,fill opacity=1 ] (170,33.97) .. controls (170,32.88) and (170.88,32) .. (171.97,32) .. controls (173.05,32) and (173.93,32.88) .. (173.93,33.97) .. controls (173.93,35.05) and (173.05,35.93) .. (171.97,35.93) .. controls (170.88,35.93) and (170,35.05) .. (170,33.97) -- cycle ;
\draw [color={rgb, 255:red, 0; green, 0; blue, 255 }  ,draw opacity=1 ]   (220.41,46.33) .. controls (221.83,29.12) and (243.4,31.93) .. (243.36,46.33) ;
\draw [color={rgb, 255:red, 0; green, 0; blue, 255 }  ,draw opacity=1 ]   (243.36,46.33) .. controls (241.94,63.53) and (220.37,60.72) .. (220.41,46.33) ;
\draw  [fill={rgb, 255:red, 0; green, 0; blue, 0 }  ,fill opacity=1 ] (230,33.97) .. controls (230,32.88) and (230.88,32) .. (231.97,32) .. controls (233.05,32) and (233.93,32.88) .. (233.93,33.97) .. controls (233.93,35.05) and (233.05,35.93) .. (231.97,35.93) .. controls (230.88,35.93) and (230,35.05) .. (230,33.97) -- cycle ;
\draw  [fill={rgb, 255:red, 0; green, 0; blue, 0 }  ,fill opacity=1 ] (230,57.97) .. controls (230,56.88) and (230.88,56) .. (231.97,56) .. controls (233.05,56) and (233.93,56.88) .. (233.93,57.97) .. controls (233.93,59.05) and (233.05,59.93) .. (231.97,59.93) .. controls (230.88,59.93) and (230,59.05) .. (230,57.97) -- cycle ;
\draw [color={rgb, 255:red, 0; green, 0; blue, 255 }  ,draw opacity=1 ]   (281.41,46.33) .. controls (282.83,29.12) and (304.4,31.93) .. (304.36,46.33) ;
\draw [color={rgb, 255:red, 0; green, 0; blue, 255 }  ,draw opacity=1 ]   (304.36,46.33) .. controls (302.94,63.53) and (281.37,60.72) .. (281.41,46.33) ;
\draw  [fill={rgb, 255:red, 0; green, 0; blue, 0 }  ,fill opacity=1 ] (291,33.97) .. controls (291,32.88) and (291.88,32) .. (292.97,32) .. controls (294.05,32) and (294.93,32.88) .. (294.93,33.97) .. controls (294.93,35.05) and (294.05,35.93) .. (292.97,35.93) .. controls (291.88,35.93) and (291,35.05) .. (291,33.97) -- cycle ;
\draw  [fill={rgb, 255:red, 0; green, 0; blue, 0 }  ,fill opacity=1 ] (283,53.97) .. controls (283,52.88) and (283.88,52) .. (284.97,52) .. controls (286.05,52) and (286.93,52.88) .. (286.93,53.97) .. controls (286.93,55.05) and (286.05,55.93) .. (284.97,55.93) .. controls (283.88,55.93) and (283,55.05) .. (283,53.97) -- cycle ;
\draw  [fill={rgb, 255:red, 0; green, 0; blue, 0 }  ,fill opacity=1 ] (302.39,50.3) .. controls (302.39,49.21) and (303.27,48.33) .. (304.36,48.33) .. controls (305.45,48.33) and (306.33,49.21) .. (306.33,50.3) .. controls (306.33,51.38) and (305.45,52.26) .. (304.36,52.26) .. controls (303.27,52.26) and (302.39,51.38) .. (302.39,50.3) -- cycle ;

\draw (39,34) node [anchor=north west][inner sep=0.75pt]    {$\sum _{D} L( D) A( D) =$};
\draw (193.4,40.27) node [anchor=north west][inner sep=0.75pt]    {$+$};
\draw (166,18) node [anchor=north west][inner sep=0.75pt]    {$\Sigma _{\eta _{2}}$};
\draw (334.4,40.27) node [anchor=north west][inner sep=0.75pt]    {$+\dotsc $};
\draw (254.4,40.27) node [anchor=north west][inner sep=0.75pt]    {$+$};
\draw (226,18) node [anchor=north west][inner sep=0.75pt]    {$\Sigma _{\eta _{2}}$};
\draw (235.93,58.97) node [anchor=north west][inner sep=0.75pt]    {$\Sigma _{\eta _{2}}$};
\draw (287,18) node [anchor=north west][inner sep=0.75pt]    {$\Sigma _{\eta _{2}}$};
\draw (271.93,58.97) node [anchor=north west][inner sep=0.75pt]    {$\Sigma _{\eta _{2}}$};
\draw (308,38) node [anchor=north west][inner sep=0.75pt]    {$\Sigma _{\eta _{2}}$};
\draw (135.4,84.27) node [anchor=north west][inner sep=0.75pt]    {$=\ \frac{N}{2}\int _{k}\sum _{n=1}^{\infty }\frac{( -1)^{n+1}}{n}\left(\frac{\Sigma _{\eta _{2}}}{k^{2}}\right)^{n} =\frac{N}{2}\int _{k}\ln\left( 1+\frac{\Sigma _{\eta _{2}}}{k^{2}}\right) .$};

\end{tikzpicture}
\end{center}
The second sum allows us to mark a single quartic interaction and attach loops and interactions on top of it,
\begin{center}
    \tikzset{every picture/.style={line width=0.75pt}} 

\begin{tikzpicture}[x=0.75pt,y=0.75pt,yscale=-1,xscale=1]

\draw [color={rgb, 255:red, 0; green, 0; blue, 255 }  ,draw opacity=1 ]   (266.41,71.78) .. controls (267.83,54.57) and (289.4,57.38) .. (289.36,71.78) ;
\draw [color={rgb, 255:red, 0; green, 0; blue, 255 }  ,draw opacity=1 ]   (289.36,71.78) .. controls (287.94,88.98) and (266.37,86.17) .. (266.41,71.78) ;
\draw [color={rgb, 255:red, 0; green, 0; blue, 255 }  ,draw opacity=1 ]   (266.41,47.78) .. controls (267.83,30.57) and (289.4,33.38) .. (289.36,47.78) ;
\draw [color={rgb, 255:red, 0; green, 0; blue, 255 }  ,draw opacity=1 ]   (289.36,47.78) .. controls (287.94,64.98) and (266.37,62.17) .. (266.41,47.78) ;
\draw  [fill={rgb, 255:red, 0; green, 0; blue, 0 }  ,fill opacity=1 ] (276,59.41) .. controls (276,58.33) and (276.88,57.45) .. (277.97,57.45) .. controls (279.05,57.45) and (279.93,58.33) .. (279.93,59.41) .. controls (279.93,60.5) and (279.05,61.38) .. (277.97,61.38) .. controls (276.88,61.38) and (276,60.5) .. (276,59.41) -- cycle ;
\draw [color={rgb, 255:red, 0; green, 0; blue, 255 }  ,draw opacity=1 ]   (313.41,71.78) .. controls (314.83,54.57) and (336.4,57.38) .. (336.36,71.78) ;
\draw [color={rgb, 255:red, 0; green, 0; blue, 255 }  ,draw opacity=1 ]   (336.36,71.78) .. controls (334.94,88.98) and (313.37,86.17) .. (313.41,71.78) ;
\draw [color={rgb, 255:red, 0; green, 0; blue, 255 }  ,draw opacity=1 ]   (313.41,47.78) .. controls (314.83,30.57) and (336.4,33.38) .. (336.36,47.78) ;
\draw [color={rgb, 255:red, 0; green, 0; blue, 255 }  ,draw opacity=1 ]   (336.36,47.78) .. controls (334.94,64.98) and (313.37,62.17) .. (313.41,47.78) ;
\draw  [fill={rgb, 255:red, 0; green, 0; blue, 0 }  ,fill opacity=1 ] (323,59.41) .. controls (323,58.33) and (323.88,57.45) .. (324.97,57.45) .. controls (326.05,57.45) and (326.93,58.33) .. (326.93,59.41) .. controls (326.93,60.5) and (326.05,61.38) .. (324.97,61.38) .. controls (323.88,61.38) and (323,60.5) .. (323,59.41) -- cycle ;
\draw    (309.8,30.19) -- (320.11,35.37) ;
\draw [shift={(322.21,36.43)}, rotate = 26.68] [color={rgb, 255:red, 0; green, 0; blue, 0 }  ][line width=0.75]      (0, 0) circle [x radius= 3.35, y radius= 3.35]   ;
\draw    (328.45,24.01) -- (323.27,34.33) ;
\draw [shift={(322.21,36.43)}, rotate = 116.68] [color={rgb, 255:red, 0; green, 0; blue, 0 }  ][line width=0.75]      (0, 0) circle [x radius= 3.35, y radius= 3.35]   ;

\draw [color={rgb, 255:red, 0; green, 0; blue, 255 }  ,draw opacity=1 ]   (383.41,71.78) .. controls (384.83,54.57) and (406.4,57.38) .. (406.36,71.78) ;
\draw [color={rgb, 255:red, 0; green, 0; blue, 255 }  ,draw opacity=1 ]   (406.36,71.78) .. controls (404.94,88.98) and (383.37,86.17) .. (383.41,71.78) ;
\draw [color={rgb, 255:red, 0; green, 0; blue, 255 }  ,draw opacity=1 ]   (383.41,47.78) .. controls (384.83,30.57) and (406.4,33.38) .. (406.36,47.78) ;
\draw [color={rgb, 255:red, 0; green, 0; blue, 255 }  ,draw opacity=1 ]   (406.36,47.78) .. controls (404.94,64.98) and (383.37,62.17) .. (383.41,47.78) ;
\draw  [fill={rgb, 255:red, 0; green, 0; blue, 0 }  ,fill opacity=1 ] (393,59.41) .. controls (393,58.33) and (393.88,57.45) .. (394.97,57.45) .. controls (396.05,57.45) and (396.93,58.33) .. (396.93,59.41) .. controls (396.93,60.5) and (396.05,61.38) .. (394.97,61.38) .. controls (393.88,61.38) and (393,60.5) .. (393,59.41) -- cycle ;
\draw    (379.8,30.19) -- (390.11,35.37) ;
\draw [shift={(392.21,36.43)}, rotate = 26.68] [color={rgb, 255:red, 0; green, 0; blue, 0 }  ][line width=0.75]      (0, 0) circle [x radius= 3.35, y radius= 3.35]   ;
\draw    (398.45,24.01) -- (393.27,34.33) ;
\draw [shift={(392.21,36.43)}, rotate = 116.68] [color={rgb, 255:red, 0; green, 0; blue, 0 }  ][line width=0.75]      (0, 0) circle [x radius= 3.35, y radius= 3.35]   ;

\draw [color={rgb, 255:red, 0; green, 0; blue, 255 }  ,draw opacity=1 ]   (403.41,34.78) .. controls (404.83,17.57) and (426.4,20.38) .. (426.36,34.78) ;
\draw [color={rgb, 255:red, 0; green, 0; blue, 255 }  ,draw opacity=1 ]   (426.36,34.78) .. controls (424.94,51.98) and (403.37,49.17) .. (403.41,34.78) ;
\draw [color={rgb, 255:red, 0; green, 0; blue, 255 }  ,draw opacity=1 ]   (423.41,22.78) .. controls (424.83,5.57) and (446.4,8.38) .. (446.36,22.78) ;
\draw [color={rgb, 255:red, 0; green, 0; blue, 255 }  ,draw opacity=1 ]   (446.36,22.78) .. controls (444.94,39.98) and (423.37,37.17) .. (423.41,22.78) ;
\draw [color={rgb, 255:red, 0; green, 0; blue, 255 }  ,draw opacity=1 ]   (399.41,89.78) .. controls (400.83,72.57) and (422.4,75.38) .. (422.36,89.78) ;
\draw [color={rgb, 255:red, 0; green, 0; blue, 255 }  ,draw opacity=1 ]   (422.36,89.78) .. controls (420.94,106.98) and (399.37,104.17) .. (399.41,89.78) ;
\draw [color={rgb, 255:red, 0; green, 0; blue, 255 }  ,draw opacity=1 ]   (367.41,89.78) .. controls (368.83,72.57) and (390.4,75.38) .. (390.36,89.78) ;
\draw [color={rgb, 255:red, 0; green, 0; blue, 255 }  ,draw opacity=1 ]   (390.36,89.78) .. controls (388.94,106.98) and (367.37,104.17) .. (367.41,89.78) ;
\draw    (424.41,71.94) -- (422.17,83.26) ;
\draw [shift={(421.71,85.57)}, rotate = 101.18] [color={rgb, 255:red, 0; green, 0; blue, 0 }  ][line width=0.75]      (0, 0) circle [x radius= 3.35, y radius= 3.35]   ;
\draw    (435.34,88.26) -- (424.02,86.02) ;
\draw [shift={(421.71,85.57)}, rotate = 191.18] [color={rgb, 255:red, 0; green, 0; blue, 0 }  ][line width=0.75]      (0, 0) circle [x radius= 3.35, y radius= 3.35]   ;

\draw    (455.83,12.5) -- (448.53,21.43) ;
\draw [shift={(447.04,23.25)}, rotate = 129.26] [color={rgb, 255:red, 0; green, 0; blue, 0 }  ][line width=0.75]      (0, 0) circle [x radius= 3.35, y radius= 3.35]   ;
\draw    (457.8,32.05) -- (448.86,24.74) ;
\draw [shift={(447.04,23.25)}, rotate = 219.26] [color={rgb, 255:red, 0; green, 0; blue, 0 }  ][line width=0.75]      (0, 0) circle [x radius= 3.35, y radius= 3.35]   ;

\draw [color={rgb, 255:red, 0; green, 0; blue, 0 }  ,draw opacity=1 ]   (483.41,72.78) .. controls (484.83,55.57) and (506.4,58.38) .. (506.36,72.78) ;
\draw [color={rgb, 255:red, 0; green, 0; blue, 0 }  ,draw opacity=1 ]   (506.36,72.78) .. controls (504.94,89.98) and (483.37,87.17) .. (483.41,72.78) ;
\draw [color={rgb, 255:red, 0; green, 0; blue, 0 }  ,draw opacity=1 ]   (483.41,48.78) .. controls (484.83,31.57) and (506.4,34.38) .. (506.36,48.78) ;
\draw [color={rgb, 255:red, 0; green, 0; blue, 0 }  ,draw opacity=1 ]   (506.36,48.78) .. controls (504.94,65.98) and (483.37,63.17) .. (483.41,48.78) ;
\draw  [fill={rgb, 255:red, 0; green, 0; blue, 0 }  ,fill opacity=1 ] (493,60.41) .. controls (493,59.33) and (493.88,58.45) .. (494.97,58.45) .. controls (496.05,58.45) and (496.93,59.33) .. (496.93,60.41) .. controls (496.93,61.5) and (496.05,62.38) .. (494.97,62.38) .. controls (493.88,62.38) and (493,61.5) .. (493,60.41) -- cycle ;

\draw (146,47.45) node [anchor=north west][inner sep=0.75pt]    {$\sum _{D} V( D) A( D) =$};
\draw (294.4,53.72) node [anchor=north west][inner sep=0.75pt]    {$+$};
\draw (420.4,53.72) node [anchor=north west][inner sep=0.75pt]    {$+\dotsc \ =$};
\draw (342.4,53.72) node [anchor=north west][inner sep=0.75pt]    {$+...+$};
\draw (244.4,115.72) node [anchor=north west][inner sep=0.75pt]    {$=\frac{N\lambda _{\phi }}{8}\left(\int _{k}\frac{1}{k^{2} +\Sigma _{\eta _{2}}}\right)^{2} =\frac{N\lambda _{\phi }}{8}\left(\frac{2\Sigma _{\eta _{2}}}{\lambda _{\phi }} -(\phi ^{2} +\alpha \chi ^{2} )\right)^{2} ,$};

\end{tikzpicture}
\end{center}
where the last equality follows from the self-consistency equation for $\Sigma_{\eta_2}$ \eqref{eq: bicon_sigma_eta2_eq}. Summing and simplifying the three contributions from \eqref{eq: effective_potential_sum_diagrams}, we obtain 
\begin{equation} \label{eq: bicon_effective_potential_lo_preirlimit}
    V(\phi, \chi) = N\Bigg[ \frac{1}{2}\int_k \ln \left(k^2 + \Sigma_{\eta_2}\right) + \frac{1}{2}\Sigma_{\eta_2}(\phi^2 + \alpha\chi^2) + \frac{1}{8}\left(\lambda_\chi-\frac{\lambda_{\phi\chi}^2}{\lambda_\phi}\right)\chi^4 - \frac{\Sigma_{\eta_2}^2}{2\lambda_\phi} \Bigg] + O(N^0). 
\end{equation}
We have included the field-independent massless Gaussian determinant $\frac{N}{2}\int_k \ln k^2$ that arises from the definition \eqref{eq: effective_action_path_integral_definition}. While its effect is irrelevant here, at $T>0$, it introduces a $T$-dependent constant. In the IR limit $\phi, \chi \ll \Lambda^{\frac{d-2}{2}}$, the self-energy is given by \eqref{eq: sigma_eta2_lo} and remains finite as $\lambda_\phi\sim \Lambda^{4-d} \rightarrow \infty$. Consequently, the term $-\Sigma_{\eta_2}^2/(2\lambda_\phi)$ vanishes, while the integral term reduces to 
\begin{equation*}
    \frac{1}{2}\int_k \ln \left(k^2 + \Sigma_{\eta_2}\right) = - \frac{\Gamma\left(-\frac{d}{2}\right)}{2(4\pi)^{d/2}} \Sigma_{\eta_2}^{d/2}. 
\end{equation*}
Substituting $\Sigma_{\eta_2}$ from \eqref{eq: sigma_eta2_lo} and simplifying, we obtain the IR effective potential
\begin{equation} \label{eq: bicon_effective_potential_lo}
    V(\phi,\chi) = N \nu_d \left(\phi^2+\alpha\chi^2\right)^{\frac{d}{d-2}} + \frac{N}{8}\left(\lambda_{\chi}-\frac{\lambda_{\phi\chi}^2}{ \lambda_\phi}\right) \chi^4 + O(N^{0}),\quad  \phi^2+\alpha\chi^2 \geq 0,
\end{equation}
where 
\begin{equation}
    \nu_d = \left(\frac{(4\pi)^{d/2}}{-\Gamma(1-\frac{d}{2})}\right)^{\frac{2}{d-2}} \frac{d-2}{2d}.
\end{equation}
For $\phi^2+\alpha\chi^2 < 0$, the self-consistency equation \eqref{eq: sigma_eta2_lo} admits no solution for $\Sigma_{\eta_2}$, and therefore the effective potential restricted to field-space-localized states is not defined in this region. In $d=3$, the expression above can be formally continued into the region $\phi^2+\alpha\chi^2 < 0$ and remains real, but this continuation does not correspond to any solution of the original self-consistency equation and is therefore unphysical. Rather, the absence of a solution means that no linear source can make a field-space-localized state centered around the given $(\phi,\chi)$ background into a stable vacuum. See the footnote on page \pageref{footnote: field-space-localized_states} and~\cite{Coleman:1974jh} for an analogous discussion in the $\phi^4$ $O(N)$ theory in $d=3$.

We have thus obtained the IR effective potential of the biconical model \eqref{eq: bicon_uv_action} for generic values of the quartic couplings. We now determine conditions on the values of these couplings under which the theory is scale-invariant in the IR. A necessary condition is that the effective potential be invariant under the scaling transformation, 
\begin{equation}
    V(\phi,\chi) \rightarrow s^{-d} V(\phi s^{\frac{d-2}{2}+\gamma_\phi},\, \chi s^{\frac{d-2}{2}+\gamma_\chi}),
\end{equation}
where $\gamma_\phi, \gamma_\chi$ are the anomalous dimensions of the fields $\phi$ and $\chi$, respectively. Equivalently, the effective potential must satisfy
\begin{equation} \label{eq: scale_variation}
    \hat \delta_s  V(\phi,\chi) \stackrel{\text{CFT}}{=} 0, \qquad \hat \delta_s = - d + \left(\frac{d-2}{2}+\gamma_\phi\right) \phi \partial_\phi + \left(\frac{d-2}{2}+\gamma_\chi\right) \chi \partial_\chi.
\end{equation}
As shown in Section \ref{sec: anomalous_dimensions}, both anomalous dimensions are $O(N^{-1})$, so at LO the fields scale with their classical dimensions. For $d<4$, the $\chi^4$ term in the effective potential \eqref{eq: bicon_effective_potential_lo} violates scale invariance unless its coefficient vanishes. Consequently, the LO condition for the theory to have a scale-invariant IR limit is
\begin{equation}
\lambda_{\phi\chi}^2=\lambda_\phi\lambda_\chi,
\end{equation}
provided $\lambda_\phi > 0$ and $\lambda_\chi \geq 0$ for a consistent $1/N$ expansion and UV stability. Imposing this condition, the leading-order effective potential at a critical point takes the form
\begin{equation}
    V(\phi,\chi) = N \nu_d \left(\phi^2+\alpha\chi^2\right)^{\frac{d}{d-2}}  + O(N^{0}), \quad  \phi^2+\alpha\chi^2 \geq 0.
\end{equation}
At this order, no further condition fixes $\alpha$, which may therefore take any real value. The LO analysis thus yields a one-parameter family of conformal theories labeled by $\alpha$. For $\alpha < 0$, the effective potential also develops a flat direction $\phi^2 = - \alpha \chi^2$. Both properties are features of the exact $N=\infty$ limit. At NLO, the conformal manifold is reduced to isolated points, while the flat direction is lifted. We show later that a critical theory with negative $\alpha$ exhibits PSSB, so $\alpha < 0$ is the main focus of the present study.

\subsection{$d=3$ and the $\chi^6$ term} \label{sec: d3_chi6_addition}

At first sight, the $d=3$ theory appears to be on the same footing as the $3<d<4$ theories. At zero temperature, the first essential difference indeed arises only at NLO. However, finite temperature exposes the distinction already at LO: for negative $\alpha$, the $d=3$ theory at $N=\infty$ in its present form has no stable homogeneous minimum of the thermal effective potential.

At $T\neq 0$, the $d$-dimensional Euclidean integrals are replaced by finite-temperature sum-integrals
\begin{equation*}
\sum_K \equiv T\sum_{n\in\mathbb{Z}}\int\frac{d^{d-1}k}{(2\pi)^{d-1}}, \quad K = (k_0 , \vec{k}), \quad k_0 = 2\pi n T.    
\end{equation*}
Hence, the self-consistency equation for $\Sigma_{\eta_2}$ \eqref{eq: bicon_sigma_eta2_eq} is modified by a finite-temperature contribution. To simplify notation, we denote the LO self-energy by $m^2 \equiv \Sigma_{\eta_2}^{\text{LO}}$. In the IR limit,
\begin{equation}
    0 = \sum_K \frac{1}{K^2 + m^2} + \rho \,\,\,\,\,\stackrel{d=3}{\Rightarrow}\,\,\,\,\,  0 = - \frac{m}{4\pi} - \frac{T}{2\pi} \ln (1 - e^{-m/T}) + \rho,
\end{equation}
where we define $\rho \equiv \phi^2 + \alpha \chi^2$. In contrast to the $T=0$ solution \eqref{eq: sigma_eta2_lo}, which does not exist for negative values of $\rho$, this equation has a positive $m>0$ solution for every finite value of $\rho$. The limit $m \rightarrow 0$ is reached only as $\rho \rightarrow -\infty$, which for $\alpha <0$ corresponds to $\chi^2 \rightarrow \infty$ at fixed $\phi^2$. On the other hand, using the identity given in \eqref{eq: two_point_function_effective_potential} for large-$N$ rescaled fields, we have $\Sigma_{\eta_2}(0) = 2\partial_{\phi^2} V(\phi, \chi)/N$. The LO thermal effective potential $V_{\text{LO}}(\phi,\chi)$ is obtained by repeating the vacuum-diagram computation with the replacement $\int_k\rightarrow\sum_K$. However, we do not need its explicit form here; it is sufficient to note that, at this order, its field dependence enters only through $\rho$. Therefore,
\begin{equation}
    m^2  = 2\partial_{\rho} V_{\text{LO}}(\rho)/N.
\end{equation}
We want to find the minimum $(\bar\phi,\bar \chi)$ of the thermal effective potential. The $\chi$-stationarity condition gives
\begin{equation}
    0 = \frac{1}{N}\frac{\partial V_{\text{LO}}(\phi, \chi)}{\partial \chi} =\frac{1}{N} \frac{\partial V_{\text{LO}}(\rho)}{\partial \rho} \frac{\partial \rho}{\partial \chi} = \alpha \chi m^2. 
\end{equation}
Note that because $m$ is defined for all values of $\rho$, so is the thermal effective potential, justifying this differentiation. Since $m^2 > 0$ for every finite value of $\rho$, the stationarity condition with $\alpha \neq 0$ forces $\bar \chi = 0$. However, 
\begin{equation}
    \frac{1}{N}\frac{\partial^2 V_{\text{LO}}(\phi, \chi)}{\partial \chi^2}\Bigg|_{\chi = 0} = \alpha m^2,
\end{equation}
which is negative for $\alpha < 0$. Thus $\bar \chi = 0$ is unstable, while no other stationary point exists at finite $\chi$. Hence, for $\alpha < 0$, the $d=3$ theory at LO in its present form has no stable homogeneous minimum of the thermal effective potential.

This is an artifact of the strict $N=\infty$ limit. As the NLO analysis will reveal, a $\chi^6$ interaction is generated along the RG flow toward the IR, and this interaction cures the pathology. To see this, consider introducing a $g/(6!N^2)\chi^6$ term to the UV action \eqref{eq: bicon_uv_action}. At LO, following the large-$N$ field rescaling, this merely contributes an $N g\chi^6/6!$ term to $U(\phi,\chi)$, leaving the rest of the LO vacuum-diagram analysis unaffected. Consequently, the $\chi$-stationarity condition is modified as follows:
\begin{equation}
    0 = \frac{1}{N}\frac{\partial V_{\text{LO}}(\phi, \chi)}{\partial \chi} = \alpha \chi m^2 + \frac{g}{5!} \chi^5 \,\,\,\,\stackrel{\bar \chi \neq0}{\Rightarrow} \,\,\,\, \bar\chi^2 = \sqrt{5!}\,\sqrt{\frac{-\alpha}{g}} \,\,m. 
\end{equation}
This solution corresponds to stable minima for $\alpha < 0$ and $g > 0$.
We will return to this analysis once the NLO computation fixes these couplings to specific values. Meanwhile, the important lesson is that a theory at $N=\infty$ that appears well-behaved at zero temperature can reveal a pathology when probed at nonzero temperature. 

Hence, from now on we extend the UV action \eqref{eq: bicon_uv_action} by including the $\chi^6$ term,
\begin{equation} \label{eq: bicon_uv_action_chi6}
S[\phi,\chi]= \int d^d x \left\{
\frac12(\partial\phi)^2
+\frac12(\partial\chi)^2
+\frac{\lambda_\phi}{8N}\,(\phi^2)^2
+\frac{\lambda_{\phi\chi}}{4N}\,\phi^2\,\chi^2
+\frac{\lambda_\chi}{8N}\,\chi^4 + \frac{g}{6!N^2} \chi^6 \right\},
\end{equation}
where the UV coupling scales as $g \sim \Lambda^{6-2d}$. In the shifted UV action \eqref{eq: bicon_shited_action}, following the large-$N$ field rescaling, this generates additional interactions
\begin{equation}
\begin{split}
\Delta \tilde U(\eta,\xi)
= N \frac{g}{6!} \chi^6 + N^{\frac12}\frac{g \chi^5}{5!}\xi+\frac{g\chi^4}{4!2!}\xi^2
+N^{-\frac12}\frac{g \chi^3}{3! 3!}\,\xi^3\\
+\,N^{-1}\frac{g\chi^2}{4!2!}\xi^4 + N^{-\frac32}\frac{g\chi}{5!}\xi^5 + N^{-2}\frac{g}{6!}\xi^6.
\end{split}
\end{equation}
The critical LO effective potential at $T=0$ becomes
\begin{equation}
    V(\phi,\chi) = N \nu_d \left(\phi^2+\alpha\chi^2\right)^{\frac{d}{d-2}} + N \frac{g}{6!} \chi^6 \delta_{d,3}  + O(N^{0}), \quad  \phi^2+\alpha\chi^2 \geq 0.
\end{equation}
In the IR limit $\phi, \chi \ll \Lambda^{\frac{d-2}{2}}$, the $g\chi^6$ term vanishes for $3<d<4$, but survives in $d=3$. Consequently, unlike the $3<d<4$ case, where the LO potential has a flat direction for $\alpha<0$, the $d=3$ theory at $T=0$ has a unique vacuum at $\phi=\chi=0$ for $g>0$. In addition, the conformal manifold in $d=3$ is two-dimensional, parametrized by both $\alpha$ and $g$. The NLO computation fixes these values, reducing the manifold to isolated points.

\subsection{Self-energies at LO in $1/N$} \label{sec: self_energies_lo}
We now return to $T=0$ and compute the remaining LO self-energies needed for the NLO effective potential. The LO self-energy $\Sigma_{\eta_2}$ was already obtained in \eqref{eq: sigma_eta2_lo}; we turn to the $\eta_1$, $\xi$, and mixed $\eta_1 \xi$ self-energies. 

Diagrammatically, the self-energy $\Sigma_{\xi}(p)$ is given by
\begin{center}
    \tikzset{every picture/.style={line width=0.75pt}} 

\begin{tikzpicture}[x=0.75pt,y=0.75pt,yscale=-1,xscale=1]

\draw [color={rgb, 255:red, 255; green, 0; blue, 0 }  ,draw opacity=1 ]   (192.25,34.33) -- (156.25,34.33) ;
\draw    (162.41,21.33) .. controls (163.83,4.12) and (185.4,6.93) .. (185.36,21.33) ;
\draw    (185.36,21.33) .. controls (183.94,38.53) and (162.37,35.72) .. (162.41,21.33) ;
\draw [color={rgb, 255:red, 255; green, 0; blue, 0 }  ,draw opacity=1 ]   (277.48,34.45) -- (214.92,34.45) ;
\draw [color={rgb, 255:red, 255; green, 0; blue, 0 }  ,draw opacity=1 ]   (236.38,24.63) -- (244.54,32.79) ;
\draw [shift={(246.2,34.45)}, rotate = 45] [color={rgb, 255:red, 255; green, 0; blue, 0 }  ,draw opacity=1 ][line width=0.75]      (0, 0) circle [x radius= 3.35, y radius= 3.35]   ;
\draw [color={rgb, 255:red, 255; green, 0; blue, 0 }  ,draw opacity=1 ]   (256.03,24.63) -- (247.86,32.79) ;
\draw [shift={(246.2,34.45)}, rotate = 135] [color={rgb, 255:red, 255; green, 0; blue, 0 }  ,draw opacity=1 ][line width=0.75]      (0, 0) circle [x radius= 3.35, y radius= 3.35]   ;
\draw [color={rgb, 255:red, 255; green, 0; blue, 0 }  ,draw opacity=1 ]   (330.36,34.78) -- (295.52,34.78) ;
\draw    (334.01,34.85) .. controls (335.04,22.29) and (350.79,24.34) .. (350.76,34.85) ;
\draw    (350.76,34.85) .. controls (349.73,47.41) and (333.98,45.36) .. (334.01,34.85) ;

\draw [color={rgb, 255:red, 255; green, 0; blue, 0 }  ,draw opacity=1 ]   (431.6,34.85) -- (396.76,34.85) ;
\draw [line width=3]  [dash pattern={on 7.88pt off 4.5pt}]  (394.54,35.6) -- (351.01,35.6) ;
\draw [color={rgb, 255:red, 255; green, 0; blue, 0 }  ,draw opacity=1 ]   (406.58,25.02) -- (398.42,33.19) ;
\draw [shift={(396.76,34.85)}, rotate = 135] [color={rgb, 255:red, 255; green, 0; blue, 0 }  ,draw opacity=1 ][line width=0.75]      (0, 0) circle [x radius= 3.35, y radius= 3.35]   ;
\draw [color={rgb, 255:red, 255; green, 0; blue, 0 }  ,draw opacity=1 ]   (320.54,24.96) -- (328.7,33.12) ;
\draw [shift={(330.36,34.78)}, rotate = 45] [color={rgb, 255:red, 255; green, 0; blue, 0 }  ,draw opacity=1 ][line width=0.75]      (0, 0) circle [x radius= 3.35, y radius= 3.35]   ;

\draw (98,13.27) node [anchor=north west][inner sep=0.75pt]  [font=\normalsize]  {$\Sigma _{\xi }( p) =$};
\draw (196.4,15.27) node [anchor=north west][inner sep=0.75pt]    {$+$};
\draw (278.4,15.27) node [anchor=north west][inner sep=0.75pt]    {$+$};
\draw (443.2,11) node [anchor=north west][inner sep=0.75pt]    {$+\ O\left( N^{-1}\right) ,$};

\end{tikzpicture}
\end{center}
where the red line denotes the propagation of the $\xi$ field, the red double-spine denotes the quadratic $\xi^2$ interaction, and the red single-spine denotes the cubic $\xi \eta^2$ interaction. Algebraically, this reads
\begin{equation}
    \Sigma_\xi (p) = \frac{\lambda _{\phi \chi }}{2}\int _{k}\frac{1}{k^{2} +\Sigma _{\eta _{2}}} +\left(\frac{\lambda _{\phi \chi }}{2} \phi ^{2} +\frac{3\lambda _{\chi }}{2} \chi ^{2} + \frac{g}{4!}\chi^4\right) -( \lambda _{\phi \chi } \chi )^{2}\frac{\Pi _{\Sigma _{\eta _{2}}}( p)}{1+\lambda _{\phi } \Pi _{\Sigma _{\eta _{2}}}( p)} +\ O\left( N^{-1}\right),
\end{equation}
where $\Pi_{\Sigma_{\eta_2}}(p)$ is the bubble defined in \eqref{eq: bubble} with $m^2 = \Sigma_{\eta_2}$. Using equation \eqref{eq: bicon_sigma_eta2_eq}, we express the above integral in terms of $\Sigma_{\eta_2}$. Expanding the last fraction in the IR limit $\lambda_a \sim \Lambda^{4-d} \rightarrow \infty$ and collecting terms, we find
\begin{align}
    \Sigma_\xi (p) & =\frac{\lambda_{\phi\chi}}{\lambda_\phi} \Sigma_{\eta_2} + \frac{3}{2} \left(\lambda_\chi- \frac{\lambda_{\phi\chi}^2}{\lambda_\phi}\right)\chi^2+ \frac{g}{4!}\chi^4 + \frac{\lambda_{\phi\chi}^2}{\lambda_\phi^2} \frac{1}{\Pi_{\Sigma_{\eta_2}}(p)} \chi^2 + O(1/\lambda_a) + O(N^{-1}) \nonumber \\ &\stackrel{\text{IR, criticality}}{=} \alpha \,\Sigma_{\eta_2} + \frac{\alpha^2 \chi^2}{\Pi_{\Sigma_{\eta_2}}(p)} + \frac{g}{4!}\chi^4 \delta_{d,3} + O(N^{-1}),
\end{align}
where in the last equality we used the criticality condition $\lambda_{\phi\chi}^2 = \lambda_\phi \lambda_\chi$, and the fact that in the IR limit $g \sim \Lambda^{6-2d}$ survives only at $d=3$. 

Since the background field $\phi$ breaks the symmetry between $\eta_1$ and the remaining $\eta_{2\dots N}$ components \eqref{eq: bicon_shited_action}, their self-energies differ. Diagrammatically, the self-energy $\Sigma_{\eta_1}(p)$ is given by
\begin{center}
    \tikzset{every picture/.style={line width=0.75pt}} 

\begin{tikzpicture}[x=0.75pt,y=0.75pt,yscale=-1,xscale=1]

\draw    (198.59,37.39) .. controls (196.92,39.06) and (195.26,39.06) .. (193.59,37.39) .. controls (191.92,35.72) and (190.26,35.72) .. (188.59,37.39) .. controls (186.92,39.06) and (185.26,39.06) .. (183.59,37.39) .. controls (181.92,35.72) and (180.26,35.72) .. (178.59,37.39) .. controls (176.92,39.06) and (175.26,39.06) .. (173.59,37.39) .. controls (171.92,35.72) and (170.26,35.72) .. (168.59,37.39) .. controls (166.92,39.06) and (165.26,39.06) .. (163.59,37.39) -- (162.59,37.39) -- (162.59,37.39) ;
\draw    (168.74,24.39) .. controls (170.16,7.18) and (191.73,9.99) .. (191.69,24.39) ;
\draw    (191.69,24.39) .. controls (190.28,41.59) and (168.7,38.78) .. (168.74,24.39) ;
\draw    (282.15,36.85) .. controls (280.48,38.52) and (278.82,38.52) .. (277.15,36.85) .. controls (275.48,35.18) and (273.82,35.18) .. (272.15,36.85) .. controls (270.48,38.52) and (268.82,38.52) .. (267.15,36.85) .. controls (265.48,35.18) and (263.82,35.18) .. (262.15,36.85) .. controls (260.48,38.52) and (258.82,38.52) .. (257.15,36.85) .. controls (255.48,35.18) and (253.82,35.18) .. (252.15,36.85) .. controls (250.48,38.52) and (248.82,38.52) .. (247.15,36.85) .. controls (245.48,35.18) and (243.82,35.18) .. (242.15,36.85) .. controls (240.48,38.52) and (238.82,38.52) .. (237.15,36.85) .. controls (235.48,35.18) and (233.82,35.18) .. (232.15,36.85) .. controls (230.48,38.52) and (228.82,38.52) .. (227.15,36.85) .. controls (225.48,35.18) and (223.82,35.18) .. (222.15,36.85) -- (219.59,36.85) -- (219.59,36.85) ;
\draw    (337.87,36.85) .. controls (336.2,38.52) and (334.54,38.52) .. (332.87,36.85) .. controls (331.2,35.18) and (329.54,35.18) .. (327.87,36.85) .. controls (326.2,38.52) and (324.54,38.52) .. (322.87,36.85) .. controls (321.2,35.18) and (319.54,35.18) .. (317.87,36.85) .. controls (316.2,38.52) and (314.54,38.52) .. (312.87,36.85) .. controls (311.2,35.18) and (309.54,35.18) .. (307.87,36.85) -- (306.59,36.85) -- (306.59,36.85) ;
\draw    (437.43,36.85) .. controls (435.76,38.52) and (434.1,38.52) .. (432.43,36.85) .. controls (430.76,35.18) and (429.1,35.18) .. (427.43,36.85) .. controls (425.76,38.52) and (424.1,38.52) .. (422.43,36.85) .. controls (420.76,35.18) and (419.1,35.18) .. (417.43,36.85) .. controls (415.76,38.52) and (414.1,38.52) .. (412.43,36.85) .. controls (410.76,35.18) and (409.1,35.18) .. (407.43,36.85) -- (406.15,36.85) -- (406.15,36.85) ;
\draw [line width=3.75]  [dash pattern={on 9pt off 4.5pt}]  (356.6,36.63) -- (406.15,36.85) ;
\draw    (342.08,36.81) .. controls (342.99,25.75) and (356.85,27.56) .. (356.82,36.81) ;
\draw    (356.82,36.81) .. controls (355.91,47.86) and (342.05,46.06) .. (342.08,36.81) ;

\draw [color={rgb, 255:red, 0; green, 0; blue, 0 }  ,draw opacity=1 ]   (241.04,27.02) -- (249.21,35.19) ;
\draw [shift={(250.87,36.85)}, rotate = 45] [color={rgb, 255:red, 0; green, 0; blue, 0 }  ,draw opacity=1 ][line width=0.75]      (0, 0) circle [x radius= 3.35, y radius= 3.35]   ;
\draw [color={rgb, 255:red, 0; green, 0; blue, 0 }  ,draw opacity=1 ]   (260.69,27.02) -- (252.53,35.19) ;
\draw [shift={(250.87,36.85)}, rotate = 135] [color={rgb, 255:red, 0; green, 0; blue, 0 }  ,draw opacity=1 ][line width=0.75]      (0, 0) circle [x radius= 3.35, y radius= 3.35]   ;
\draw [color={rgb, 255:red, 0; green, 0; blue, 0 }  ,draw opacity=1 ]   (328.04,27.02) -- (336.21,35.19) ;
\draw [shift={(337.87,36.85)}, rotate = 45] [color={rgb, 255:red, 0; green, 0; blue, 0 }  ,draw opacity=1 ][line width=0.75]      (0, 0) circle [x radius= 3.35, y radius= 3.35]   ;
\draw [color={rgb, 255:red, 0; green, 0; blue, 0 }  ,draw opacity=1 ]   (415.97,27.02) -- (407.81,35.19) ;
\draw [shift={(406.15,36.85)}, rotate = 135] [color={rgb, 255:red, 0; green, 0; blue, 0 }  ,draw opacity=1 ][line width=0.75]      (0, 0) circle [x radius= 3.35, y radius= 3.35]   ;

\draw (97.67,17.97) node [anchor=north west][inner sep=0.75pt]  [font=\normalsize]  {$\Sigma _{\eta _{1}}( p) =$};
\draw (202.73,19.33) node [anchor=north west][inner sep=0.75pt]    {$+$};
\draw (287.73,19.33) node [anchor=north west][inner sep=0.75pt]    {$+$};
\draw (447.73,15.67) node [anchor=north west][inner sep=0.75pt]    {$+\ O\left( N^{-1}\right) ,$};

\end{tikzpicture}
\end{center}
where the wavy line denotes the propagation of the $\eta_1$ field, the double-spine denotes the quadratic $\eta_1^2$ interaction, and the single-spine denotes the cubic $\eta_1 \eta^2$ interaction. Algebraically, this reads 
\begin{align}
    \Sigma_{\eta_1}(p) &= \ \frac{\lambda _{\phi } }{2}\int _{k}\frac{1}{k^{2} +\Sigma _{\eta _{2}}} +\left(\frac{3\lambda _{\phi }}{2} \phi ^{2} +\frac{\lambda _{\phi \chi }}{2} \chi ^{2}\right) -( \lambda _{\phi } \phi )^{2}\frac{\Pi _{\Sigma _{\eta _{2}}}( p)}{1+\lambda _{\phi } \Pi _{\Sigma _{\eta _{2}}}( p)} + O(N^{-1}) \nonumber \\ &\stackrel{\text{IR}}{=}  \,\,\Sigma_{\eta_2} + \frac{\phi^2}{\Pi_{\Sigma_{\eta_2}}(p)} + O(N^{-1}).
\end{align}
Diagrammatically, the mixed self-energy $\Sigma_{\eta_1\xi}(p)$ is given by
\begin{center}
    \tikzset{every picture/.style={line width=0.75pt}} 

\begin{tikzpicture}[x=0.75pt,y=0.75pt,yscale=-1,xscale=1]

\draw    (234.87,36.52) .. controls (233.2,38.19) and (231.54,38.19) .. (229.87,36.52) .. controls (228.2,34.85) and (226.54,34.85) .. (224.87,36.52) .. controls (223.2,38.19) and (221.54,38.19) .. (219.87,36.52) .. controls (218.2,34.85) and (216.54,34.85) .. (214.87,36.52) .. controls (213.2,38.19) and (211.54,38.19) .. (209.87,36.52) .. controls (208.2,34.85) and (206.54,34.85) .. (204.87,36.52) -- (203.59,36.52) -- (203.59,36.52) ;
\draw    (326.87,36.52) .. controls (325.2,38.19) and (323.54,38.19) .. (321.87,36.52) .. controls (320.2,34.85) and (318.54,34.85) .. (316.87,36.52) .. controls (315.2,38.19) and (313.54,38.19) .. (311.87,36.52) .. controls (310.2,34.85) and (308.54,34.85) .. (306.87,36.52) .. controls (305.2,38.19) and (303.54,38.19) .. (301.87,36.52) .. controls (300.2,34.85) and (298.54,34.85) .. (296.87,36.52) -- (295.59,36.52) -- (295.59,36.52) ;
\draw [color={rgb, 255:red, 255; green, 0; blue, 0 }  ,draw opacity=1 ]   (426.43,36.52) -- (395.15,36.52) ;
\draw [line width=3.75]  [dash pattern={on 9pt off 4.5pt}]  (345.6,36.3) -- (395.15,36.52) ;
\draw    (331.08,36.48) .. controls (331.99,25.42) and (345.85,27.23) .. (345.82,36.48) ;
\draw    (345.82,36.48) .. controls (344.91,47.53) and (331.05,45.73) .. (331.08,36.48) ;

\draw [color={rgb, 255:red, 255; green, 0; blue, 0 }  ,draw opacity=1 ]   (266.15,36.52) -- (234.87,36.52) ;
\draw [color={rgb, 255:red, 0; green, 0; blue, 0 }  ,draw opacity=1 ]   (225.04,26.69) -- (233.21,34.86) ;
\draw [shift={(234.87,36.52)}, rotate = 45] [color={rgb, 255:red, 0; green, 0; blue, 0 }  ,draw opacity=1 ][line width=0.75]      (0, 0) circle [x radius= 3.35, y radius= 3.35]   ;
\draw [color={rgb, 255:red, 255; green, 0; blue, 0 }  ,draw opacity=1 ]   (244.69,26.69) -- (236.53,34.86) ;
\draw [shift={(234.87,36.52)}, rotate = 135] [color={rgb, 255:red, 255; green, 0; blue, 0 }  ,draw opacity=1 ][line width=0.75]      (0, 0) circle [x radius= 3.35, y radius= 3.35]   ;
\draw [color={rgb, 255:red, 0; green, 0; blue, 0 }  ,draw opacity=1 ]   (317.04,26.69) -- (325.21,34.86) ;
\draw [shift={(326.87,36.52)}, rotate = 45] [color={rgb, 255:red, 0; green, 0; blue, 0 }  ,draw opacity=1 ][line width=0.75]      (0, 0) circle [x radius= 3.35, y radius= 3.35]   ;
\draw [color={rgb, 255:red, 255; green, 0; blue, 0 }  ,draw opacity=1 ]   (404.97,26.69) -- (396.81,34.86) ;
\draw [shift={(395.15,36.52)}, rotate = 135] [color={rgb, 255:red, 255; green, 0; blue, 0 }  ,draw opacity=1 ][line width=0.75]      (0, 0) circle [x radius= 3.35, y radius= 3.35]   ;

\draw (131.67,25.64) node [anchor=north west][inner sep=0.75pt]  [font=\normalsize]  {$\Sigma _{\eta _{1} \xi }( p) =$};
\draw (273.73,28) node [anchor=north west][inner sep=0.75pt]    {$+$};
\draw (436.73,24.34) node [anchor=north west][inner sep=0.75pt]    {$+\ O\left( N^{-1}\right) ,$};

\end{tikzpicture}
\end{center}
where the double-spine denotes the quadratic $\eta_1 \xi$ interaction. Algebraically, this reads
\begin{align}
    \Sigma_{\eta_1 \xi}(p) &= 
\lambda _{\phi \chi } \phi \chi -( \lambda _{\phi } \phi )( \lambda _{\phi \chi } \chi )\frac{\Pi _{\Sigma _{\eta _{2}}}( p)}{1+\lambda _{\phi } \Pi _{\Sigma _{\eta _{2}}}(p)} + O(N^{-1}) = \frac{\lambda _{\phi \chi } \phi \chi}{1+\lambda _{\phi } \Pi _{\Sigma _{\eta _{2}}}(p)}+O\left( N^{-1}\right) \nonumber\\ &\stackrel{\text{IR}}{=} \frac{\alpha \phi \chi}{\Pi _{\Sigma _{\eta _{2}}}(p)}+O\left( N^{-1}\right).
\end{align}

\section{The Effective Potential at NLO in $1/N$} \label{sec: NLO}
We now turn to the computation of the zero-temperature effective potential at NLO. To simplify notation, we denote the LO self-energy $\Sigma_{\eta_2}$ by $m^2$, which, from the results of the preceding section, is given by
\begin{equation} \label{eq: bicon_m_definition}
    m^2 = \frac{2d}{d-2}\nu_d\left(\phi^2 + \alpha \chi^2\right)^{\frac{2}{d-2}}.
\end{equation}
Several classes of vacuum diagrams contribute to the effective potential at NLO. The first class consists of necklace diagrams, built from $\eta_{i=2\dots N}$ loops,
\begin{center}
\tikzset{every picture/.style={line width=0.75pt}} 

\begin{tikzpicture}[x=0.75pt,y=0.75pt,yscale=-1,xscale=1]

\draw [color={rgb, 255:red, 0; green, 0; blue, 0 }  ,draw opacity=1 ]   (121.41,77.33) .. controls (122.83,60.12) and (144.4,62.93) .. (144.36,77.33) ;
\draw [color={rgb, 255:red, 0; green, 0; blue, 0 }  ,draw opacity=1 ]   (144.36,77.33) .. controls (142.94,94.53) and (121.37,91.72) .. (121.41,77.33) ;

\draw [color={rgb, 255:red, 0; green, 0; blue, 0 }  ,draw opacity=1 ]   (132.41,56.33) .. controls (133.83,39.12) and (155.4,41.93) .. (155.36,56.33) ;
\draw [color={rgb, 255:red, 0; green, 0; blue, 0 }  ,draw opacity=1 ]   (155.36,56.33) .. controls (153.94,73.53) and (132.37,70.72) .. (132.41,56.33) ;

\draw [color={rgb, 255:red, 0; green, 0; blue, 0 }  ,draw opacity=1 ]   (109.41,56.33) .. controls (110.83,39.12) and (132.4,41.93) .. (132.36,56.33) ;
\draw [color={rgb, 255:red, 0; green, 0; blue, 0 }  ,draw opacity=1 ]   (132.36,56.33) .. controls (130.94,73.53) and (109.37,70.72) .. (109.41,56.33) ;

\draw [color={rgb, 255:red, 0; green, 0; blue, 0 }  ,draw opacity=1 ]   (78.89,52.85) .. controls (96.09,54.27) and (93.28,75.84) .. (78.89,75.8) ;
\draw [color={rgb, 255:red, 0; green, 0; blue, 0 }  ,draw opacity=1 ]   (78.89,75.8) .. controls (61.68,74.39) and (64.49,52.81) .. (78.89,52.85) ;
\draw [color={rgb, 255:red, 0; green, 0; blue, 0 }  ,draw opacity=1 ]   (199.41,39.33) .. controls (200.83,22.12) and (222.4,24.93) .. (222.36,39.33) ;
\draw [color={rgb, 255:red, 0; green, 0; blue, 0 }  ,draw opacity=1 ]   (222.36,39.33) .. controls (220.94,56.53) and (199.37,53.72) .. (199.41,39.33) ;

\draw [color={rgb, 255:red, 0; green, 0; blue, 0 }  ,draw opacity=1 ]   (217.41,23.33) .. controls (218.83,6.12) and (240.4,8.93) .. (240.36,23.33) ;
\draw [color={rgb, 255:red, 0; green, 0; blue, 0 }  ,draw opacity=1 ]   (240.36,23.33) .. controls (238.94,40.53) and (217.37,37.72) .. (217.41,23.33) ;

\draw [color={rgb, 255:red, 0; green, 0; blue, 0 }  ,draw opacity=1 ]   (197.41,63.33) .. controls (198.83,46.12) and (220.4,48.93) .. (220.36,63.33) ;
\draw [color={rgb, 255:red, 0; green, 0; blue, 0 }  ,draw opacity=1 ]   (220.36,63.33) .. controls (218.94,80.53) and (197.37,77.72) .. (197.41,63.33) ;

\draw [color={rgb, 255:red, 0; green, 0; blue, 0 }  ,draw opacity=1 ]   (240.41,20.33) .. controls (241.83,3.12) and (263.4,5.93) .. (263.36,20.33) ;
\draw [color={rgb, 255:red, 0; green, 0; blue, 0 }  ,draw opacity=1 ]   (263.36,20.33) .. controls (261.94,37.53) and (240.37,34.72) .. (240.41,20.33) ;

\draw [color={rgb, 255:red, 0; green, 0; blue, 0 }  ,draw opacity=1 ]   (206.41,85.33) .. controls (207.83,68.12) and (229.4,70.93) .. (229.36,85.33) ;
\draw [color={rgb, 255:red, 0; green, 0; blue, 0 }  ,draw opacity=1 ]   (229.36,85.33) .. controls (227.94,102.53) and (206.37,99.72) .. (206.41,85.33) ;

\draw [color={rgb, 255:red, 0; green, 0; blue, 0 }  ,draw opacity=1 ]   (228.41,94.33) .. controls (229.83,77.12) and (251.4,79.93) .. (251.36,94.33) ;
\draw [color={rgb, 255:red, 0; green, 0; blue, 0 }  ,draw opacity=1 ]   (251.36,94.33) .. controls (249.94,111.53) and (228.37,108.72) .. (228.41,94.33) ;

\draw [color={rgb, 255:red, 0; green, 0; blue, 0 }  ,draw opacity=1 ]   (261.41,30.33) .. controls (262.83,13.12) and (284.4,15.93) .. (284.36,30.33) ;
\draw [color={rgb, 255:red, 0; green, 0; blue, 0 }  ,draw opacity=1 ]   (284.36,30.33) .. controls (282.94,47.53) and (261.37,44.72) .. (261.41,30.33) ;

\draw [color={rgb, 255:red, 0; green, 0; blue, 0 }  ,draw opacity=1 ]   (251.41,90.33) .. controls (252.83,73.12) and (274.4,75.93) .. (274.36,90.33) ;
\draw [color={rgb, 255:red, 0; green, 0; blue, 0 }  ,draw opacity=1 ]   (274.36,90.33) .. controls (272.94,107.53) and (251.37,104.72) .. (251.41,90.33) ;

\draw [color={rgb, 255:red, 0; green, 0; blue, 0 }  ,draw opacity=1 ]   (270.41,75.33) .. controls (271.83,58.12) and (293.4,60.93) .. (293.36,75.33) ;
\draw [color={rgb, 255:red, 0; green, 0; blue, 0 }  ,draw opacity=1 ]   (293.36,75.33) .. controls (291.94,92.53) and (270.37,89.72) .. (270.41,75.33) ;

\draw [color={rgb, 255:red, 0; green, 0; blue, 0 }  ,draw opacity=1 ]   (272.41,51.33) .. controls (273.83,34.12) and (295.4,36.93) .. (295.36,51.33) ;
\draw [color={rgb, 255:red, 0; green, 0; blue, 0 }  ,draw opacity=1 ]   (295.36,51.33) .. controls (293.94,68.53) and (272.37,65.72) .. (272.41,51.33) ;

\draw [color={rgb, 255:red, 0; green, 0; blue, 0 }  ,draw opacity=1 ]   (35.32,52.28) .. controls (47.75,53.23) and (45.42,71.8) .. (35.02,71.77) ;
\draw [color={rgb, 255:red, 0; green, 0; blue, 0 }  ,draw opacity=1 ]   (35.02,71.77) .. controls (22.59,70.82) and (24.92,52.25) .. (35.32,52.28) ;
\draw [color={rgb, 255:red, 0; green, 0; blue, 0 }  ,draw opacity=1 ]   (35.32,52.28) .. controls (54.47,48.84) and (50.47,77.84) .. (35.74,76.83) ;
\draw [color={rgb, 255:red, 0; green, 0; blue, 0 }  ,draw opacity=1 ]   (35.74,76.83) .. controls (17.47,76.84) and (17.68,48.66) .. (35.32,52.28) ;
\draw [color={rgb, 255:red, 0; green, 0; blue, 0 }  ,draw opacity=1 ]   (78.89,52.85) .. controls (89.47,59.84) and (88.47,68.84) .. (78.89,75.8) ;
\draw [color={rgb, 255:red, 0; green, 0; blue, 0 }  ,draw opacity=1 ]   (78.89,75.8) .. controls (71.47,69.84) and (66.47,61.84) .. (78.89,52.85) ;

\draw (95.4,58.27) node [anchor=north west][inner sep=0.75pt]    {$+$};
\draw (156.86,58.39) node [anchor=north west][inner sep=0.75pt]    {$+...+$};
\draw (301.86,58.39) node [anchor=north west][inner sep=0.75pt]    {$+...$};
\draw (51,59) node [anchor=north west][inner sep=0.75pt]    {$+$};
\draw (33.86,114.39) node [anchor=north west][inner sep=0.75pt]    {$=\frac{1}{2}\int _{k}\left[ \lambda _{\phi } \Pi _{m^{2}}( k) -\frac{1}{2} \lambda _{\phi }^{2} \Pi _{m^{2}}^{2}( k) +\frac{1}{3} \lambda _{\phi }^{3} \Pi _{m^{2}}^{3}( k) +...\right] =\frac{1}{2}\int _{k}\ln( 1+\lambda _{\phi } \Pi _{m^{2}}( k)) .$};

\end{tikzpicture}
\end{center}
In the IR limit, this contribution reduces to $\frac{1}{2} \int_k \ln \Pi_{m^2}(k)$, up to power-law $\Lambda$ terms canceled by the appropriate counterterms. The next class of diagrams consists of loops in the $(\eta_1, \xi)$ sector 
\begin{center}
\input{pics/bicon_nlo_vacuum_Gpsi_loop}
\end{center}
where the blue wavy line denotes the massless $\eta_1$ propagator $1/k^2$, the pink line denotes the massless $\xi$ propagator $1/k^2$, and the self-energy insertions are evaluated at LO. Since the massless Gaussian determinant for $\eta_1$ was already accounted for at LO \eqref{eq: bicon_effective_potential_lo_preirlimit}, we need only add the same contribution for the $\xi$ field to obtain
\begin{equation}
    \frac{1}{2} \int_k \ln \det G_{\Psi}^{-1}(k) - \frac{1}{2}\int_k \ln k^2, 
\end{equation}
where $G_\Psi$ denotes the propagator matrix in the $\Psi \equiv(\eta_1, \xi)$ sector, 
\begin{equation}
    G_\Psi(k) = \left(\begin{matrix}
        k^2+\Sigma_{\eta_1}(k) & \Sigma_{\eta_1 \xi}(k) \\
        \Sigma_{\eta_1 \xi}(k) & k^2 + \Sigma_\xi(k)
    \end{matrix} \right)^{-1}.
\end{equation}

Further NLO contributions arise from the LO vacuum diagrams evaluated in Section~\ref{sec: LO}.
First, each loop in these diagrams sums over the $\eta_{i=2\dots N}$ fields, so it has an $(N-1)$ enhancement, not $N$. Hence, in the equation \eqref{eq: effective_potential_sum_diagrams} the amplitude of a diagram is $A(D) = (1-1/N)^{L(D)} A_{\text{LO}}(D) = A_{\text{LO}}(D) - \frac{1}{N} L(D) A_{\text{LO}}(D) + O(N^{-1}).$ Therefore, the NLO contribution is 
\begin{equation}
     -\frac{1}{N} \sum_{D\in \text{LO}} L(D) A_{\text{LO}}(D)= -\frac{1}{2}\int_k \ln \left(1 + \frac{m^2}{k^2}\right).
\end{equation}
Second, the value of $\alpha = \frac{\lambda_{\phi\chi}}{\lambda_\phi}$ can be expanded in powers of $1/N$, i.e., $\alpha = \alpha_0 + \alpha_1/N + \dots$. At LO, every occurrence of $\alpha$ in the vacuum diagrams is therefore replaced by $\alpha_0$. The NLO correction proportional to $\alpha_1$ is obtained by replacing one occurrence of $\alpha_0$ by $\alpha_1/N$ in each LO vacuum diagram and summing over all possible replacements. Hence, this correction reads
\begin{equation} \label{eq: alpha1}
    \frac{\alpha_1}{N} \partial_{\alpha_0} V_{\text{LO}}(\phi,\chi) = \frac{d}{d-2} \nu_d \alpha_1 \chi^2 (\phi^2 + \alpha_0 \chi^2)^{\frac{2}{d-2}}.
\end{equation}
Recognizing this as a simple Taylor expansion of the LO potential and collecting all the above contributions, we obtain the effective potential up to NLO
\begin{multline} \label{eq: bicon_effective_potential_general_nlo} 
    V(\phi, \chi) = N \nu_d (\phi^2 + \alpha \chi^2)^{\frac{d}{d-2}} + N\frac{g}{6!} \chi^6 \delta_{d,3} + \frac{1}{2} \int_k \ln \left[\det G_{\Psi}^{-1}(k) \frac{\Pi_{m^2}(k)}{k^2 + m^2} \right] + O(N^{-1}), \\[0.1cm]
    \phi^2+\alpha_0\chi^2 \geq 0,
\end{multline}
where in the first term $\alpha = \alpha_0 + \alpha_1/N$, whereas in the NLO integral term it is sufficient to set $\alpha = \alpha_0$. The coupling $g$ can likewise be expanded as $g = g_0 + g_1/N + O(N^{-2})$. Substituting $G_\Psi^{-1}$ and the self-energies from Section \ref{sec: self_energies_lo}, we obtain 
\begin{equation} \label{eq: bicon_effective_potential_general_nlo_expanded}
    \begin{split}
    V(\phi, \chi) =\, & N \nu_d (\phi^2 + \alpha \chi^2)^{\frac{d}{d-2}} + N\frac{g}{6!} \chi^6 \delta_{d,3} \\ &+\frac{1}{2} \int_k \ln \Bigg[\Pi_{m^2}(k) (k^2 + \alpha m^2) + \frac{(\phi^2 + \alpha^2\chi^2)k^2 + \alpha(\phi^2 + \alpha\chi^2)m^2}{k^2 + m^2} \\ &\qquad\qquad\qquad\qquad\,\, +\left(\Pi_{m^2}(k)+\frac{\phi^2}{k^2+m^2}\right)\frac{g}{4!}\chi^4\delta_{d,3}\Bigg] + O(N^{-1}), 
    \end{split}
\end{equation}
where $\phi^2+\alpha_0\chi^2 \geq 0$. To analyze scale invariance at NLO, it is enough to keep the LO effective potential together with the logarithmically divergent part of the NLO contribution. As before, we discard power-law divergences. Extracting the logarithmic divergence via the large-momentum expansion (see Appendix~\ref{app: nlo_log_div}), and simplifying with \eqref{eq: bicon_m_definition}, we find
\begin{equation} \label{eq: bicon_effective_potential_nlo_log}
    \begin{split}
    V(\phi, \chi) = \,\,& N \nu_d (\phi^2 + \alpha \chi^2)^{\frac{d}{d-2}} +N\frac{g}{6!} \chi^6 \delta_{d,3}
    \\  + & \left\{
    C_1 m^d + C_2\left(b\,m^2\chi^2+\chi^6 \delta_{d,3}(c+\alpha^2 g/4!)\right)\right\} \ln \left(\frac{\Lambda }{m}\right) + O(N^0\Lambda^0, N^{-1}), 
        \end{split}
\end{equation}
where
\begin{equation*}
\begin{aligned}
    &\quad C_1 = \frac{2^{-d}(4-d)\Gamma(d-1)}{d\,\pi^{d/2}\Gamma\!\left(\frac d2\right)^3},\quad C_2 = \frac{\Gamma(d-1)\sin\!\left(\frac{\pi d}{2}\right)}{\pi \Gamma\!\left(\frac d2\right)^2}, \\ &b = 
\alpha^2(\alpha+2d-6)
-
\alpha(2d-5), \quad c = 256 (1-\alpha)^3\alpha^3/3.
\end{aligned}
\end{equation*}
Since the anomalous dimensions of the $\phi$ and $\chi$ fields, \eqref{eq: bicon_gammaphi} and \eqref{eq: bicon_gammachi}, are $O(N^{-1})$, applying the scale variation \eqref{eq: scale_variation} to the LO effective potential $V_{\text{LO}}(\phi,\chi)$ produces NLO terms. The finite NLO part of the effective potential is classically scale-invariant: the integrand in \eqref{eq: bicon_effective_potential_general_nlo_expanded} contains no fixed dimensionful scales. Its scaling variation is therefore NNLO, so we neglect it. Hence, only the logarithmically divergent NLO terms in $V(\phi,\chi)$ can cancel the NLO variation of $V_{\text{LO}}(\phi,\chi)$. Indeed, $\hat \delta_s (L(\phi,\chi) \ln(\Lambda/m)) = -L(\phi,\chi) +O(N^{-1})$ for a classically scale-invariant $L(\phi,\chi)$.

We impose scale invariance of the effective potential by requiring that $\hat \delta_s V(\phi,\chi)=0$. Since this equation must hold for arbitrary field values, the coefficients of the linearly independent functions of $\phi$ and $\chi$ that appear in it must each vanish. These conditions determine the couplings of the critical theory. In $3<d<4$, the effective potential is scale invariant when $\alpha$ takes one of the following values:
\begin{equation}
    \alpha = 0, \qquad \alpha = 1, \qquad\alpha = -\frac{(d-1)(d-2)}{2} + O(N^{-1}). 
\end{equation}
Together with the LO criticality condition $\lambda_{\phi\chi}^2 = \lambda_\phi \lambda_\chi$, these solutions have the following interpretations. For $\alpha = 0$ with a positive $\lambda_\phi$, we find $\lambda_{\phi\chi} = \lambda_\chi = 0$, meaning the theory is a decoupled product of the critical $O(N)$ model and a free massless scalar. For $\alpha = 1$, all three quartic couplings coincide, and the theory reduces to the critical $O(N+1)$ model. The remaining solution has $\lambda_{\phi\chi}<0$ and corresponds to the critical biconical model, on which we focus henceforth.

For the $d=3$ critical biconical model, the condition $\hat \delta_s V(\phi,\chi) = 0$ fixes both $\alpha$ and $g$:  
\begin{equation}
    \alpha = -1 + O(N^{-1}), \quad g = 15360 + O(N^{-1}).
\end{equation}
At NLO in $d=3$, a scale-invariant theory requires the $\chi^6$ interaction for $\alpha \neq 0,1$. Indeed, introducing this term is the only way to cancel the scale variation of the logarithmic term $\sim c\chi^6 \ln(\Lambda/m)$ in \eqref{eq: bicon_effective_potential_nlo_log}, which arises only in $d=3$.

Thus, the conformal manifold found at LO disappears: at criticality, $\alpha$ is restricted to isolated values, and in $d=3$ for $\alpha \neq 0$, $g$ is fixed as well. When $\alpha$ and $g$ take their critical values, the LO part and the logarithmically divergent NLO part of the effective potential can be combined into a form in which scale invariance is manifest:
\begin{equation} \label{eq: bicon_effective_potential_nlo_log_pretty}
\begin{split}
    V(\phi, \chi) = & \nu_d  N \left(\phi ^2 \left(\frac{\Lambda }{m}\right)^{2 \gamma_\phi }+\alpha  \chi ^2 \left(\frac{\Lambda }{m}\right)^{2 \gamma_\chi }\right)^{\frac{d}{d-2}}\\ & \hspace{3.05cm}+  N\frac{g}{6!} \chi^6  \left(\frac{\Lambda}{\chi^{\frac{2}{d-2}}}\right)^{6\gamma_\chi} \delta_{d,3} + O(N^0\Lambda^{0}, N^{-1}).
\end{split}
\end{equation}
For $3<d<4$, the LO effective potential has a flat direction along which the degenerate vacua satisfy $\phi^2 + \alpha_0 \chi^2 = 0$. Evaluating the finite NLO part of the effective potential along this direction yields (see Appendix \ref{app: integral_restricted_effective_potential} for the derivation)
\begin{align} \label{eq: bicon_effective_potential_nlo_restricted_m0}
    V(\phi,\chi)\big|_{\phi^2 + \alpha \chi^2 = 0} = \frac{\csc \left(\frac{\pi  d}{d-2}\right) \left(-4^{d-1} \pi ^{\frac{d-3}{2}} \sin \left(\frac{\pi  d}{2}\right) \Gamma \left(\frac{d-1}{2}\right)\right)^{\frac{d}{d-2}}}{2^d \pi ^{\frac{d}{2}-1} d \Gamma
   \left(\frac{d}{2}\right)} (1-\alpha)^{\frac{d}{d-2}} (\phi^2&)^{\frac{d}{d-2}}\nonumber \\ {}+ O(N^{-1}), \quad 3<d<4.&
\end{align}
The positive coefficient of $(\phi^2)^{\frac{d}{d-2}}$ lifts the LO degeneracy, leaving $\phi=\chi=0$ as the unique stable vacuum. In $d=3$, the theory admits this vacuum already at LO due to the $\chi^6$ term. The above expression has a pole as $d\rightarrow 3$, but the $g$-dependent contribution present in $d=3$ cancels it; numerical evaluation yields a finite result.

As discussed above, for $\rho=\phi^2+\alpha\chi^2<0$, the effective potential is ill-defined because no linear source can stabilize a field-space-localized state with such field averages. For the negative-$\alpha$ critical theory in dimensions $3<d<4$, configurations with $\rho>0$ are also unstable. Indeed, since $\Pi_{m^2}(0) > 0$ and $\alpha < 0$, $\det G_\Psi^{-1}(0)=\alpha m^2\bigl(m^2+\rho/\Pi_{m^2}(0)\bigr)<0$, so the logarithm in the NLO effective potential \eqref{eq: bicon_effective_potential_general_nlo} acquires an imaginary part. Therefore, for $3<d<4$, all stable field-space-localized states lie on the $\rho=0$ locus, and their energy densities are given by the restricted effective potential \eqref{eq: bicon_effective_potential_nlo_restricted_m0}.

\subsection{The $d \to 4$ limit}
In $d=4-\epsilon$, the biconical model can be studied perturbatively in the spirit of the Wilson–Fisher fixed point. We write down the beta functions and, from their common zeros, determine the fixed-point couplings. Since these couplings are $O(\epsilon)$, substituting them into the classical potential yields the effective potential at leading order in $\epsilon$. The beta functions are given in~\cite{Chai:2020zgq}: 
\begin{equation}
\begin{aligned}
    &\beta_{\lambda_{\phi}} = -\epsilon \lambda_{\phi} + \frac{(N + 8)}{16\pi^2N} \lambda_{\phi}^2  + \frac{\lambda_{\phi\chi}^2}{16\pi^2N} +O(\lambda_a^3), \,\,\,\,\,\\ &
    \beta_{\lambda_\chi} = -\epsilon {\lambda_\chi} + \frac{9 {\lambda_\chi}^2}{16\pi^2N} + \frac{1}{16\pi^2} \lambda_{\phi\chi}^2  +O(\lambda_a^3),  \\
    &\beta_{\lambda_{\phi\chi}} = - \epsilon \lambda_{\phi\chi} + \frac{(N + 2)}{16\pi^2N} \lambda_{\phi} \lambda_{\phi\chi}  + \frac{1}{4\pi^2N} \lambda_{\phi\chi}^2 + \frac{3}{16\pi^2N} {\lambda_\chi} \lambda_{\phi\chi}  +O(\lambda_a^3).
\end{aligned}
\end{equation}
The fixed-point couplings of the critical biconical model at leading order in $\epsilon$ and at NLO in $1/N$ are 
\begin{equation}
    \lambda_{\phi}^* = 16 \pi^2 \varepsilon\left(1 - \frac{17}{N}\right), \quad \lambda_\chi^* = 16 \pi^2 \varepsilon\left(9 - \frac{873}{N}\right), \quad \lambda_{\phi\chi}^* = 16 \pi^2 \varepsilon\left(-3 + \frac{267}{N}\right).
\end{equation}
Then the effective potential of the theory reads
\begin{align} \label{eq: bicon_effective_potential_epsilon}
    V(\phi,\chi) = &\,\,N\left(\frac{\lambda_\phi^*}{8} \phi^4 + \frac{\lambda_{\phi\chi}^*}{4} \phi^2 \chi^2 + \frac{\lambda_\chi^*}{8} \chi^4\right) + O(\epsilon^2, N^{-1}) \\ =  & \,\, 2\pi^2 \epsilon  N(\phi^2 - 3\chi^2)^2 + \,\, 2\pi^2 \epsilon \left[-17(\phi^2 - 3\chi^2)^2 + 576 \chi^4  + 432 \chi^2 (\phi^2 - 3\chi^2)\right] \nonumber\\[0.1cm] & \hspace{9.75cm}+O(\epsilon^2, N^{-1}). \nonumber
\end{align}
We now compare this expression with our general-$d$ result \eqref{eq: bicon_effective_potential_general_nlo_expanded}. The leading terms in the $1/N$ expansion agree, since $\alpha_0=-3+O(\epsilon)$ and $\nu_d = 2\pi^2 \epsilon + O(\epsilon^2)$. The contributions from the anomalous dimensions can be neglected at this order, as $\gamma_\phi, \gamma_\chi \sim \epsilon^2$. Since the general-$d$ NLO analysis does not determine $\alpha_1$, we compare the two expressions on loci where the $\alpha_1$ contribution \eqref{eq: alpha1} vanishes:
$\phi^2=-\alpha_0\chi^2$ and
$\chi=0$. On the first locus, substituting $d=4-\epsilon$ into the analytical result
\eqref{eq: bicon_effective_potential_nlo_restricted_m0} and expanding to leading order in $\epsilon$, we obtain
\begin{equation} 
    V_{\text{NLO}}(\phi, \chi)\Big|_{\phi^2 + \alpha_0 \chi^2 = 0}  = 128 \pi^2 \epsilon\phi^{4} + O(\epsilon^2) = 2 \pi^2 \epsilon \cdot  576 \chi^{4} + O(\epsilon^2),
\end{equation}
in agreement with the weak-coupling result \eqref{eq: bicon_effective_potential_epsilon}. On the second locus, $\chi = 0$, the calculation in Appendix \ref{app: integral_restricted_effective_potential_chi0_d4} gives
\begin{equation}
    V_{\text{NLO}}(\phi, \chi = 0)  =-2 \pi^2 \epsilon \cdot 17\phi^4 + O(\epsilon^2),
\end{equation} 
again in agreement with the weak-coupling result. Although the $\chi=0$ locus is unstable for $\phi \neq 0$, the imaginary part in the effective potential begins only at $O(\epsilon^2)$. 
Thus, two of the three NLO terms in \eqref{eq: bicon_effective_potential_epsilon} are reproduced by the general-$d$ calculation, while the remaining term takes precisely the form of the $\alpha_1$ contribution in \eqref{eq: alpha1}. Determining $\alpha_1$ for general $d$ requires an NNLO analysis of the effective potential, a task we leave for future work. 

We have thus verified that our general-$d$ effective potential has the correct $d\rightarrow 4$ limit. In~\cite{Komargodski:2024zmt}, the NLO expression for the effective potential evaluated on the locus $\phi^2+\alpha\chi^2=0$ is missing a $d$-dependent prefactor and therefore does not reproduce this limit.

\section{Persistent Spontaneous Symmetry Breaking} \label{sec: PSSB}
Having identified the critical biconical model, we are now ready to investigate its thermal phase structure. We introduce a finite temperature $T$ by compactifying the Euclidean time direction on a circle of circumference $\beta = 1/T$ and imposing periodic boundary conditions on the bosonic fields. Consequently, the temporal momentum is restricted to the discrete bosonic Matsubara frequencies, and the $d$-dimensional Euclidean momentum integrals are replaced by finite-temperature sum-integrals,
\begin{equation*}
\sum_K \equiv T\sum_{n\in\mathbb{Z}}\int\frac{d^{d-1}k}{(2\pi)^{d-1}}, \quad K = (k_0 , \vec{k}), \quad k_0 = 2\pi n T.    
\end{equation*}
Accordingly, the LO self-consistency equation \eqref{eq: bicon_sigma_eta2_eq} for the $\eta_2$ self-energy $\Sigma_{\eta_2}^{\text{LO}} \equiv m^2$ becomes
\begin{equation} \label{eq: bicon_sigma_eta2_eq_thermal}
    m^2 = \frac{\lambda_\phi}{2}\sum_K \frac{1}{K^2 + m^2} + \frac{\lambda_\phi}{2}\rho, \qquad \rho \equiv \phi^2 + \alpha \chi^2.
\end{equation}
Throughout this section, we set $\alpha$ and $g$ to their critical values determined by the NLO analysis,
\begin{equation}
\alpha=-(d-1)(d-2)/2 + O(N^{-1}), \quad g = [15360 + O(N^{-1})]\, \delta_{d,3}.    
\end{equation}

\subsection{$3<d<4$}
Using the identity \eqref{eq: two_point_function_effective_potential} for large-$N$ rescaled fields, we find $\Sigma_{\eta_2}(0) = 2\partial_{\phi^2} V(\phi, \chi)/N$. The LO thermal effective potential $V_{\text{LO}}(\phi,\chi)$ is obtained by repeating the vacuum-diagram computation with the replacement $\int_k\rightarrow\sum_K$. For now, we do not need its explicit form; it is sufficient to note that, at this order, its field dependence enters only through $\rho$. Therefore,
\begin{equation} \label{eq: mfromV_largeNscaled}
    m^2  = 2\partial_{\rho} V_{\text{LO}}(\rho)/N.
\end{equation}
The thermal expectation values of the fields are determined by the minimum $(\bar\phi, \bar\chi)$ of the thermal effective potential.
The $\chi$- and $\phi$-stationarity conditions give
\begin{equation}
\begin{aligned}
    0 &= \frac{1}{N}\frac{\partial V_{\text{LO}}(\phi, \chi)}{\partial \chi} = \frac{1}{N}\frac{\partial V_{\text{LO}}(\rho)}{\partial \rho} \frac{\partial \rho}{\partial \chi} = \alpha \chi m^2,  \\ 
    0 &= \frac{1}{N}\frac{\partial V_{\text{LO}}(\phi, \chi)}{\partial \phi} = \frac{1}{N}\frac{\partial V_{\text{LO}}(\rho)}{\partial \rho} \frac{\partial \rho}{\partial \phi} = \phi m^2. 
\end{aligned}
\end{equation}
For $m>0$, stationarity forces $\phi=\chi=0$. We find that
$\frac{1}{N}\frac{\partial^2 V_{\text{LO}}}{\partial \chi^2}\big|_{\phi=\chi=0} =\alpha m^2 < 0$, revealing that this configuration is unstable. Consequently, any stable minimum must satisfy $m = 0$. Substituting this condition into the thermal self-consistency equation \eqref{eq: bicon_sigma_eta2_eq_thermal}, we obtain
\begin{equation} \label{eq: flat_direction_at_T}
    \bar \rho = - \sum_K \frac{1}{K^2} = - a_d T^{d-2},\quad a_d = \frac{\zeta (d-2) \Gamma
   \left(\frac{d-2}{2}\right)}{2 \pi^{d/2}} > 0.
\end{equation}
To investigate the stability of these stationary points, we evaluate the LO thermal effective potential in their neighborhood. 
In the vicinity $\rho \gtrapprox\bar \rho$ we have $m\approx 0$, so we can expand the sum-integral from \eqref{eq: bicon_sigma_eta2_eq_thermal} in $m/T\ll 1$. Isolating the zero Matsubara mode, we obtain
\begin{equation}
\begin{split}
    \sum_K \frac{1}{K^2 + m^2} = \,\,&T\int\frac{d^{d-1}k}{(2\pi)^{d-1}} \frac{1}{k^2 + m^2} + \sum_{K:\,\, n\neq0} \frac{1}{K^2 +m^2} \\ = \,\,& T \frac{\Gamma(\frac{3-d}{2})}{(4\pi)^{\frac{d-1}{2}}} m^{d-3} + a_d T^{d-2} (1+O(m^2/T^2)).
\end{split}
\end{equation}
Substituting this result into \eqref{eq: bicon_sigma_eta2_eq_thermal} and solving for $m^2$ yields
\begin{equation} 
    m^2 \approx \left(\frac{(4\pi)^{\frac{d-1}{2}}}{-\Gamma(\frac{3-d}{2})}\left(\frac{\rho}{T} + a_d T^{d-3}\right)\right)^{\frac{2}{d-3}}, \quad 0\leq \rho +a_d T^{d-2}\ll T^{d-2}.
\end{equation}
Using \eqref{eq: mfromV_largeNscaled}, the effective potential in the region $0\leq \phi^2 + \alpha\chi^2+a_d T^{d-2}\ll T^{d-2}$ is  
\begin{equation} \label{eq: VLO_atT_3d4}
    V_{\text{LO}}(\phi,\chi) \approx -N \frac{\Gamma(d/2)\zeta(d)}{\pi^{d/2}} T^d + N \gamma_d T^{-\frac{2}{d-3}} \left(\phi^2 + \alpha\chi^2+a_d T^{d-2}\right)^{\frac{d-1}{d-3}},
\end{equation}
where the integration constant is the $\frac{N}{2}\sum_K \ln K^2$ term arising from \eqref{eq: bicon_effective_potential_lo_preirlimit} evaluated at $m^2=\Sigma_{\eta_2}^{\text{LO}}=0$, and
\begin{equation*}
    \gamma_d = \frac{d-3}{2(d-1)}\left(\frac{(4\pi)^{\frac{d-1}{2}}}{-\Gamma(\frac{3-d}{2})}\right)^{\frac{2}{d-3}} > 0, \qquad 3 < d<4.
\end{equation*}
The form of the effective potential \eqref{eq: VLO_atT_3d4} near its minima demonstrates that at LO, the flat direction persists at finite temperatures, deforming from a pair of intersecting straight lines $\chi = \pm \phi/\sqrt{-\alpha}$ at $T=0$ into a hyperbola at $T>0$.
However, the NLO correction lifts this degeneracy, restricting the minima of the potential to $\phi = 0$. Taking the zero-temperature NLO effective potential \eqref{eq: bicon_effective_potential_general_nlo_expanded}, replacing $\int_k \to \sum_K$ and setting $m=0$, we find that this correction takes the form
\begin{equation}
    V(\phi, \chi)\big|_{m=0} = \text{const}(T)+\frac{1}{2} \sum_K \ln \left[\Pi(K) K^2 + x\right] + O(N^{-1}),  
\end{equation}
where we denote $x \equiv \phi^2 + \alpha^2 \chi^2$, and $\Pi(K) = \frac{1}{2} \sum_P \frac{1}{(P-K)^2 P^2}$ is the thermal bubble at zero mass. After subtracting the power-law divergences, numerical evaluation detailed in Appendix \ref{app: positivity_derivative_VNLO_m0} reveals that for $x\geq 0$
\begin{equation}
    \frac{\partial}{\partial x} \frac{1}{2} \sum_K \ln \left[\Pi(K) K^2 + x\right] = \frac{1}{2} \sum_K \frac{1}{\Pi(K) K^2 + x} > 0.
\end{equation}
Hence, the NLO contribution $V(\phi, \chi)\big|_{m=0}$ monotonically grows with $x$. Constraining $x$ to the flat direction \eqref{eq: flat_direction_at_T}, we have $x = \phi^2 (1-\alpha) - \alpha a_d T^{d-2}$. Since $\alpha < 0$, the thermal effective potential is minimized at $\phi = 0$, leading to the following thermal expectation values
\begin{equation}
    \bar\phi^2 = 0 + O(N^{-1}), \quad \bar\chi^2 = \frac{\pi ^{-d/2} \zeta (d-2) \Gamma
   \left(\frac{d-2}{2}\right)}{(d-2)(d-1)} T^{d-2} + O(N^{-1}), \quad 3<d<4.
\end{equation}
This result agrees with~\cite{Komargodski:2024zmt} up to a factor of $1/2$. As $d\rightarrow 4$, the formula above reproduces the leading $d=4-\epsilon$ result from \cite{Chai:2020onq}, namely $\bar\chi^2 = \frac{T^2}{36} + O(\epsilon,N^{-1})$.

\subsection{$d=3$}
Turning now to the $d=3$ case, the self-consistency equation for the self-energy \eqref{eq: bicon_sigma_eta2_eq_thermal} yields
\begin{equation}
    m^2 = - \lambda_\phi \left(\frac{m}{8\pi} + \frac{T \ln (1 - e^{-m/T})}{4\pi}\right) +  \frac{\lambda_{\phi}}{2}\rho.
\end{equation}
In the IR limit, $\lambda_\phi \to \infty$, the solution for $m$ takes the form
\begin{equation} \label{eq: m_T_d3}
    m = 2T\,\text{arcsinh}\left(\frac{1}{2}e^{2\pi \rho/T}\right).
\end{equation}
Remarkably, the introduction of finite temperature grants us access to the otherwise forbidden $\rho < 0$ region. In the $\rho<0$ domain, $m$ is exponentially suppressed $\sim T e^{2\pi \rho/T}$ as $\rho/T \to -\infty$, and in the limit $T \to 0$, it tends to zero. Consequently, the LO effective potential in this limit reads $\lim_{T\to 0}V_{\text{LO}}(\phi,\chi) = \frac{N}{2}\int m^2 d\rho + N\frac{g}{6!}\chi^6 = N\nu_3 \rho^{3} \theta(\rho) + N\frac{g}{6!}\chi^6.$

Returning to the search for the minimum of the thermal effective potential, we impose the $\chi$-stationarity condition 
\begin{equation}
     0 = \frac{1}{N}\frac{\partial V_{\text{LO}}(\phi, \chi)}{\partial \chi} = \alpha \chi m^2 + \frac{g}{5!}\chi^5.
\end{equation}
The configuration $\chi = 0$ is not a stable solution since $\frac{1}{N}\frac{\partial^2 V_{\text{LO}}}{\partial \chi^2}\big\vert{}_{\chi=0} =\alpha m^2 < 0$.
Assuming $\chi \neq 0$ and substituting the LO critical values $\alpha = -1$ and $g = 15360$ into the stationarity equation, we obtain 
\begin{equation} \label{eq: chi2_T_d3}
    \chi^2 = \frac{1}{8\sqrt{2}} m. 
\end{equation}
The $\phi$-stationarity condition yields 
\begin{equation}
     0 = \frac{1}{N}\frac{\partial V_{\text{LO}}(\phi, \chi)}{\partial \phi} = \phi m^2.
\end{equation}
Since equation \eqref{eq: m_T_d3} gives $m>0$ for every finite field configuration, the $\phi$-stationarity condition forces $\phi = 0$. Hence, substituting \eqref{eq: chi2_T_d3} into \eqref{eq: m_T_d3} with $\phi$ set to zero, we find 
\begin{equation}
        8 \sqrt{2}\, \chi^2 = 2T\,\text{arcsinh}\left(\frac{1}{2}e^{-2\pi \chi^2/T}\right).
\end{equation}
Solving this equation numerically, we obtain the thermal expectation values 
\begin{equation}
    \bar \phi = 0, \quad \bar \chi^2 \approx 0.0596\, T + O(N^{-1}),
\end{equation}
where $\bar \phi$ must remain zero to all orders in $1/N$, as the Coleman--Mermin--Wagner theorem forbids finite-temperature spontaneous breaking of continuous symmetries in $d=2+1$. We verify that this configuration corresponds to a stable minimum by evaluating the Hessian of the effective potential:
\begin{align}
    &\frac{1}{N}\frac{\partial^2 V_{\text{LO}}(\phi,\chi)}{\partial \phi^2}\Bigg|_{\bar\phi, \bar\chi} = m^2 \approx 0.455 \, T^2 > 0, \quad \frac{1}{N}\frac{\partial^2 V_{\text{LO}}(\phi,\chi)}{\partial \phi \partial \chi}\Bigg|_{\bar\phi, \bar\chi} = 0, \nonumber \\ & \frac{1}{N}\frac{\partial^2 V_{\text{LO}}(\phi,\chi)}{\partial \chi^2}\Bigg|_{\bar\phi, \bar\chi} = \alpha m^2 + \alpha \chi \partial_\chi  m^2+ \frac{g}{4!}\chi^4\approx 2.478\, T^2 >0.
\end{align}
In summary, using the critical couplings determined by the NLO zero-temperature analysis, we find that for any nonzero temperature and $3\leq d<4$, the thermal effective potential possesses two degenerate $\mathbb{Z}_2$-related minima. As a result, the $\chi$ field acquires a non-vanishing thermal expectation value. We computed this expectation value at LO, and while $O(N^{-1})$ corrections may modify its magnitude, they cannot make it vanish for sufficiently large $N$. Thus, for large finite $N$, the $\mathbb{Z}_2$ symmetry remains spontaneously broken at arbitrarily high temperatures.

\section{Beta Functions} \label{sec: beta_functions}
In this section, we return to $T=0$ and conduct a more comprehensive study of the infrared RG flow of the biconical model. We enlarge the coupling space to include all relevant and weakly irrelevant deformations of the IR fixed points and derive the corresponding beta functions. By weakly irrelevant, we mean couplings that are marginal at LO in $1/N$ but become irrelevant at NLO. Although strongly irrelevant couplings, namely those that are already irrelevant at LO, can be included in our analysis, we largely omit this sector, retaining only a few such couplings. We conclude this section by identifying the critical surface of the biconical model, calculating the relevant scaling exponents, and comparing the results with the FRG numerics. 

\subsection{Introducing more couplings}
To investigate the RG flow comprehensively, we broaden the space of couplings. The UV potential takes the form 
\begin{equation} \label{eq: UV_potential_extended}
    U(\phi,\chi) = N \left[\mathcal E + \frac{1}{2}\mu_{\phi}^2 \phi^2 + \frac{1
}{2}\mu_{\chi}^2 \chi^2 +\frac{\lambda_\phi}{8}\,(\phi^2)^2
+\frac{\lambda_{\phi\chi}}{4}\,\phi^2\,\chi^2
+\frac{\lambda_\chi}{8}\,\chi^4 + \frac{g}{6!} \chi^6\right],
\end{equation}
where $\mathcal E$ is the vacuum-energy density, and $\mu_\phi^2$, $\mu_\chi^2$ are the quadratic mass terms for $\phi$ and $\chi$ fields, respectively. 

Could any interaction omitted in \eqref{eq: UV_potential_extended} become relevant or weakly irrelevant at the interacting IR fixed point? To answer this, consider the massless fixed-point correlation functions. At separated points,  
\begin{equation*}
\langle \chi(x) \chi(y)\rangle_{\text{LO}} \sim |x-y|^{-2\Delta_\chi}, \qquad   \langle \lambda_\phi\phi^2(x) \lambda_\phi\phi^2(y)\rangle_{\text{LO}} \sim |x-y|^{-2\Delta_{\lambda_\phi\phi^2}}. 
\end{equation*}
At LO, the $\chi$ field is free with scaling dimension $\Delta_{\chi} = \frac{d-2}{2}$, whereas the $\lambda_\phi\phi^2$ singlet acquires $\Delta_{\lambda_\phi\phi^2}=2$ through bubble resummation. Due to large-$N$ factorization, interactions between these operators are $1/N$ suppressed, so the dimensions of composite operators built from these constituents and derivatives are additive at LO. Consequently, all operators beyond those in \eqref{eq: UV_potential_extended}, including higher-order operators such as $(\lambda_\phi\phi^2)^3$, $(\lambda_\phi\phi^2)^2\chi^2$, $(\lambda_\phi\phi^2)\chi^4$, $\dots$, and derivative interactions possess LO scaling dimensions $\Delta_{\mathcal O}>d$, rendering them strongly irrelevant. In particular, writing $\lambda_\phi(\phi^2)^2=\lambda_\phi^{-1}(\lambda_\phi\phi^2)^2$ shows that $\lambda_\phi^{-1}$ couples to an operator of dimension $4>d$ and hence vanishes at the IR fixed point.

We define the following convenient coupling combinations 
\begin{equation}
\mathcal E,\quad r =2\mu_\phi^2/\lambda_\phi, \quad v = \mu_\chi^2 - \alpha \mu_\phi^2,\quad \kappa = \lambda_\chi - \frac{\lambda_{\phi\chi}^2}{\lambda_\phi}, \quad  h = \frac{1}{\lambda_\phi}, \quad \alpha, \quad g.   
\end{equation}
Following the logic of the main text, we recompute the effective potential in the presence of the additional couplings. Due to the $\tfrac{1}{2} \mu_\phi^2 \eta^2$ term in the shifted UV action, the equation for the $\Sigma_{\eta_2}$ self-energy \eqref{eq: bicon_sigma_eta2_eq} acquires an additional $\mu_\phi^2$ term on the right-hand side. Hence,
\begin{equation}
    m^2\equiv \Sigma_{\eta_2}^{\text{LO}} = \frac{2d}{d-2} \nu_d \left(\phi^2 + \alpha \chi^2 + r\right)^{\frac{2}{d-2}}, \quad \phi^2 + \alpha \chi^2 + r\geq 0.   
\end{equation}
Repeating the derivation for the LO effective potential \eqref{eq: effective_potential_sum_diagrams}, one obtains
\begin{align} 
    V(\phi,\chi) = \,\, & U(\phi,\chi)+ \sum_D A(D) \nonumber\\= \,\,& U(\phi,\chi)+ N\Bigg[\frac{1}{2}\int_k \ln \left(k^2 + m^2\right) - \frac{\lambda_\phi}{8} \left(\frac{2m^2}{\lambda_\phi} - (\phi^2+\alpha \chi^2 + r)\right)^2\Bigg] + O(N^0).
\end{align}
Taking the IR limit and simplifying yields
\begin{equation} \label{eq: bicon_effective_potential_lo_extended_couplings}
    V(\phi,\chi) = N \left[ \mathcal E - \frac{1}{8} \lambda_\phi r^2+  \nu_d \left(\phi^2+\alpha\chi^2 +r\right)^{\frac{d}{d-2}} + \frac{1}{2} v\chi^2 + \frac{\kappa}{8}\chi^4 + \frac{g}{6!} \chi^6 \right] + O(N^{0}).
\end{equation}
We absorb the constant term $- \frac{1}{8} \lambda_\phi r^2$ into the vacuum energy by redefining $\mathcal E \rightarrow \mathcal E + \frac{1}{8} \lambda_\phi r^2$. The $\Sigma_\xi (p)$ self-energy changes to
\begin{equation}
    \Sigma_\xi (p) = \alpha \,m^2 +  \frac{\alpha^2 \chi^2}{\Pi_{m^2}(p)}  + \left(\frac{g}{4!} \chi^4 + \frac{3}{2}\kappa \chi^2+v\right) + O(N^{-1}),
\end{equation}
while the expressions for the remaining self-energies calculated in Section \ref{sec: self_energies_lo} remain unchanged. Thus, the modifications in self-energies are fully captured by the new $m^2$ and the shift $\frac{g}{4!} \chi^4 \rightarrow \frac{g}{4!} \chi^4 + \frac{3}{2}\kappa \chi^2+v$. Using the expression for the NLO logarithmic divergence of the effective potential derived in Appendix \ref{app: nlo_log_div} for a general $m$ and performing the shift, one obtains 
\begin{multline}
    L(\phi,\chi;u)=B_1 m^d +B_2 \bigg[m^2\bigl(s_1\phi^2+s_2\chi^2\bigr)   + \delta_{d,3}            \frac{4}{3\pi^3} \left(m-4\pi(\phi^2+\alpha^2 \chi^2)\right)^3 + \\ +\alpha^2\left(\frac{g}{4!}\chi^6+\frac{3}{2}\kappa \chi^4+v\chi^2\right)\bigg],
\end{multline}
where $u = (\mathcal E, r, v, \kappa, \alpha, g)$ and the coefficients read
\begin{equation*}
    B_1 =
    \frac{\pi^{-\frac{d+1}{2}}\Gamma\!\left(\frac{d+1}{2}\right)}
    {\Gamma\!\left(\frac d2-1\right)\Gamma\!\left(\frac d2+1\right)}, \,\,\,\, B_2 = \frac{\Gamma(d-1)\sin\!\left(\frac{\pi d}{2}\right)}
    {\pi\,\Gamma\!\left(\frac d2\right)^2},\,\,\,\, s_1 = 2d-5,
    \,\,\,\,
    s_2 = \alpha^2(\alpha+2d-6).
\end{equation*}
Note that, aside from discarding scheme-dependent power-law divergences, we have not yet performed any renormalization. Hence, we write down the effective potential and explicitly label fields and couplings with $B$ as bare quantities defined at the cutoff scale $\Lambda$: 
\begin{multline}
    V_B(\phi_B, \chi_B) =N \left[ \mathcal E_B +  \nu_d \left(\phi_B^2+\alpha_B\chi_B^2 +r_B\right)^{\frac{d}{d-2}} + \frac{1}{2} v_B\chi_B^2 + \frac{\kappa_B}{8}\chi_B^4 + \frac{g_B}{6!} \chi_B^6 \right] +\\+ L(\phi_B,\chi_B; u_B) \ln \left(\frac{\Lambda }{m_B}\right)  +F(\phi_B,\chi_B; u_B) + O(N^{-1}),
\end{multline}
where $F(\phi_B,\chi_B; u_B)$ is the finite NLO part of the effective potential, whose explicit form is irrelevant to the following discussion. Both $L$ and $F$ are $O(N^0)$ and cutoff-independent. 

Recall that setting $\lambda_{\phi\chi,B}^2 = \lambda_{\phi}^B \lambda_\chi^B \Leftrightarrow \kappa_B = 0$ in Section \ref{sec: LO} was necessary to make the massless IR theory scale-invariant at LO. To calculate the beta functions and scaling exponents, we relax this condition.

Coupling $h_B$ drops out of the IR effective potential, or in other words, it is irrelevant for the IR physics. One can retain the $h_B$ coupling in the effective potential before taking the IR limit. At LO, this corresponds to modifying $m^2$ and adding a term $-\frac{h_B}{2} m^4$ to the effective potential \eqref{eq: bicon_effective_potential_lo_extended_couplings}. However, there are no logarithmic divergences to renormalize at this order, so in the minimal subtraction scheme we set the renormalized coupling to be $h = h_B + O(N^{-1})$, and the beta function is
\begin{equation}
    \beta_{\bar h} = (4-d) \bar{h} +O(N^{-1}), \quad h = \mu^{d-4} \bar h. 
\end{equation}
Dimensional analysis and regularity of the $1/N$ expansion around the large-$N$ fixed point imply that every monomial in a dimensionful beta function of a strongly irrelevant coupling contains at least one strongly irrelevant coupling. Hence, the locus on which all strongly irrelevant couplings vanish is RG invariant order by order in $1/N$. We therefore set $h=0$, just as we have done for the other strongly irrelevant couplings.

\subsection{Renormalization}
We choose to renormalize the theory at some scale $\mu \ll \Lambda$ using the minimal subtraction scheme. The $p^2 \ln \Lambda$ divergences in the self-energies from Section \ref{sec: anomalous_dimensions} dictate the relation between the bare and renormalized fields
\begin{equation}
\begin{gathered}
    \phi_B = Z_\phi^{1/2} \phi, \quad \chi_B = Z_\chi^{1/2} \chi, \qquad\\ Z_\phi = 1-2\gamma_\phi \ln \frac{\Lambda}{\mu} + O(N^{-2}), \quad Z_\chi = 1-2\gamma_\chi \ln \frac{\Lambda}{\mu} + O(N^{-2}).
    \end{gathered}
\end{equation}
We express the bare effective potential $V_B(\phi_B, \chi_B)$ in terms of the renormalized fields and couplings, yielding a renormalized potential $V(\phi, \chi)$ that is free of the cutoff dependence. At LO, there are no divergences to renormalize, hence, 
\begin{multline}
   \text{LO:}\quad \phi_B = \phi, \quad \chi_B = \chi, \quad u_B = u, \quad  \\ V_{\text{LO}}(\phi, \chi) =N \left[ \mathcal E +  \nu_d \left(\phi^2+\alpha\chi^2 +r\right)^{\frac{d}{d-2}} + \frac{1}{2} v\chi^2 + \frac{\kappa}{8}\chi^4 + \frac{g}{6!} \chi^6 \right]. 
\end{multline}
At NLO, the bare effective potential contains a logarithmic cutoff dependence demanding renormalization 
\begin{equation}
    \text{NLO:}\quad \phi_B = \phi (1 -\gamma_\phi \ln \Lambda/\mu), \quad \chi_B = \chi (1 -\gamma_\chi \ln \Lambda/\mu), \quad u_B = u + \frac{1}{N} \delta u(u,\mu,\Lambda), 
\end{equation}
where $\delta u_i$ are the coupling counterterms. If the $u_i$ coupling space is closed under renormalization, the NLO logarithmic cutoff dependence must be entirely absorbed by the field renormalizations and the coupling counterterms. Then the effective potential can be rewritten in terms of the renormalized fields and couplings as
\begin{multline}
    V_B(\phi_B, \chi_B) = V(\phi, \chi) =N \left[ \mathcal E  +  \nu_d \left(\phi^2+\alpha\chi^2 +r\right)^{\frac{d}{d-2}} + \frac{1}{2} v\chi^2 + \frac{\kappa}{8}\chi^4 + \frac{g}{6!} \chi^6 \right] +\\+ L(\phi,\chi; u) \ln \left(\frac{\mu }{m}\right)  +F(\phi,\chi; u) + O(N^{-1}).
\end{multline}
The effective potential is independent of the renormalization scale 
\begin{equation}
    0=\mu \frac{d}{d\mu} V_B(\phi_B, \chi_B) = \mu \frac{d}{d\mu} V(\phi, \chi).
\end{equation}
Expanding the total derivative in terms of partial derivatives yields the Callan--Symanzik RG equation. Satisfying this equation proves that the $u_i$ coupling space is closed under renormalization.
\begin{equation}
    \mu \frac{d}{d\mu} V(\phi, \chi) = \left(\mu \frac{\partial}{\partial \mu} - \gamma_\phi \phi\frac{\partial}{\partial \phi} - \gamma_\chi \chi\frac{\partial}{\partial \chi}+\beta_{u_i} \frac{\partial}{\partial u_i}\right) V(\phi, \chi) = 0, 
\end{equation}
where $u_i$ denote the components of $u = (\mathcal E, r, v, \kappa, \alpha, g)$, and $\beta_{u_i}$ are the corresponding beta functions. Expanding to NLO, we obtain the following equation
\begin{equation}
    L(\phi,\chi; u) + \left(- \gamma_\phi \phi\frac{\partial}{\partial \phi} - \gamma_\chi \chi\frac{\partial}{\partial \chi} +\beta_{u_i} \frac{\partial}{\partial u_i}\right) V_{\text{LO}}(\phi,\chi) = 0 + O(N^{-1}).
\end{equation}
Since the RG equation must hold for arbitrary field values, the coefficients of the linearly independent functions of $\phi$ and $\chi$ appearing in it must vanish separately. Setting these coefficients to zero determines all the beta functions. We note that without the $g$ coupling, the RG equation cannot be satisfied in $d=3$: the logarithmic divergence prefactor $L(\phi,\chi;u)|_{g=0}$ contains a term proportional to $\chi^6 \delta_{d,3}$, which remains uncompensated in the absence of a $g\chi^6$ interaction. This implies that the $\chi^6$ operator is generated by the RG flow, and must therefore be included to close the coupling space under renormalization.

The dimensionless couplings are defined as
\begin{equation}
    \mathcal E=\mu^d \bar{\mathcal E},\qquad
    r=\mu^{d-2}\bar r ,\qquad
    v=\mu^2 \bar v,\qquad
    \kappa=\mu^{4-d}\bar\kappa,\qquad
    g=\mu^{6-2d}\bar g.
\end{equation}
Then the beta functions take the following form
{\allowdisplaybreaks
\begin{align}
    \beta_{\bar{\mathcal E}}
    &=
    -d\,\bar{\mathcal E}
    +\delta_{d,3}\frac{1024}{3\pi^2N}\bar r^3
    +O(N^{-2}), \\ \beta_{\bar v}
    &=
    -2\bar v
    -\frac{2C_d}{N}\,
    \alpha
    \left[
        \alpha \bar v
        -64d(\alpha-1)\bar r^2\,\delta_{d,3}
    \right]
    +O(N^{-2}),
    \\
    \beta_{\bar\kappa}
    &=
    -(4-d)\bar\kappa
    -\frac{2C_d}{N}\,
    \alpha^2
    \left[
        (d+2)\bar\kappa
        +256d(\alpha-1)^2\bar r\,\delta_{d,3}
    \right]
    +O(N^{-2}),\\
    \beta_{\bar r}
    &=
    -(d-2)\bar r
    +\frac{(d-1)(d-2)}{N}C_d\,\bar r
    +O(N^{-2}),
    \\
    \beta_{\bar g}
    &=
    -(6-2d)\bar g
    -\frac{6C_d}{N}\,
    \alpha^2
    \left[
        (d+1)\bar g
        -2560d\,\alpha(\alpha-1)^3\delta_{d,3}
    \right]
    +O(N^{-2}), \\
    \beta_\alpha
    &=
    -\frac{2C_d}{N}\,
    \alpha(\alpha-1)
    \left[
        \alpha+\frac{(d-1)(d-2)}{2}
    \right]
    +O(N^{-2}),
\end{align}}
where 
\begin{equation*}
    C_d \equiv
    \frac{4\Gamma(d-1)\sin\left(\frac{\pi d}{2}\right)}
    {\pi d\,\Gamma\left(\frac d2\right)^2}.
\end{equation*}
Our expression for the $\beta_\alpha$ function, alongside the $\beta_g$ function at $d=3$, is in agreement with the results from \cite{Komargodski:2024zmt}. The biconical fixed point corresponds to the following critical coupling values 
\begin{equation}
\begin{aligned}
    &\qquad\qquad\qquad \mathcal E^* = 0,\quad r^* = 0,\quad v^* = 0,\quad \kappa^* = 0, \\[0.1cm] &\alpha^* = -\frac{(d-1)(d-2)}{2} + O(N^{-1}), \quad g^* = [15360 +O(N^{-1})] \times\delta_{d,3} .
    \end{aligned}
\end{equation}
Scaling exponents $\theta_i$ are minus the eigenspectrum of the stability matrix $$B_{ij} = \frac{\partial \beta_{\bar u_i}}{\partial \bar u_j} \Bigg|_{\bar u = \bar u^*},$$ where $\bar u_i$ are the dimensionless couplings and $\bar u_i^*$ are their fixed point values. The number of all possible couplings is infinite; however, we focus on the relevant directions corresponding to positive $\theta_i$, and the coupling space we consider is sufficient for this. The coupling $\alpha$ is weakly irrelevant, while $g$ is weakly irrelevant at $d=3$ and strongly irrelevant at any fixed~$3<d<4$. Disregarding the relevant vacuum energy direction, we obtain three relevant scaling exponents 
\begin{equation}
\begin{aligned}
    &\theta_1 = 2+\frac{1}{N}\frac{2 (d-1) (d-2)^2 \sin \left(\frac{\pi  d}{2}\right) \Gamma (d)}{\pi  d\, \Gamma \left(\frac{d}{2}\right)^2}+O(N^{-2}), \\
    &\theta_2 = (d-2)-\frac{1}{N}\frac{2^{d+2} \sin \left(\frac{\pi  d}{2}\right) \Gamma \left(\frac{d+1}{2}\right)}{\pi ^{3/2} d\, \Gamma \left(\frac{d}{2}-1\right)} + O(N^{-2}),\\
    &\theta_3 = (4-d)+\frac{1}{N}\frac{2^d \left(d^2+d-2\right) (d-2)^2 \sin \left(\frac{\pi  d}{2}\right) \Gamma \left(\frac{d+1}{2}\right)}{\pi ^{3/2} d \,\Gamma \left(\frac{d}{2}\right)} + O(N^{-2}).
    \end{aligned}
\end{equation}
For $3<d<4$, the $\theta_{1,2,3}$ scaling exponents correspond to pure $v,\, r,\, \kappa$ relevant deformations, respectively. Without changing the dynamics of the theory, one can introduce the Hubbard--Stratonovich field $\sigma$ by $\mathcal{L} \rightarrow\mathcal{L} - \frac{N}{2 \lambda_\phi} (\sigma -\frac{\lambda_\phi}{2}\phi^2-\mu_\phi^2-\frac{\lambda_{\phi\chi}}{2}\chi^2)^2$. The action then contains a $\frac{N}{2} r \sigma$ term, and $d-\theta_2$ coincides with the scaling dimension of $\sigma$ in the critical $O(N)$ model~\cite{Goykhman:2019kcj}. At $d=3$, infinitesimal $r$ and $\kappa$ deformations mix, as seen from $\beta_{\bar \kappa}$. 

We can compare $\gamma_\phi, \gamma_\chi$ and $\theta_{1,2,3}$ with those obtained from the numerical FRG method applied to the biconical model \cite{Hawashin:2024dpp, SmolkinYung:2025parity}.
Table~\ref{tab: largeN_vs_FRG} compares our large-$N$ results with the FRG numerics and the $\epsilon$-expansion. We find reasonable agreement with the FRG and recover the $\epsilon$-expansion near $d=4-\epsilon$.

Dimensional analysis of the dimensionful beta functions and regularity of the $1/N$ expansion around the large-$N$ fixed point imply that the surface $\kappa=v=r=0$ is invariant under the RG flow, order by order in $1/N$. For sufficiently large $N$ and $\alpha\neq0$, the coupling $g$ is irrelevant in the IR; thus, any finite $g\geq 0$ taken in the UV flows to its IR critical value. The beta function $\beta_\alpha$ has one unstable fixed point at $\alpha = 0$ and two stable fixed points at $\alpha = -\frac{(d-2)(d-1)}{2}$ and $\alpha = 1$. Consequently, for large finite $N$, setting $\lambda_{\phi\chi}^2 = \lambda_{\phi}\lambda_\chi$ ($\kappa = 0$) in the massless ($v=r=0$) UV action \eqref{eq: bicon_uv_action} is sufficient to guarantee a flow to an IR critical theory. Taking $\lambda_{\phi\chi} > 0$ drives the flow to the critical $O(N+1)$ model. If $\lambda_{\phi\chi} = 0$ and $\lambda_\phi >0$, the system decouples into a critical $O(N)$ model and a free massless scalar. Finally, if we choose $\lambda_{\phi\chi} < 0$, the UV theory flows to the critical biconical model. Notably, for large finite $N$, the UV action need not contain the microscopic $\chi^6$ interaction, as it will be dynamically generated during the RG flow in $d=3$.   
\begin{table*}[h]
\centering
\renewcommand{\arraystretch}{1.12}
\setlength{\tabcolsep}{6pt}
\begin{tabular}{ccccccccc}
\toprule
$d$ & $N$ & Method 
& $\gamma_\phi$ & $\gamma_\chi$ 
& $\theta_1$ & $\theta_2$ & $\theta_3$
\\
\midrule

\multirow{2}{*}{$3$} & \multirow{2}{*}{$100$}
& Large-$N$ & $0.00135$ & $0.00135$ & $1.989$ & $1.011$ & $0.946$ 
\\
& & FRG       & $0.00121$ & $0.00121$ & $1.992$ & $1.010$ & $0.947$ 
\\

\midrule

\multirow{2}{*}{$3.5$} & \multirow{2}{*}{$200$}
& Large-$N$ & $0.00025$ & $0.00089$ & $1.986$ & $1.508$ & $0.422$  
\\
& & FRG       & $0.00027$ & $0.00070$ & $1.993$ & $1.506$ & $0.442$ 
\\

\midrule

\multirow{2}{*}{$4-\epsilon$} & \multirow{2}{*}{$N$}
& Large-$N$
& \multirow{2}{*}{$\displaystyle O(\epsilon^2)$}
& \multirow{2}{*}{$\displaystyle O(\epsilon^2)$}
& \multirow{2}{*}{$\displaystyle 2-\frac{18\epsilon}{N}$}
& \multirow{2}{*}{$\displaystyle 2-\epsilon+\frac{6\epsilon}{N}$}
& \multirow{2}{*}{$\displaystyle\epsilon
  -\frac{108\epsilon}{N}$}
\\
& & $\epsilon$-expansion & & & & &
\\

\bottomrule
\end{tabular}
\caption{Anomalous dimensions and relevant scaling exponents at NLO in the large-$N$ expansion, compared with numerical FRG results in the LPA$^{\prime}8$ truncation. For $d=4-\epsilon$, we recover the leading-order results of the $\epsilon$-expansion.
}
\label{tab: largeN_vs_FRG}
\end{table*}

\section{Discussion} \label{sec: discussion}

In this paper, we studied the $O(N)\times \mathbb{Z}_2$ biconical model in $3 \leq d < 4$ spacetime dimensions at leading and next-to-leading order in the $1/N$ expansion. We identified the model's IR fixed points, computed the beta functions governing their relevant and weakly irrelevant deformations, and determined the relevant scaling exponents of the critical biconical model. The agreement of our results with those obtained from the $\epsilon$-expansion and numerical FRG provides an independent consistency check of our analysis.

At nonzero temperature, a critical theory with no flat directions has no dimensionful scales other than $T$. Scaling analysis therefore fixes the form of the thermal expectation value of a scaling operator $\chi$
\begin{equation}
    \langle \chi \rangle_T = a_\chi T^{\Delta_\chi},
\end{equation}
where $\Delta_\chi$ is the scaling dimension of $\chi$, and $a_\chi$ is a dimensionless, temperature-independent coefficient. For the critical biconical model, our leading-order calculation gives $a_\chi\neq 0$. Although $O(N^{-1})$ corrections may modify its value, it remains nonzero for sufficiently large $N$, establishing spontaneous $\mathbb{Z}_2$ breaking at arbitrarily high temperatures. For the $O(N)$ field, by contrast, the leading-order calculation gives $a_\phi=0$, consistent with the Coleman--Mermin--Wagner theorem in $d=3$.

The NLO analysis of the critical biconical model qualitatively changes the picture found at $N=\infty$. At LO, imposing scale invariance leaves a continuous family of critical theories parametrized by $\alpha=\lambda_{\phi\chi}/\lambda_\phi$, and for $\alpha<0$ the effective potential possesses a flat direction. At NLO, however, scale invariance selects
\begin{equation}
\alpha=-\frac{(d-1)(d-2)}{2}+O(N^{-1}),
\end{equation}
thereby eliminating the conformal manifold and leaving a unique critical theory with negative $\alpha$. The flat direction is lifted as well, yielding a unique minimum at the origin at $T=0$ and a pair of $\mathbb Z_2$-related minima at $T>0$.

The case $d=3$ requires special treatment. At LO, the $\chi^6$ interaction is necessary for the existence of a homogeneous stable minimum of the thermal effective potential when $\alpha<0$. Independently, the NLO analysis reveals that the $\chi^6$ operator is dynamically generated under the RG flow toward the IR, meaning it must be included for proper renormalization of the theory. Scale invariance fixes the sextic coupling to $g=15360+O(N^{-1})$. This leads to the absence of a flat direction in $d=3$ already at LO.

The critical biconical model has three independent relevant directions. These correspond to the mass-like deformations $v$ and $r$, together with the quartic coupling combination $\kappa=\lambda_\chi-\lambda_{\phi\chi}^2/\lambda_\phi$. By contrast, the coupling $\alpha$ is weakly irrelevant, while the sextic coupling $g$ is weakly irrelevant at $d=3$ and becomes strongly irrelevant as we approach $d=4$. Thus, reaching the critical theory requires tuning onto a codimension-three critical surface, whereas the remaining couplings approach their fixed-point values within the corresponding basin of attraction. Consequently, for the IR physics of the massless ($v=r=0$) theory \eqref{eq: bicon_uv_action} to be described by the critical biconical model, two simple conditions must be met: $\lambda_{\phi\chi}^2 = \lambda_\phi \lambda_\chi$ and $\lambda_{\phi\chi} < 0$.

More generally, our calculation suggests a practical strategy for studying RG flows in large-$N$ theories. Applications include completing the analysis of the $O(N)\times O(M)$ biconical model~\cite{Chai:2020hnu} by determining its critical coupling values, providing an analytic treatment of the model exhibiting persistent spacetime-parity breaking~\cite{SmolkinYung:2025parity}, and addressing other large-$N$ theories, including further candidates for persistent spontaneous symmetry breaking.

\section*{Acknowledgments}

We thank Alexei Yung for insightful discussions. This work was supported in part by BSF Grant No. 2022113, ISF Grant No. 2526/25, NSF-BSF Grant No. 2022726, and Israel’s Council for Higher Education.

\pagebreak

\appendix

\section*{Appendix}
\addcontentsline{toc}{section}{Appendix}

\section{Large-momentum expansion of $\Pi_{m^2}(k)$} \label{app: Pi_large_k_expansion}
To identify the logarithmic and power-law divergences in the NLO effective potential, we evaluate the large-momentum expansion of the bubble integral:
\begin{equation}
    \Pi_{m^2} (k) = \frac{1}{2} \int_p \frac{1}{p^2 + m^2} \frac{1}{(p+k)^2 + m^2}, \quad \int_p \equiv \int \frac{d^d p}{(2\pi)^d}.
\end{equation}
This integral can be evaluated explicitly, yielding
\begin{multline}
    \Pi_{m^2}(k) = \frac{1}{2} \frac{\Gamma(2-\frac{d}{2})}{(4\pi)^{d/2}} \int_0^1 dx \left[m^2 + x(1-x)k^2\right]^{\frac{d}{2}-2} = \\=\frac{1}{2 (4\pi)^{d/2}} \Gamma\left(2 - \frac{d}{2}\right) \left(m^2 + \frac{k^2}{4}\right)^{\frac{d}{2} - 2} \,_2F_1\left(\frac{1}{2},2-\frac{d}{2}, \frac{3}{2}, \frac{k^2}{k^2 + 4m^2}\right),
\end{multline}
where $k \equiv |k|$. Inverting and expanding this expression at large momentum $k\gg m$, we obtain for $3\leq d<4$
\begin{equation}
    \begin{aligned}
\frac{1}{\Pi_{m^2}(k)}
={}&
A_0 k^{4-d}
+A_1 m^{d-2}k^{6-2d}
+A_2 m^2k^{2-d}
+A_3 m^{2d-4}k^{8-3d} + 
\\
&+
A_4 m^dk^{4-2d}
+A_5 m^4k^{-d}
+A_6 m^{3d-6}k^{10-4d}
+A_7 m^{2d-2}k^{6-3d}
+\cdots ,
\end{aligned}
\end{equation}
where the expansion terms are not strictly ordered by magnitude, as their relative scaling varies across $3 \leq d < 4$. The expansion coefficients are given by
\[
A_0
=\frac{2(4\pi)^{d/2}\Gamma(d-2)}
{\Gamma\!\left(2-\frac d2\right)\Gamma\!\left(\frac d2-1\right)^2}, \qquad
A_1=\frac{8(4\pi)^{d/2}\Gamma(d-2)^2}
{(d-2)\,\Gamma\!\left(2-\frac d2\right)\Gamma\!\left(\frac d2-1\right)^4},
\]
\[
A_2=-\frac{4(d-3)(4\pi)^{d/2}\Gamma(d-2)}
{\Gamma\!\left(2-\frac d2\right)\Gamma\!\left(\frac d2-1\right)^2}, \quad A_3=\frac{32(4\pi)^{d/2}\Gamma(d-2)^3}
{(d-2)^2\,\Gamma\!\left(2-\frac d2\right)\Gamma\!\left(\frac d2-1\right)^6},
\]
\[A_4=-\frac{32(d^2-3d+1)(4\pi)^{d/2}\Gamma(d-2)^2}
{d(d-2)\,\Gamma\!\left(2-\frac d2\right)\Gamma\!\left(\frac d2-1\right)^4}, \quad
A_5=
\frac{
4(d-3)(d-1)(4\pi)^{d/2}\Gamma(d-2)
}{
\Gamma\!\left(2-\frac d2\right)
\Gamma\!\left(\frac d2-1\right)^2
},
\]
\[A_6=
-\frac{
256(4\pi)^{d/2}\Gamma(d-2)^4
}{
(d-2)^4
\Gamma\!\left(1-\frac d2\right)
\Gamma\!\left(\frac d2-1\right)^8
}, \quad A_7=
\frac{
128(3d^2-9d+4)
(4\pi)^{d/2}\Gamma(d-2)^3
}{
d(d-2)^3
\Gamma\!\left(1-\frac d2\right)
\Gamma\!\left(\frac d2-1\right)^6
}.\]
We note that the expansion contains an additional term $\propto m^{4 d - 8} k^{12 - 5 d}$, which is subleading for $3<d<4$ but becomes of the same order as the $A_7 m^{2d-2}k^{6-3d}$ term at $d=3$. However, the coefficients $A_0$ through $A_6$ are fully sufficient for our purposes.

\section{Log divergence of $V_{\text{NLO}}$} \label{app: nlo_log_div}
We consider the logarithmically divergent part of the following integral, which arises from the $T=0$ NLO effective potential \eqref{eq: bicon_effective_potential_general_nlo_expanded} 
\begin{multline} 
    \frac{1}{2} \int_k \ln \Bigg[\Pi_{m^2}(k) (k^2 + \alpha m^2) + \frac{(\phi^2 + \alpha^2\chi^2)k^2 + \alpha(\phi^2 + \alpha\chi^2)m^2}{k^2 + m^2} \\ + \left(\Pi_{m^2}(k)+\frac{\phi^2}{k^2+m^2}\right)\frac{g}{4!}\chi^4\Bigg].
\end{multline}
Inserting the large-momentum expansion of $\Pi_{m^2}(k)$ from Appendix \ref{app: Pi_large_k_expansion} into the expression above, we expand the integrand in Mathematica for large $k$ and isolate the terms proportional to $k^{-d}$. This yields the logarithmically divergent part
\begin{equation} 
    L(\phi,\chi; \alpha, g) \ln \left(\frac{\Lambda }{m}\right),
\end{equation}
where
\begin{equation}
\begin{split}
    L(\phi,\chi;\alpha, g)=\,\,& B_1 m^d +B_2 \bigg[m^2\bigl(s_1\phi^2+s_2\chi^2\bigr) \\ &+ \delta_{d,3}            \frac{4}{3\pi^3} \left(m-4\pi(\phi^2+\alpha^2 \chi^2)\right)^3+\alpha^2\frac{g}{4!}\chi^6\bigg],
    \end{split}
\end{equation}
with the coefficients
\begin{equation}
    B_1 =
    \frac{\pi^{-\frac{d+1}{2}}\Gamma\!\left(\frac{d+1}{2}\right)}
    {\Gamma\!\left(\frac d2-1\right)\Gamma\!\left(\frac d2+1\right)}, \,\,\,\, B_2 = \frac{\Gamma(d-1)\sin\!\left(\frac{\pi d}{2}\right)}
    {\pi\,\Gamma\!\left(\frac d2\right)^2},\,\,\,\, s_1 = 2d-5,
    \,\,\,\,
    s_2 = \alpha^2(\alpha+2d-6).
\end{equation}
Substituting the relation $m^2 = \frac{2d}{d-2}\nu_d\left(\phi^2 + \alpha \chi^2\right)^{\frac{2}{d-2}}$ from \eqref{eq: bicon_m_definition} and taking $g\to g \delta_{d,3}$ yields the simplified expression \eqref{eq: bicon_effective_potential_nlo_log}. The unsimplified version remains useful when the relation for $m^2$ is modified, such as by the introduction of the quadratic mass terms $\mu_\phi^2$ and $\mu_\chi^2$ (see Section \ref{sec: beta_functions}).

\section{$V(\phi, \chi)\big|_{\phi^2 + \alpha \chi^2 = 0}$ at $3<d<4$} \label{app: integral_restricted_effective_potential}
The $T=0$ NLO effective potential \eqref{eq: bicon_effective_potential_general_nlo_expanded} restricted to the $\phi^2 + \alpha \chi^2 = 0$ locus is given by the following integral, where $x = \phi^2 + \alpha^2 \chi^2$: 
\begin{equation}
    I(x) = \frac{1}{2}\int_k \ln\left( x+ \Pi_0 (k) k^2\right) = \frac{1}{2}\int_k \ln\left( x+ k^{d-2}/A_0\right), \quad A_0=\frac{2(4\pi)^{d/2}\Gamma(d-2)}
{\Gamma\!\left(2-\frac d2\right)\Gamma\!\left(\frac d2-1\right)^2}.
\end{equation}
Differentiating $I(x)$ with respect to $x$ yields
\begin{equation}
    \begin{split}
    \frac{\partial I (x)}{\partial x} = \,\,& \frac{1}{2} \int_k \frac{1}{x+k^{d-2}/A_0} = \frac{1}{2} \frac{S_{d-1}}{(2\pi)^d} \int_0^\infty dk \frac{k^{d-1}}{x + k^{d-2}/A_0} \\[0.1cm] = \,\,&\frac{1}{2} \frac{S_{d-1}}{(2\pi)^d} \frac{A_0^{\frac{d}{d-2}}}{d-2} x^{\frac{2}{d-2}}\int_0^\infty dt \frac{t^{\frac{d}{d-2}-1}}{1+t},    
    \end{split}
\end{equation}
where $S_{d-1} = \frac{2\pi^{d/2}}{\Gamma(d/2)}$, and in the last equality, we changed the integration variable to $t = k^{d-2}/(A_0x)$. Invoking the analytic continuation of the integral identity
\begin{equation}
    \int_0^\infty dt \frac{t^{q-1}}{1+t} = \pi \csc (\pi q),
\end{equation}
we obtain
\begin{equation}
\begin{aligned}
    &\frac{\partial I (x)}{\partial x} = \frac{1}{2} \frac{S_{d-1}}{(2\pi)^d} \frac{A_0^{\frac{d}{d-2}}}{d-2} \pi \csc \left(\frac{\pi d}{d-2}\right) x^{\frac{2}{d-2}} \\&\Rightarrow I(x) = \frac{1}{2} \frac{S_{d-1}}{(2\pi)^d} \frac{A_0^{\frac{d}{d-2}}}{d} \pi \csc \left(\frac{\pi d}{d-2}\right) x^{\frac{d}{d-2}},
    \end{aligned}
\end{equation}
where we omit an $x$-independent constant of integration. Substituting the explicit expression for $A_0$, we find
\begin{equation}
    I(x) = \frac{\csc \left(\frac{\pi  d}{d-2}\right) \left(-4^{d-1} \pi ^{\frac{d-3}{2}} \sin \left(\frac{\pi  d}{2}\right) \Gamma \left(\frac{d-1}{2}\right)\right)^{\frac{d}{d-2}}}{2^{d} \pi ^{\frac{d}{2}-1}d \Gamma
   \left(\frac{d}{2}\right)} x^{\frac{d}{d-2}},
\end{equation}
where $x=(1-\alpha)\,\phi^2$ on the locus $ \phi^2 + \alpha \chi^2 = 0$.

\section{$V(\phi, \chi)\big|_{\chi = 0}$ at $d\rightarrow 4$} \label{app: integral_restricted_effective_potential_chi0_d4}
We evaluate the $T=0$ NLO effective potential \eqref{eq: bicon_effective_potential_general_nlo_expanded} at $\chi=0$ in the limit $d=4-\epsilon$ as $\epsilon\rightarrow 0$. The LO contribution simplifies to $N \nu_d \phi^{\frac{2d}{d-2}} \rightarrow N2\pi^2 \epsilon \phi^4$. The NLO contribution reads
\begin{equation}
\begin{split}
    V_{\text{NLO}}(\phi, \chi = 0) &=\frac{1}{2} \int_k \ln \Bigg[\Pi_{m^2}(k) (k^2 + \alpha m^2) + \frac{(\phi^2 + \alpha^2\chi^2)k^2 + \alpha(\phi^2 + \alpha\chi^2)m^2}{k^2 + m^2}\Bigg] \Bigg|_{\chi = 0} \\ &=\frac{1}{2}\int_k \ln (k^2 + \alpha m^2)  +\ln\left(\Pi_{m^2}(k)+\frac{\phi^2}{k^2 + m^2}\right).
    \end{split}
\end{equation}
In the limit $\epsilon \rightarrow 0$ with $\chi = 0$, we obtain
\begin{equation}
    m^2 = 8\pi^2 \epsilon \phi^2 + O(\epsilon^2),\quad \Pi_{m^2}(k) = \frac{1}{16\pi^2 \epsilon} + O(\epsilon^0),
\end{equation}
where we used the definition of $m^2$ \eqref{eq: bicon_m_definition} and the evaluation of $\Pi_{m^2}(k)$ from Appendix \ref{app: Pi_large_k_expansion}. Hence,
\begin{equation}
    \ln\left(\Pi_{m^2}(k)+\frac{\phi^2}{k^2 + m^2}\right) = \ln \frac{1}{16\pi^2 \epsilon} + \ln \left(1 + \frac{2}{p^2 + 1}\right) = \text{const} + \ln \left(\frac{p^2 + 3}{p^2 + 1}\right), 
\end{equation}
where $p = k/m$. The constant can be dropped since it yields only a pure power-law divergence. Changing the integration variable to $p = k/m$, we find
\begin{equation}
\begin{split}
    V_{\text{NLO}}(\phi, \chi = 0)  \stackrel{\epsilon\rightarrow 0}{=} & \,\,\frac{m^d}{2}\int_p \ln (p^2 + \alpha)  + \ln(p^2 + 3) - \ln (p^2+1)  \\ = & \,\,- \frac{\Gamma(-\frac{d}{2})}{2(4\pi)^{d/2}}m^d (\alpha^{d/2}+3^{d/2}-1) \stackrel{\epsilon\rightarrow 0}{=} -34 \pi^2 \epsilon \phi^4,
\end{split}    
\end{equation}
where $\alpha = -3 + O(\epsilon)$. Note that the imaginary part appears only at $O(\epsilon^2)$. In the above, we used the standard integral 
\begin{equation}
    \frac{1}{2} \int_k \ln(k^2 + m^2) =- \frac{\Gamma\left(-\frac{d}{2}\right)}{2(4\pi)^{d/2}} (m^2)^{d/2}.
\end{equation}

\section{$\Pi_{m^2}(\Omega, q)$ at $T\neq 0$}
We consider the finite-temperature bubble 
\begin{equation}
\begin{split}
    \Pi(\Omega, q) = & \,\,\frac{1}{2} \sum_K \frac{1}{(K-P)^2 + m^2} \frac{1}{K^2 + m^2} \\ \equiv & \,\,\frac{1}{2} T \sum_{n\in\mathbb{Z}} \int \frac{d^{d-1} k}{(2\pi)^{d-1}}  \frac{1}{(\omega_n-\Omega)^2 + (k-q)^2 + m^2} \frac{1}{\omega_n^2 +  k^2 + m^2}, 
    \end{split}
\end{equation}
where $P = (\Omega, q)$, $K = (\omega_n, k)$, and the Matsubara frequencies are $\omega_n = 2\pi n T$. While the finite-temperature behavior of $\Pi(\Omega, q)$ in $d=3$ was analyzed in Appendix B of \cite{Diatlyk:2023msc}, we extend this analysis here to general dimensions. Evaluating the Matsubara sum, we obtain for $\Omega \in 2\pi T \mathbb{Z}$
\begin{multline}
    \Pi(\Omega, q) = \int\frac{d^{d-1}k}{(2\pi)^{d-1}} \frac{1}{4\omega_k \omega_{k+q}}\Bigg(\frac{\omega_{k+q}+\omega_k}{\Omega^2+(\omega_{k+q}+\omega_k)^2} (n_k+ n_{k+q}+1) \\ +\frac{\omega_{k+q} - \omega_k}{\Omega^2+(\omega_{k+q}-\omega_k)^2}(n_k-n_{k+q}) \Bigg),
\end{multline}
where $\omega_k = \sqrt{k^2 + m^2}$, $n_k = 1/(e^{\omega_k/T}-1)$ is the Bose-Einstein distribution. 
We split this expression into a non-thermal and a purely thermal part,
\begin{equation}
    \Pi(\Omega, q) = \Pi_{T=0}(\Omega, q) + \delta\Pi_T(\Omega, q).
\end{equation}
The non-thermal part is the one that does not contain the Bose-Einstein distribution. All other terms that involve $n_k$ or $n_{k+q}$ constitute the purely thermal part. By performing consecutive changes of integration variables, $k\to k-q$ followed by $k\to -k$, the terms proportional to $n_{k+q}$ can be combined with the $n_k$ terms, yielding 
\begin{equation}
\begin{aligned}
    &\Pi_{T=0}(\Omega, q) =  \int\frac{d^{d-1}k}{(2\pi)^{d-1}} \frac{1}{4\omega_k \omega_{k+q}}\frac{\omega_{k+q}+\omega_k}{\Omega^2+(\omega_{k+q}+\omega_k)^2}, \\ &\qquad\,\,\,\,\, \delta\Pi_T(\Omega, q) = \int\frac{d^{d-1}k}{(2\pi)^{d-1}}\frac{n_k}{\omega_k} \frac{A}{A^2 + 4\Omega^2 \omega_k^2},
    \end{aligned}
\end{equation}
where we have defined $A \equiv \Omega^2 + \omega_{k+q}^2 - \omega_k^2$. 
The non-thermal part is evaluated in Appendix \ref{app: Pi_large_k_expansion}, namely
\begin{equation} \label{eq: Pi0_raw}
    \Pi_{T=0}(P) =\frac{1}{2 (4\pi)^{d/2}} \Gamma\left(2 - \frac{d}{2}\right) \left(m^2 + \frac{P^2}{4}\right)^{\frac{d}{2} - 2} \,_2F_1\left(\frac{1}{2},2-\frac{d}{2}, \frac{3}{2}, \frac{P^2}{P^2 + 4m^2}\right),
\end{equation}
which is a function of $P^2=\Omega^2 + q^2$. For the thermal part, applying the identity
$$A \equiv \Omega^2 + \omega_{k+q}^2 - \omega_k^2 = \Omega^2 + q^2 + 2  k\cdot q = P^2 + 2q\cdot k$$
allows us to rewrite the integral as 
\begin{equation} \label{eq: deltaPi_xy_representation}
    \delta \Pi_T(\Omega, q) = \frac{1}{P^2} \int\frac{d^{d-1}k}{(2\pi)^{d-1}}\frac{n_k}{\omega_k} \frac{1+x}{(1+x)^2 + y}, \quad x \equiv \frac{2q\cdot k}{P^2},\,\,\, y \equiv \frac{4\Omega^2 \omega_k^2}{P^4},
\end{equation}
where $q\cdot k$ is the scalar product of two $(d-1)$-dimensional vectors. For the large-momentum expansion, we use the formula above. However, $\delta \Pi_T(\Omega, q)$ can also be rewritten in a form more useful for numerical evaluations by using the identity
\begin{equation}
    \frac{A}{A^2 +4 \Omega^2 \omega^2} = \Re \frac{1}{A - 2i\Omega \omega} = \Re \frac{1}{P^2 + 2qk\cos\theta -i2\Omega \omega},
\end{equation}
where $\theta$ is the angle between $k$ and $q$. Furthermore, in $D=d-1$ dimensions, the angular integral yields
\begin{equation}
    \int d\Omega_{D-1} \frac{1}{a + b \cos \theta} = S_{D-2} \int_{-1}^1 d \mu (1-\mu^2)^{\frac{D-3}{2}} \frac{1}{a+b\mu} = \frac{S_{D-1}}{a}\,\,_2 F_1\left(1, \frac{1}{2}, \frac{D}{2}, \frac{b^2}{a^2}\right). 
\end{equation}
Hence, setting $a = P^2 - 2i\Omega \omega$ and $b = 2qk$, we find 
\begin{multline} \label{eq: deltaPiT_for_numerics}
    \delta \Pi_T(\Omega, q) = \int\frac{d^{d-1}k}{(2\pi)^{d-1}}\frac{n}{\omega} \frac{A}{A^2 + 4\Omega^2 \omega^2} = \\ =\frac{S_{d-2}}{(2\pi)^{d-1}} \int_0^\infty dk k^{d-2} \frac{n}{\omega} \Re \left[\frac{1}{P^2 -i2\Omega \omega} \,\,_2 F_1\left(1, \frac{1}{2}, \frac{d-1}{2}, \frac{4q^2 k^2}{(P^2 -i2\Omega \omega)^2}\right)\right],\\ \quad \omega = \sqrt{k^2 + m^2},\quad n = \frac{1}{e^{\omega/T}-1}.
\end{multline}

\subsection{Large-momentum expansion} \label{app: Pi_large_k_expansion_T}
We consider the large-momentum $P=(\Omega, q)$ behavior of \eqref{eq: deltaPi_xy_representation}. Since the integration momentum $k$ in the thermal part is effectively bounded by the temperature scale $T$, we can perform a Taylor expansion of the integrand taking $P \gg k$, which yields
\begin{equation}
    \frac{1+x}{(1+x)^2 + y} = \frac{1}{1+x + y/(1+x)} = 1 - x + x^2 - y -x^3 + 3xy + O\left(P^{-4}\right).
\end{equation}
Since odd powers of $x \sim q \cdot k$ vanish upon integration, we discard them. Substituting the remaining expansion terms into the momentum integral yields
\begin{equation}\label{eq: thermal_expansion_three_terms}
\begin{aligned}
    &\frac{1}{P^2}\int\frac{d^{d-1}k}{(2\pi)^{d-1}}\frac{n_k}{\omega_k} \equiv \frac{T^{d-2}}{P^2} I_0 (\Delta), \\ &\frac{1}{P^2}\int\frac{d^{d-1}k}{(2\pi)^{d-1}}\frac{n_k}{\omega_k} (-y) = -\frac{4\Omega^2}{P^6} \int\frac{d^{d-1}k}{(2\pi)^{d-1}}n_k\omega_k \equiv -\frac{4\Omega^2 T^{d}}{P^6} I_1 (\Delta),\\
    &\frac{1}{P^2}\int\frac{d^{d-1}k}{(2\pi)^{d-1}}\frac{n_k}{\omega_k}  x^2 = \frac{4q^2}{P^6}\int\frac{d^{d-1}k}{(2\pi)^{d-1}}\frac{n_k}{\omega_k} \frac{k^2}{d-1}=\\ &\qquad\qquad\qquad\qquad\qquad\qquad\qquad=\frac{4q^2 T^d}{P^6}\frac{1}{d-1} I_1(\Delta) - \frac{4q^2 T^{d-2}m^2}{P^6}\frac{1}{d-1} I_0 (\Delta),
\end{aligned}
\end{equation}
where $\Delta = m/T$, and the thermal integrals are defined as
\begin{equation}
\begin{aligned}
    & \,\,\,I_0 (\Delta) = \frac{S_{d-2}}{(2\pi)^{d-1}} \int_\Delta^\infty du \, (u^2 - \Delta^2)^{\frac{d-3}{2}} \frac{1}{e^u - 1},\\ & I_1 (\Delta) = \frac{S_{d-2}}{(2\pi)^{d-1}} \int_\Delta^\infty du\, (u^2 - \Delta^2)^{\frac{d-3}{2}} u^2 \frac{1}{e^u - 1}, 
    \end{aligned}
\end{equation}
Here, $S_{d-1} = \frac{2\pi^{d/2}}{\Gamma(d/2)}$. Combining the three terms \eqref{eq: thermal_expansion_three_terms}, we obtain
\begin{multline}
    \delta\Pi_T(\Omega,q) = \frac{T^{d-2}}{P^2} I_0(\Delta) + \frac{4(q^2/(d-1) - \Omega^2)T^d}{P^6} I_1(\Delta) \\- \frac{4 q^2 T^{d-2} m^2}{P^6} \frac{1}{d-1} I_0(\Delta) + O(P^{-6}).
\end{multline}

The $T=0$ contribution \eqref{eq: Pi0_raw} admits the following large-momentum expansion for $2<d<4$
\begin{multline}
    \Pi_{T=0}(P) = \frac{1}{2} \frac{\Gamma(2-\frac{d}{2})}{(4\pi)^{d/2}} \Bigg[\frac{\Gamma(\frac{d}{2} - 1)^2}{\Gamma(d-2)} P^{d-4} \left(1 + 2(d-3)\frac{m^2}{P^2} + 2(d-3)(d-5)\frac{m^4}{P^4}\right)  \\ - \frac{4}{d-2} \frac{m^{d-2}}{P^2} + \frac{16}{d(d-2)} \frac{m^d}{P^4}\Bigg] + O(P^{-6}).
\end{multline}
We note the overall coefficient of $P^{-2}$ in the resulting expansion for $\Pi(\Omega, q)$:
\begin{equation}
\begin{split}
    \Pi(\Omega, q) = & -\frac{\Gamma(d/2)\Gamma(1-d/2)} {4^{d-1}\pi^{\frac{d-1}{2}} \Gamma((d-1)/2)} P^{d-4} \\&+ \left(\frac{\Gamma(1-d/2)}{(4\pi)^{d/2}} m^{d-2} + T^{d-2} I_0(\Delta)\right) P^{-2} +\, O(P^{d-6}), \quad 2 < d<4.
    \end{split}
\end{equation}
This coefficient is exactly the finite part of the following sum-integral:
\begin{equation}
    \sum_K \frac{1}{K^2 + m^2}
    = \frac{\Gamma(1-d/2)}{(4\pi)^{d/2}} m^{d-2} + T^{d-2} I_0(\Delta).
\end{equation}
Using the self-consistency equation for $\Sigma_{\eta_2}^{\text{LO}} \equiv m^2$,
\begin{equation}
    m^2 = \frac{\lambda_\phi}{2}\sum_K \frac{1}{K^2 + m^2} + \frac{\lambda_\phi}{2} (\phi^2 + \alpha \chi^2),
\end{equation}
we see that in the IR limit $\lambda_\phi \to \infty$, the coefficient of $P^{-2}$ in the $\Pi(\Omega, q)$ large-momentum expansion is purely field-dependent, and this term takes the form $-(\phi^2 + \alpha\chi^2)P^{-2}$.

\section{Positivity of $\frac{1}{2} \sum_K \frac{1}{\Pi(K) K^2 + x}$ at $T\neq 0$} \label{app: positivity_derivative_VNLO_m0}
The positivity of 
\begin{equation}
    F(x) = \frac{1}{2} \sum_K \frac{1}{\Pi(K) K^2 + x}, \quad x\geq 0,
\end{equation}
implies the lifting of the flat direction at NLO for $3<d<4$. As demonstrated below, the sum-integral on the right is free of logarithmic divergences; therefore, following the convention used throughout this paper, $F(x)$ is defined as the finite part of the sum-integral with all pure power-law divergent terms subtracted. Any twice-differentiable function $F(x)$ admits the following representation
\begin{equation} \label{eq: F_integral}
    F(x) = F(0) + x F'(0) + \int_0^x dy\,(x-y) F''(y).
\end{equation}
In our case, 
\begin{equation}
    F(0) = \frac{1}{2} \sum_K \frac{1}{\Pi(K) K^2}, \quad F'(0) = -\frac{1}{2} \sum_K \frac{1}{(\Pi(K) K^2)^2}, \quad F''(x) = \sum_K \frac{1}{(\Pi(K) K^2 + x)^3}.
\end{equation}
Since $x\geq 0$ and $\Pi(K) \geq 0$, the integrand of $F''(x)$ is non-negative for all $K$. As confirmed shortly, $F''(x)$ is free of UV divergences, implying that no subtraction is required, and the non-negative integrand therefore gives $F''(x) > 0$. Consequently, the integral in \eqref{eq: F_integral} is positive. Therefore, to prove the positivity of $F(x)$ for all $x \ge 0$, it is sufficient to show that $F(0) \geq 0$ and $F'(0) \geq 0$. We evaluate $F(0)$ and $F'(0)$ numerically as functions of $d$. However, both quantities contain power-law divergences that must be canceled by appropriate counterterms in the UV action; in our numerical evaluation, we identify and explicitly subtract these divergences. Using the large-momentum expansion from Appendix \ref{app: Pi_large_k_expansion_T}, we have for $m^2=0$
\begin{equation}
\begin{split}
    \Pi(K) = A_0^{-1} K^{d-4} + T^{d-2} I_0 (0)K^{-2} + O(K^{-4}),\qquad \\ K^2 \Pi(K) + x = A_0^{-1} K^{d-2} + x+ T^{d-2} I_0 (0)+ O(K^{-2}).
    \end{split}
\end{equation}
where 
\begin{equation}
    I_0 (0) = \frac{1}{2} \pi ^{-d/2} \zeta (d-2) \Gamma \left(\frac{d-2}{2}\right).
\end{equation}
Hence, expanding the integrand of $F(x)$ yields
\begin{equation}
    \frac{1}{\Pi(K) K^2 + x} = A_0 K^{2-d} - A_0^2 K^{4-2d}\left(x+ T^{d-2} I_0 (0)\right) + \dots ,
\end{equation}
where the terms on the right contribute to power-law divergences and dots denote UV-convergent terms. The divergences are at most linear in $x$, so they do not contribute to $F''(x)$. One might be concerned that the subleading divergence appears temperature-dependent, while temperature cannot affect UV divergences. However, as noted at the end of Appendix \ref{app: Pi_large_k_expansion_T}, the $K^{-2}$ prefactor in the large-momentum expansion of $\Pi(\Omega, q)$ is actually a purely field-dependent expression. Specifically, the self-consistency equation for $m$ shows that $T^{d-2}I_0(0)\big\vert{}_{m=0}=-(\phi^2 + \alpha \chi^2)$.

\begin{figure}[h]
    \centering
    \includegraphics[width=0.5\linewidth]{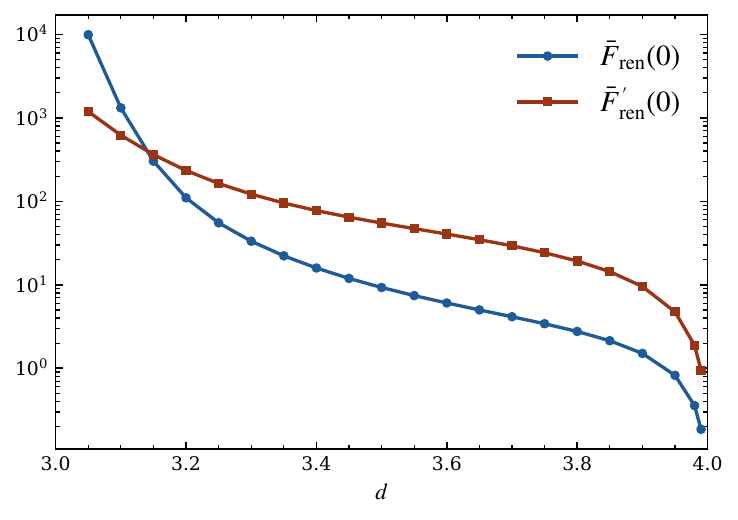}
    \caption{Finite, renormalized values of the dimensionless coefficients $\bar{F}(0)$ and $\bar{F}'(0)$ as functions of the dimension $d$. Both quantities are found to be positive across the interval $3 < d < 4$. A hard cutoff $\Lambda/T = 2000$ was used; further increasing $\Lambda$ does not produce a significant change in the results.}
    \label{fig: F0_Fprime0_vs_d}
\end{figure}

First, we factor out the temperature dependence by defining $F(0) = T^2 \bar F(0)$ and $F'(0) = T^{4-d} \bar F'(0)$. We then evaluate these dimensionless quantities numerically using the thermal bubble $\Pi(K)$ from \eqref{eq: Pi0_raw} and \eqref{eq: deltaPiT_for_numerics}. To regularize the sum-integrals, we implement a hard momentum cutoff $K^2 \leq \Lambda^2$, and by explicitly subtracting the pure power-law $\Lambda$ divergences, we extract the finite renormalized values. We plot $\bar F(0)$ and $\bar F'(0)$ across $3< d < 4$ in Figure \ref{fig: F0_Fprime0_vs_d}. The graph indicates that both values remain strictly positive throughout this range, thereby demonstrating that $F(x) > 0$ for all $x\geq 0$.

\bibliographystyle{JHEP.bst}
\bibliography{bibliography.bib}

\end{document}